\documentclass[final,authoryear,3p,times,11pt]{elsarticle}
\usepackage{float}
\usepackage{graphicx, color,colortbl}
\usepackage{enumerate}
\usepackage[colorlinks]{hyperref}
\usepackage{amsmath,amssymb,amsthm,bm}
\usepackage{multicol}
\usepackage[dvipsnames]{xcolor}
\usepackage{ulem}
\usepackage{tikz}
\usetikzlibrary{automata,positioning}
\usepackage{afterpage}
\usepackage{calc}
\usepackage{ifthen}
\usepackage{algorithm}
\usepackage{algpseudocode}
\usepackage{caption}
\usepackage{subcaption}
\usepackage{siunitx}
\usepackage{booktabs}
\usepackage[nameinlink,capitalise]{cleveref}
\usepackage{multirow}
\setcitestyle{numbers,square,sort&compress}
\newcommand{\eqn}[2]{\begin{equation}
\label{#1}
{#2}
\end{equation}}

\newcommand{\bs}[1]{\boldsymbol{#1}}

\edef\svtheparindent{\the\parindent}
\usepackage{parskip}
\parindent=\svtheparindent\relax

\newcommand{\boldface}[1]{\boldsymbol{#1}}  % italic (slanted)

\newcommand{\bft}{\boldface{t}}
\newcommand{\bfu}{\boldface{u}}
\newcommand{\bfv}{\boldface{v}}

\newcommand{\bfF}{\boldface{F}}

\newcommand{\bfI}{\boldface{I}}

\newcommand{\bfP}{\boldface{P}}
\newcommand{\bfQ}{\boldface{Q}}

\newcommand{\bftheta}{\boldsymbol{\theta}}

\newcommand{\partderiv}[2]{\frac{\partial #1}{\partial #2}}

\newcommand{\Rset}{\mathbb{R}}

\newlength{\boxwidth}
\def\dd{\;\!\mathrm{d}}

\def\btheorem{\begin{theorem}}
\def\etheorem{\end{theorem}}
\def\blemma{\begin{lemma}}
\def\elemma{\end{lemma}}
\def\bproposition{\begin{proposition}}
\def\eproposition{\end{proposition}}
\def\bcorollary{\begin{corollary}}
\def\ecorollary{\end{corollary}}
\def\bdefinition{\begin{definition}}
\def\edefinition{\end{definition}}
\def\bexample{\begin{example}}
\def\eexample{\end{example}}
\def\bremark{\begin{remark}}
\def\eremark{\end{remark}}

\newcommand{\be}{\begin{equation}}
\newcommand{\ee}{\end{equation}}
\newcommand{\beq}{\begin{eqnarray}}
\newcommand{\eeq}{\end{eqnarray}}
\newcommand{\bem}{\begin{multline}}
\newcommand{\eem}{\end{multline}}
\newcommand{\ba}{\begin{align}}
\newcommand{\ea}{\end{align}}

\newcommand{\norm}[1]{\left\lVert#1\right\rVert}

\definecolor{ForestGreen}{RGB}{34,139,34}
\definecolor{InternationalOrange}{rgb}{1.0, 0.31, 0.0}
\definecolor{WineRed}{RGB}{139,0,0}

\journal{Experimental Mechanics}

\begin{document}

\begin{frontmatter}

\title{Discovery of Hyperelastic Constitutive Laws from Experimental Data with EUCLID}

\author[eth]{Arefeh Abbasi \corref{cor1}}
\author[roma3]{Maurizio Ricci \corref{cor1}}
\author[eth]{Pietro Carrara}
\author[eth]{Moritz Flaschel}
\author[tud]{Siddhant Kumar}
\author[roma3]{Sonia Marfia}
\author[eth]{Laura De Lorenzis \corref{cor2}}

%\corref{cor2}
\cortext[cor1]{These authors contributed equally.}
\cortext[cor2]{Correspondence: L. De Lorenzis, ldelorenzis@ethz.ch}

\address[eth]{Department of Mechanical and Process Engineering, ETH Z\"{u}rich, 8092 Z\"{u}rich, Switzerland \\ e-mail: abbasia@ethz.ch, pcarrara@ethz.ch, moritz.flaschel@fau.de, ldelorenzis@ethz.ch}
\address[roma3]{Department of Civil, Computer Science and Aeronautical Technologies Engineering, Roma Tre University, Italy \\ e-mail: maurizio.ricci@uniroma3.it, sonia.marfia@uniroma3.it}
% \address[wias]{Weierstrass Institute for Applied Analysis and Stochastics, Germany \\ e-mail: moritz.flaschel@wias-berlin.de}
\address[tud]{Department of Materials Science and Engineering, Delft University of Technology, 2628 CD Delft, The Netherlands \\ e-mail: sid.kumar@tudelft.nl}

%%%%%%%%%%%%%%%%%%%%%%%%%%%%%%%%%%%%%%%%%%%%%%%%%%%%%%%%%%%%%%%
\begin{abstract}
\textbf{Background}:
Determining accurate constitutive laws from experimental data remains a key challenge in mechanics, particularly when the material behavior is nonlinear and the dataset is limited or noisy. Traditional approaches rely on identifying parameters of preselected material models, which separates the model selection and the calibration tasks leading to a potentially long and tedious trial-and-error procedure.
\textbf{Objective}:
This work aims to assess the performance of EUCLID (Efficient Unsupervised Constitutive Law Identification and Discovery), a recently proposed framework for automated discovery of constitutive laws, when applied to experimental data.
\textbf{Methods}:
Mechanical tests are conducted on natural rubber specimens of varying geometrical complexity. Both global (force–elongation) and local (full-field displacement) data are collected. Constitutive laws are obtained via two routes: (i) conventional parameter identification in a priori selected models, and (ii) EUCLID, which integrates model selection and parameter identification in a unified model discovery pipeline.
\textbf{Results}:
The two approaches are compared in terms of predictive accuracy, generalization to unseen geometries, and robustness to experimental noise. The coverage of the material state space achieved by each dataset is quantified, and the relative performance of different datasets and models is analyzed.
\textbf{Conclusions}:
EUCLID enables automated and data-driven discovery of constitutive laws, offering improved flexibility compared to conventional identification methods, and showing strong potential for reliable material modeling based on experimental data.

\end{abstract}

\end{frontmatter}
%%%%%%%%%%%%%%%%%%%%%%%%%%%%%%%%%%%%%%%%%%%%%%%%%%%%%%%%%%%%%%%%%%%%%%%%%%%
\section{Introduction}
\label{sct:intro}

Material models are an essential ingredient of any engineering simulation aiming at the prediction of mechanical phenomena. Thus, their accurate determination is of great relevance across a wide range of engineering applications. For hyperelasticity, where the material response is specified via a strain energy density function, a variety of models have been proposed, ranging from phenomenological to physically- and statistically-based approaches \citep{Ogden2004,marckmann_comparison_2006,Ricker2023}. Among the most widely used are the generalized Mooney-Rivlin \citep{Mooney1940,Rivlin1948}, the Gent-Thomas \citep{gent_forms_1958} and the generalized Ogden \citep{Ogden1972} models.
The conventional approach to finding a suitable material model requires pre-selecting its functional form (a task often referred to as \textit{model selection}) and determining its unknown parameters (i.e., \textit{parameter identification}) by minimization of a suitable objective function, which quantifies the discrepancy between experimentally measured quantities and their counterparts as predicted by the model.

Different parameter identification approaches differ primarily in the choice of this objective function. Some of them rely on \textit{global} measured quantities (such as forces and elongations); in order to convert these quantities into stress and strain measurements, they require simple experiments, e.g., uniaxial tensile, pure shear, or equibiaxial tensile tests, interpreted via analytical ideal models. Other methods rely on \textit{local} measurements, i.e. full displacement fields obtained through digital image or volume correlation, along with reaction forces. Among these are Finite Element Model Updating (FEMU), the Equilibrium Gap Method (EGM), the Virtual Fields Method (VFM), and the constitutive equation gap method (CEGM). For comprehensive overviews we refer to \cite{avril2008overview,Roux2020,fuhg2024review, Roemer}.

 %Also, systematic errors may arise if ideal loading conditions are imperfectly applied \citep{pierron2012virtual, pierron2021towards}, while a-priori assumptions on the constitutive law functional form introduce uncertainties, approximations and subjective bias in interpreting the material data \citep{carrara2020data, carrara2021data}.

The a priori selection of a functional form for a material model introduces a subjective bias and potentially leads to costly trial-and-error procedures until a suitable form is found. Recognizing this issue, recent research in the field has challenged the conventional paradigm with different ideas. Model-free approaches give up constitutive laws altogether and replace them with discrete material observations \citep{kirchdoerfer2016data, conti2018data, nguyen2018data,carrara2020data}. Other approaches substitute the interpretable functional form with a more flexible, but not interpretable ansatz based on neural networks or other parameterized architectures from machine learning \citep{huang2020, zhong2022explainable, thakolkaran2022nn, xu2023small, fuhg2024review}, often augmented through the enforcement of physics constraints \citep{fuhg2024review}. 
A third approach is known as EUCLID, which stands for Efficient Unsupervised Constitutive Law Identification and Discovery \citep{flaschel2021unsupervised}. This method unifies model selection and parameter identification (whereby the combination of the two tasks is denoted as \textit{model discovery}) by automatically selecting the most suitable material model from a wide library of possible candidates through sparse regression. This eliminates subjective bias and the need for a trial-and-error procedure, while preserving interpretability \citep{flaschel2021unsupervised,flaschel2022discovering,marino2023automated,flaschel2023automated,flaschel2023,joshi2022bayesian}. Most recently, the automatic creation of the material model library through grammars has also been pursued \citep{kissas_language_2024} and alternatives to the LASSO algorithm for sparse regression have been explored \citep{linka_best--class_2024,flaschel_non-smooth_2025,urrea}. 

In principle, the choice of the ansatz for model selection, i.e. for the potential functional form of the model (e.g. a single function, a neural network, or a material model library) and the choice of the objective function for the parameter identification (e.g. based on \textit{global} or \textit{local} data) are completely independent and can be combined in multiple ways. Early studies on EUCLID advocate the use of \textit{local} data, i.e. displacement fields and the corresponding reaction forces, and adopt for parameter identification the same objective function of the EGM (in turn a special case of the VFM). Also, these early studies use data generated synthetically through finite element (FE) simulations with the addition of noise. A more recent contribution applies EUCLID to experimental data on human brain tissues  \citep{flaschel2023tissue}. In this study, the objective function is based on \textit{global} data obtained from uniaxial compression and torsion tests.

In this work, we assess the performance of EUCLID to discover the hyperelastic constitutive law of natural rubber. An experimental campaign is performed on different specimen geometries, generating both \textit{global} and \textit{local} data. Our main objective is twofold: on the one hand, we aim to compare the conventional paradigm (manual model selection and parameter identification) with the new paradigm (material model discovery based on EUCLID) on new experimental data. On the other hand, we intend to assess the influence of using global vs. local data on the performance of both paradigms. By using different specimen shapes, we also assess the ability of the obtained models to generalize to unseen geometries and stress states as well as the accuracy in reproducing both local and global behaviors. Additionally, we quantify the experimental noise and investigate the coverage of the material state space achieved with the global vs. local datasets.
%The role of the specimen geometry in covering the state space of the material is also investigated along with the ability of the EUCLID approach to automate the selection of optimal hyperelastic models without requiring time-consuming iterative procedures. 
%This comparison, based on both supervised and unsupervised experimental data, highlights the advantages of the EUCLID framework over traditional fixed-form constitutive law calibration.
%
%Complementary numerical studies were also conducted to investigate the influence of data quality and quantity on the accuracy of the discovered material model. Moreover, the role of geometric complexity was examined, revealing its impact on the reliability of identification and the potential to reduce the number of required experiments.
%Finally, FE simulations using the material model identified by EUCLID examined the versatility and generalizability of the framework, supporting its potential to tackle key challenges in the data-driven discovery of constitutive laws.

The remainder of this paper is structured as follows. In Section~\ref{euc}, we present an overview of EUCLID, outlining the structure of the material model library and the methodologies used for discovery from global and from local data. Section~\ref{sec:Exp_results} describes the experimental campaign and the data processing. Section~\ref{sec:Num_results} discusses the performance of both the conventional paradigm and EUCLID based on the experimental data. Finally, the key findings and their implications are summarized in section~\ref{sec:con}.

%%%%%%%%%%%%%%%%%%%%%%%%%%%%%%%%%%%%%%%%%%%%%%%%%%%%%%%%%%%%%%%%%%%%%%%%%%%%%%%%%%%%%%%%%%%
\section{Overview of EUCLID}
\label{euc}

This section provides a brief overview of EUCLID as proposed in \cite{flaschel2021unsupervised} and \cite{flaschel2023automated}.

%%%%%%%%%%%%%%%%%%%%%%%%%%%%%%%%%%%%%%%%%%%
\subsection{Problem setting and model library} 
\label{sec:problem_setting}
Let us consider the undeformed domain $\Omega\subset\Rset^2$  with boundary $\partial\Omega$, representative of a specimen made of an incompressible, isotropic and hyperelastic material. The test setup complies with plane-stress conditions and involves a quasi-static loading procedure. The undeformed boundary has unit outward normal $\bs{N}$ and is composed of a Dirichlet boundary $\partial\Omega_u$, with imposed displacements $\bs{\hat u}$, and a complementary Neumann boundary $\partial\Omega_f$, with imposed tractions $\bs{\hat t}=\bs{0}$ (for a displacement-controlled test).
In the following, the displacement and the deformation gradient are indicated as $\bs{u}$ and $\mathbf{F}=\mathbf{I}+\partial{\bs{u}}/\partial{\bs{X}}$, respectively, with $\bs{X}$ as the reference coordinate, while the right Cauchy–Green deformation tensor is $\mathbf{C} = \mathbf{F}^\mathsf{T} \mathbf{F}$. Also, $\mathbf{P}$ stands for the first Piola-Kirchhoff stress tensor, which is related to the deformation gradient by the hyperelastic constitutive law $\mathbf{P} = \partial W(\mathbf{F}) / \partial \mathbf{F}$, where $W(\mathbf{F})$ is the hyperelastic strain energy density. A sufficient condition for objectivity is that the strain energy density depends on $\mathbf{F}$ through $\mathbf{C}$, and for isotropic materials we can further express it as a function of the invariants of $\mathbf{C}$, i.e. 
\eqn{eq:invariants}{I_1(\mathbf{C}) = \mathrm{tr}(\mathbf{C})\,, \quad I_2(\mathbf{C}) = \frac{1}{2}[\mathrm{tr}(\mathbf{C})^2 - \mathrm{tr}(\mathbf{C}^2)]\,\quad  \text{and} \quad I_3(\mathbf{C}) = J^2=\det(\mathbf{C})\,,}
\noindent where $\mathrm{tr}(\bullet)$ and $\det(\bullet)$ stand respectively for trace and determinant of $(\bullet)$, while $J = \det(\mathbf{F})$ is the scalar Jacobian. Assuming incompressibility, we further impose $J = 1$.

During the test, the specimen is monitored by recording its displacement field using a digital image correlation (DIC) system, which measures the displacement vector of $n$ points in the domain $\Omega$  along the in-plane directions $X_i$, with $i=1,2$. The discrete set of measured displacement degrees of freedom is denoted as $D = \{ (d, i) : d = 1,...,n; i = 1,2 \}$ and is partitioned into two subsets: the set of free degrees of freedom, $D_{\text{free}}$, and the set of degrees of freedom affected by Dirichlet boundary conditions, $D_{\text{disp}}$, such that $D_{\text{free}} \cup D_{\text{disp}} = D$. The reaction forces along either of the in-plane directions are measured by means of $n_{\ell}$ load cells connected to portions of the fixed boundary, whose degrees of freedom are collected in the subsets $D_{\alpha}\subseteq D_{\text{disp}}$ with $\alpha=1,...,n_{\ell}$. Each subset $D_{\alpha}$ corresponds to a single coordinate direction, i.e. it is related to $X_i$ with either $i=1$ or $i=2$. %\footnote{To simplify the formulation, we assume that all the load cells are aligned with one of the coordinate directions, however it is straightforward to remove this assumption.}. 
Also, the imposed machine displacement $\bs{\hat u}$ is recorded during the tests.

Given the above set of experimental data, we aim at determining the hyperelastic material model that best describes the behavior of the tested material. To this end, EUCLID relies on a wide library of hyperelastic models, i.e. of strain energy density functions, from which it performs at the same time model selection and parameter identification. The behavior of a large class of rubbers can be described by means of the generalized Mooney-Rivlin (GMR) and Gent-Thomas (GT) models \citep{Mooney1940,Rivlin1948,gent_forms_1958}, which read
\begin{equation}
\label{eq:GMR_GT}
W_{GMR}(I_1,\,I_2)=\sum_{j=1}^{n_{GMR}}\sum_{i=0}^j\theta_{i,j}\left[(I_1-3)^i(I_2-3)^{j-i}\right]
\end{equation}
and
\begin{equation}
W_{GT}(I_1,\,I_2)=\theta_{GT,1}(I_1-3)+\theta_{GT,2} \ln \left(I_2 / 3\right)
\label{eq:GT}
\end{equation}
respectively. 
%In \eqref{eq:GMR_GT}, $\theta_{i,j}$  are the material parameters related to the GMR model, while $(\theta_{GT,1},\,\theta_{GT,2})$ are those related to the GT one. 
We collectively denote the set of models belonging to this class as $W_I(I_1,\,I_2;\, \bs{\theta}_{I})$, where the symbol $;$ is used to separate the variables (i.e., $I_1,\,I_2$) from the vector collecting the material parameters $\bs{\theta}_I$. Exploiting the linearity in $\bs{\theta}_I$, we conveniently rewrite this set of models as
\begin{equation}\label{eq:ansatz}
W_I(I_1,I_2;\, \bs{\theta}_I) = \bfQ_I^T (I_1,I_2) \,\bftheta_I,
\end{equation}
where $\bfQ_I$ is a vector containing a set of $n_I$ linear and nonlinear basis functions defined as 
\begin{equation}\label{eq:features}
\bfQ_I (I_1,I_2)= 
\underbrace{\left[(I_1-3)^i(I_2-3)^{j-i} : j\in \{1,\dots,n_{GMR}\}, i\in\{0,\dots,j\}\right]^T}_{\text{generalized Mooney-Rivlin features}}
\oplus 
\underbrace{\left[ \ln \left(I_2 / 3\right)\right]}_{\text{Gent-Thomas model}},
\end{equation}
\noindent where $\oplus$ denotes vector concatenation. The first group of functions expresses the GMR models  %\citep{hartmann_numerical_2001, marckmann_comparison_2006, bower_applied_2009}, 
while, considering that the term $\theta_{GT,1}(I_1-3)$ is included in the GMR expression for $i=j=1$, the logarithmic feature introduces the GT model. 
In this work we use $n_\text{GMR}=5$, resulting in a total of $n_I=21$ candidate models and related parameters, i.e. $\bs{\theta}_I \in\Rset^{n_I}$.

Another large class of isotropic hyperelastic energy densities is described by the Ogden models. In this case, the strain energy density is expressed as a function of the eigenvalues of the right stretch tensor $\mathbf{U}=\sqrt{\mathbf{F}^\mathsf{T} \mathbf{F}}$, namely of the principal stretches $\lambda_1, \lambda_2, \lambda_3$ \citep{Ogden1972,holzapfel2002nonlinear}. The Ogden strain energy density reads
\begin{equation}\label{eq:ogden}
{W}_O(\lambda_1,\, \lambda_2,\, \lambda_3) = \sum_{i=1}^{n_\lambda} \frac{2\mu_i}{\beta_i^2} \left( \lambda_1^{\beta_i} + \lambda_2^{\beta_i} + \lambda_3^{\beta_i} - 3 \right)\,.
\end{equation}
\noindent where $n_{\lambda}$ denotes the number of terms in the series and $\mu_i$ and $ \beta_i$ are the material parameters. The Ogden models are jointly denoted as $W_\lambda(\lambda_1, \,\lambda_2, \,\lambda_3;\,\bs{\theta}_{\lambda})$, where the vector $\bs{\theta}_{\lambda}$ stores their parameters. 
The nonlinear dependence of \eqref{eq:ogden} on the material parameters prevents its reduction to an expression similar to \eqref{eq:ansatz}. To circumvent this, we follow  \cite{flaschel2023tissue}  and we proceed to fix $n_{\lambda}=500$ uniformly distributed candidates for $\beta_{i}$ selected across the range $\beta_i\in [-50, 50]$. The parametrized strain energy density  can be thus written by means of the following set of basis functions and related coefficients 
\begin{equation}
\label{eq:features_ogden}\begin{split}
{W}_\lambda(\lambda_1, \lambda_2, \lambda_3;\,\bs{\theta}_{\lambda}) = \bfQ^T_\lambda&(\lambda_1, \lambda_2, \lambda_3) \boldsymbol{\theta}_\lambda\,, \quad \text{with} \\
\bs{Q}_\lambda(\lambda_1,\, \lambda_2,\, \lambda_3)=\left[
\lambda_1^{ \beta_i} + \lambda_2^{ \beta_i} + \lambda_3^{ \beta_i} - 3\ :\  i\in\left\{1,\dots,n_{\lambda}\right\} \right]&\quad\text{and}\quad \bs{\theta}_{\lambda}=\left[\theta_{\lambda,i} = \frac{2\mu_i}{ \beta_i^2}\ :\ i\in\left\{1,\dots,n_{\lambda}\right\} \right]\,.\end{split}
\end{equation}

To capture a wide range of material behaviors, we allow the strain energy density to include different terms from the two previous classes, leading to the generic expression
\begin{equation}
  W = W_I(I_1,\, I_2;\,\bs{\theta}_{I}) + W_\lambda(\lambda_1,\, \lambda_2,\, \lambda_3;\,\bs{\theta}_{\lambda}) - p (J - 1)\,,  
  \label{eq:W}
\end{equation}
\noindent where $p$ is a Lagrange multiplier enforcing incompressibility and interpreted as the hydrostatic pressure. We can rewrite \eqref{eq:W} as
\begin{equation}
  W = \bfQ^T \bftheta - p (J - 1)\,,\quad\text{with}\quad
\boldsymbol{\theta} = 
\begin{bmatrix}
\boldsymbol{\theta}_I \\
\boldsymbol{\theta}_\lambda
\end{bmatrix}, \quad
\bfQ = 
\begin{bmatrix}
\bfQ_I(I_1, I_2) \\
\bfQ_\lambda(\lambda_1, \lambda_2, \lambda_3)
\end{bmatrix}\,,  
  \label{eq:WW}
\end{equation}
with the feature vector $\bs{Q}\in\Rset^{n_f}$ and the parameter vector $\bs{\theta}\in\Rset^{n_f}$.
The library \eqref{eq:WW} contains a total of $n_f=n_I + n_\lambda= 521$  possible terms. We further require that $\theta_i\ge0\ \forall i$, which is a sufficient (but not necessary) condition for material stability \citep{hartmann2001parameter}. This choice ensures a non-negative strain energy density and prevents unphysical non-monotonic responses in the simple homogeneous deformation paths used for identification.% We emphasize that this condition represents a practical constraint to promote physically admissible behavior within the explored deformation range.}

%Substituting \eqref{eq:ansatz} and \eqref{eq:features_ogden} into \eqref{eq:W}, we obtain
%%
%\begin{equation}\label{eq: final W}
%W = \bfQ^T \boldsymbol{\theta}  - p (J - 1)\,,\quad\text{with}\quad
%\boldsymbol{\theta} = 
%\begin{bmatrix}
%\boldsymbol{\theta}_I \\
%\boldsymbol{\theta}_\lambda
%\end{bmatrix}, \quad
%\bfQ = 
%\begin{bmatrix}
%\bfQ_I(I_1, I_2) \\
%\bfQ_\lambda(\lambda_1, \lambda_2, \lambda_3)
%\end{bmatrix}.
%\end{equation}
%%

The expression for the hydrostatic pressure $p$ in \eqref{eq:WW} can be obtained starting from the first Piola-Kirchhoff stress tensor
\be
\label{eq:stress}
P_{ij} = \partderiv{W}{ F_{ij}} = \partderiv{\bfQ^T}{F_{ij}}\bftheta - p JF_{ij}^{-T},
\ee
where we used $\partderiv{J}{F_{ij}} = J F_{ij}^{-T}$. Then, we enforce plane-stress conditions imposing $P_{33} = 0$ and, finally, we solve for the hydrostatic pressure $p$ by evaluating \eqref{eq:stress} at $i=j=3$. Using the incompressibility condition $J= 1$ we obtain
\be
\label{eq:hydrostaticPressure}
p = \underbrace{\frac{1}{F_{11}F_{22}-F_{12}F_{21}}}_{\displaystyle F_{33}}\partderiv{\bfQ^T}{F_{33}}\bftheta\,.
\ee
By substituting \eqref{eq:ansatz}, \eqref{eq:features_ogden} and \eqref{eq:hydrostaticPressure} in \eqref{eq:WW}, we obtain the parameterized expression of the adopted library of hyperelastic energy densities. Using then  \eqref{eq:hydrostaticPressure} in  \eqref{eq:stress} yields the in-plane components of the first Piola–Kirchhoff stress tensor under plane-stress conditions
%
%\begin{equation}\label{eq: final W}
%W = \left(\bfQ^T - \frac{1}{F_{11}F_{22}-F_{12}F_{21}}\partderiv{\bfQ^T}{F_{33}} \right) \boldsymbol{\theta}
%\end{equation}

\be
\label{eq:planeStressStress}
P_{ij} = \left( \partderiv{\bfQ^T}{F_{ij}} - \partderiv{\bfQ^T}{ F_{33}} F_{ij}^{-T} \cdot \frac{1}{F_{11}F_{22}-F_{12}F_{21}} \right) \bftheta
\ee
for $i,j=1,2$. To make the dependence of $\bfQ$ on kinematic quantities explicit, we note that $\boldsymbol{Q}^T \boldsymbol{\theta} = \boldsymbol{Q}_I^T \boldsymbol{\theta}_I+ \boldsymbol{Q}_\lambda^T \boldsymbol{\theta}_\lambda $. Applying the chain rule, we obtain
\begin{equation}
\partderiv{\bfQ^T}{F_{ij}}\bftheta= \left(\frac{\partial Q_I^T}{\partial I_a} \frac{\partial I_a}{\partial F_{i j}}\right) \boldsymbol{\theta}_I +\left(\frac{\partial Q_\lambda^T}{\partial \lambda_b} \frac{\partial \lambda_b}{\partial F_{i j}}\right)\boldsymbol{\theta}_\lambda\,, \quad\text{with}\quad a \in \{1,2\},\  b \in \{1,2,3\}\,,
\end{equation}
where we adopt the Einstein summation convention.

\subsection{Equilibrium and boundary conditions} 
\label{sec:equilibrium}
Adopting the library introduced in Sect.~\ref{sec:problem_setting}, the parameter identification procedure aims to determine the set of parameters $\bs{\theta}=\bs{\theta}_{\text{opt}}$ that best represents the observed material behavior while fulfilling equilibrium and boundary constraints. To this end, we write equilibrium in weak form as
\be\label{eq:weak_form}
\int_\Omega \bfP\colon\nabla \bfv \dd V = \int_{\partial\Omega_f}\hat \bft \cdot \bfv \dd S=0 \quad \forall \   \text{admissible} \ \bfv\,,
\ee
\noindent where $\bs{v}$ is a sufficiently regular test function vanishing on the Dirichlet boundary. 
%while $\hat \bft=\bf{P}\cdot\bf{N}$ is the vector of the external boundary traction, 
%$\mathcal{W}^{int}$ and $\mathcal{W}^{ext}$ represent the virtual internal and external works, respectively. 
%Here, $\mathcal{W}^{ext}=0$ holds due to the assumed homogeneous Neumann boundary conditions, i.e. $\bs{\hat t}=\bs{0}$.
We then discretize \eqref{eq:weak_form} in space by approximating the displacement field and its gradient using a mesh of linear triangular finite elements with nodes coinciding with the $n$ points used for the DIC analysis, i.e.
\begin{align}
\label{eq:discretization_F}
\bfu(\mathbf{X}) = \sum_{k=1}^{n}N_{k}(\mathbf{X}) \,\bfu_{k}, \quad 
\bfF(\mathbf{X}) = \bfI+ \sum_{k=1}^{n}\bfu_{k} \otimes \nabla N_{k}(\mathbf{X}),
\end{align}
where $\nabla$ is the gradient operator and $\bfu_{k}$ and $N_{k}:\Omega \rightarrow \Rset$ are respectively the vector collecting the observed displacements and the shape function associated with the $k$-th node. Adopting the Bubnov-Galerkin approach, the test function can be discretized similarly to the displacement field as %The vectors $\bfu_{e,k}$ introduced in the $e$-th element are linked to the vector $\bold{u}_z$ with $z=1,\dots,n$ of the whole grid through the connectivity matrix.
\be\label{eq:testFunctions}
\bfv(\mathbf{X}) = \sum_{k=1}^{n}N_{k}(\mathbf{X}) \,\bfv_{k}\,,
\ee
\noindent where $\bfv_{k}$ are the nodal values of the test function. Using \eqref{eq:planeStressStress} along with \eqref{eq:discretization_F}-\eqref{eq:testFunctions} in \eqref{eq:weak_form} and considering that equilibrium should hold for any admissible $\bfv_{k}$ we obtain the following balance equation at each free degree of freedom
\be\label{eq:fk}
\int_{\Omega} \left[\left( \partderiv{\bfQ^T}{F_{ij}} - \partderiv{\bfQ^T}{ F_{33}} F_{ij}^{-T} \cdot \frac{1}{F_{11}F_{22}-F_{12}F_{21}} \right) \bftheta\right] \partderiv{N_k}{X_j} \dd V=0\,, \quad \forall (k,\,i)\in D_{\text{free}}\,.
\ee

The data recorded by the load cells are used as global boundary constraints. In particular, each load cell gives the experimentally measured reaction force $R_\alpha$ associated to a subset of degrees of freedom $D_{\alpha}$, which leads to the following set of constraint equations
\eqn{eq:reaction}{R_{\alpha}=\sum_{(d, i)\,\in\, D_{\alpha}}\int_{\Omega} \left[\left( \partderiv{\bfQ^T}{F_{ij}} - \partderiv{\bfQ^T}{ F_{33}} F_{ij}^{-T} \cdot \frac{1}{F_{11}F_{22}-F_{12}F_{21}} \right) \bftheta\right] \partderiv{N_d}{X_j} \dd V\,,\quad \forall \alpha=1,\dots,n_{\ell}\,.}

\subsection{Constitutive law discovery} 
\label{inv_problem}

In this section, we recall the basic ideas behind EUCLID, separately for the case of local data (from DIC) and global data (from simple tests). 
%In general, the aim of this procedure is finding the set of parameters $\bs{\theta}=\bs{\theta}_{\text{opt}}$ that (i) satisfies the equilibrium condition, (ii) fulfills the boundary constraints and (iii) correctly reproduces the experimental behavior in terms of global load-displacement curves. 
In identification from DIC, a constitutive law is inferred using only quantities directly measured during the test (i.e., kinematic data from DIC analyses and global reaction forces) and a cost function associated to equilibrium and global boundary constraints.  In identification from simple tests, equilibrium and global boundary constraints are used to obtain a set of stress-strain pairs, while the cost function quantifies the mismatch between the stresses stemming from the measured forces and those predicted by the constitutive law.

\subsubsection{Discovery from local data}
\label{sct:unsup_id}

In the approach based on local data, we start by exploiting the linearity with respect to $\bs{\theta}$  of the equilibrium condition  \eqref{eq:fk}, which is rearranged as
\be\label{eq:system}
 \bold A_{\text{free}} \bftheta = \bs{0}\,,
\ee
where $\bs{A}_{\text{free}}\in \Rset^{|D_{\text{free}}|\times n_f}$ is a matrix resulting from the assembly of elemental submatrices obtained via numerical integration.
A similar procedure can be applied to the system arising from the global boundary constraints \eqref{eq:reaction}, leading to
\be\label{eq:system_fixed}
   \bold A_{\ell}\bftheta = \bold R_\ell\,,
\ee
\noindent where $\bs{A}_{\ell}\in \Rset^{n_\alpha\times n_f}$ is obtained through numerical integration and assembling the elemental contributions, while $\bs{R}_{\ell}\in \Rset^{n_\alpha}$ is a vector storing the reaction forces $R_{\alpha}\ \text{for} \ \alpha=1,\dots,n_{\ell}$. 

In the system arising from \eqref{eq:system} and \eqref{eq:system_fixed}, the two matrices $\bs{A}_{\text{free}}$ and $\bs{A}_{\ell}$ depend on the local kinematic data obtained with the DIC analysis, while $\bold R_\ell$ includes the static information measured by the load cells. Note that a set of equations \eqref{eq:system} and \eqref{eq:system_fixed} can be written for each load step (or DIC snapshot) $s\in[1,\,n_{LS}]$ recorded during the test. The collective fulfillment of both \eqref{eq:system} and \eqref{eq:system_fixed} for all $n_{LS}$ load steps is not feasible as
%would lead thus to the optimal set of parameters $\bs{\theta}_{\text{opt}}$ ensuring a material model kinematically and statically compatible with the experimental data collected. However, a direct solution of the problem is prevented by the fact that 
\eqref{eq:system} and \eqref{eq:system_fixed} are in general highly overdetermined. In principle, we may aim at computing the optimal set of parameters by solving the following regression problem
\be\label{eq:eqb_test}
%\bftheta_{\text{opt}} = 
\min_{\bftheta \geq 0} 
\left(
\sum_{s=1}^{n_{LS}}
\left(
 \underbrace{\left\|\bold A_{\text{free},s} \bftheta\right\|^2}_{\text{Equilibrium}}
+\eta  \underbrace{\left\|\bold A_{\ell,s} \bftheta - \bold R_{\ell,s}\right\|^2}_{\text{Boundary constraint}}
\right)
\right),
\ee
\noindent where the constraint $\bs{\theta}\ge\bs{0}$ guarantees stability and $\eta>0$ is a regularization parameter weighting the contribution of inner and boundary terms to ensure that both terms contribute comparably to the solution. In the following, we adopt $\eta=20$ to reflect the approximate ratio of internal to boundary nodes in the DIC-derived meshes, as suggested in \cite{flaschel2021unsupervised}  (see also Sect.~\ref{sec:Exp_results}). 
%The optimization problem \eqref{eq:eqb_test} can be then solved in a constrained least-square sense through a fixed-point iterative approach as proposed by \cite{tibshirani_regression_1996,flaschel2021unsupervised}.
%
%Therefore, \eqref{eq:eqb} can be reformulated using the full-field displacement data from DIC and the global force data from the load cell, both measured at various time steps during the experimental test, resulting in:
%%
%\be\label{eq:eqb_test}
%\bftheta^{\text{opt}} = 
%\arg\min_{\bftheta \geq 0} 
%\left(
%\sum_{l=1}^{n_{LS}}
%\left(
% \left\|\bold A^{\text{bulk},l} \bftheta - \bold f^{\text{bulk},l}\right\|^2
%+\lambda_r  \left\|\bold A^{\text{bnd},l} \bftheta - \bold f^{\text{bnd},l}\right\|^2
%\right)
%\right),
%\ee
%%
%where $l=1,\dots,n_{LS}$ and $n_{LS}$ denotes the number of load steps supplied to EUCLID.

However, \eqref{eq:eqb_test} is still highly ill-posed due to our large ansatz space (model library) with potential high collinearity among different terms; in the best case, it would yield non-unique solutions that strongly depend on the noise in the experimental data. In addition, the solution vector would likely contain a large number of non-zero parameters leading to a very complicated model of limited interpretability. To overcome these issues and obtain a robust and interpretable description of the material behavior, EUCLID deploys the LASSO (Least Absolute Shrinkage and Selection Operator) regularization technique \citep{tibshirani_regression_1996}. Accordingly, we solve the following sparse regression problem
\be\label{eq:objective}
\bftheta^*_{\lambda} = 
\arg\min_{\bftheta \geq 0} 
\left(
\sum_{s=1}^{n_{LS}}
\left(
\left\|\bold A_{\text{free},s} \bftheta\right\|^2
+\eta \left\|\bold A_{\ell,s} \bftheta - \bold R_{\ell,s}\right\|^2
\right)
+ \lambda_{p}||\bs{\theta}||_1
\right),
\ee
where $\lambda_{p} > 0$ is a penalty parameter that promotes sparsity in the solution and $||(\bullet)||_1$ stands for the $L_1$-norm of $(\bullet)$. Increasing $\lambda_{p}$ leads to a stronger sparsity-promoting regularization, resulting in a material model with fewer active features. The resulting optimization problem, being non-smooth due to the $L_1$-norm term, is solved through a fixed-point iterative scheme in which a sequence of weighted least-squares problems is solved until convergence. This strategy provides a stable and efficient handling of the sparsity-promoting regularization within EUCLID. 
%As demonstrated in previous studies \cite{flaschel2022discovering, flaschel2023automated, marino2023automated}, the choice of $\lambda_p$ also influences the accuracy of the procedure. 

To determine the optimal value of $\lambda_{p}$, we perform a Pareto analysis following the approach in \cite{flaschel2023automated}. This involves solving \eqref{eq:objective} for a wide range of values $\lambda_{p}^{r}$ with $r=1,\,...,\,n_r$, leading to different candidate solutions $\bftheta^*_r=\bftheta^*_{\lambda_{p}^{r}}$.
The accuracy of each solution can be estimated by computing the mean square error (MSE)
\be\label{eq:MSE}
\text{MSE}_r = \frac{1}{n_{LS}}
\sum_{s=1}^{n_{LS}}
\left(
 \left\|\bold A_{\text{free},s} \bftheta^*_r\right\|^2
+\eta \left\|\bold A_{\ell,s} \bftheta^*_r - \bold R_{\ell,s}\right\|^2
\right)\,\quad \forall r=1,\,...,\,n_r\,.
\ee
On the other hand, the complexity of the obtained model can be evaluated by means of the  model complexity parameter (MCP) 
\be\label{eq:MCP}
\text{MCP}=||\bs{\theta}_r^*||_1\,\quad \forall r=1,\,...,\,n_r\,.
\ee
For small values of $\lambda_{p}$, the solution tends to be dense and highly accurate, with high MCP and low MSE values. Conversely, large $\lambda_{p}$ values yield lower MCP values indicating sparsity but higher MSE reflecting reduced accuracy. In the limit case of very large $\lambda_{p}$, all parameters are suppressed and MCP = 0 while MSE reaches its maximum. To balance sparsity and accuracy, a threshold to select the most appropriate $\lambda_{p}$ is introduced as
\be\label{eq:MSE_th}
\text{MSE}_{\text{th}} = \text{MSE}_{\text{min}} + \gamma\left(\text{MSE}_{\text{max}} - \text{MSE}_{\text{min}}\right),
\ee
where $\text{MSE}_{\min}$ and $\text{MSE}_{\max}$ are the minimum and maximum MSE values obtained across all solutions, and $0 < \gamma \ll 1$ is a scalar parameter. Only solutions with $\text{MSE}_r < \text{MSE}_{th}$ are considered, and among them, the sparsest one (i.e., the one with the lowest MCP) is selected as the final candidate $\bftheta^*_R$ along with the corresponding penalty parameter $\lambda_{p}^{R}$. While the LASSO regularization  promotes sparsity, it also fictitiously reduces the values of the non-vanishing parameters, therefore, a final refinement step is performed by solving the original unregularized problem \eqref{eq:eqb_test}, considering a reduced model library with only the terms related to the non-zero parameters in $\bftheta^*_R$. This step recovers the correct magnitude of the retained parameters and further reduces the MSE.

In some simple cases the sparsity-promoting regularization allows for closed-form solutions with some parameters identical to zero \citep{flaschel_non-smooth_2025}, however this does not apply to our regularized discovery problem; 
the obtained solution vector $\bftheta^{\text{opt}}$ may include terms that are nearly but not exactly zero and that, compared to other terms, have a negligible impact on the overall solution. To address this, a threshold value $\theta^{\text{th}}$ is introduced, set to $10^{-6}$ in this work. Any parameter $\theta^{\text{opt}}_i$ satisfying $\left|\theta^{\text{opt}}_i \right|< \theta^{\text{th}}$ is considered inactive and set to zero, \textit{i.e.}, $\theta^{\text{opt}}_i = 0$. 
%This procedure simplifies the solution vector by removing insignificant terms, enhancing interpretability, and focusing the model on influential features for an efficient representation. 

%%%%%%%%%%%%%%%%%%%%%%%%%%%%%%%%%%%%%%%%%%%%%%%%%%%%%%%%%%%%%%%%%%%%%%%%%%%%%%%%%
\subsubsection{Discovery from global data} 
\label{inv_problem_supervised}

For discovery based on global data, we adopt the procedure proposed by \cite{flaschel2023automated}. This approach aims at minimizing the discrepancies between a set of  stress-stretch experimental data and their analytical counterpart obtained from the constitutive model ansatz. Since, in general, stresses cannot be directly measured, they need to be computed from measured forces, which is only feasible for simple tests such as uniaxial tension (UT) or pure shear (PS). 
%However, this might lead to systematic errors and poor coverage of the state space of the material as mentioned in Sect.~\ref{sct:intro}.
% to unpredictable systematic errors in case the assumed ideal conditions are not perfectly respected \citep{pierron2012virtual, pierron2021towards}. Another hypothesis often required in order to obtain reliable stress estimates is a homogeneous distribution of the stress-strain states within the region of interest (ROI) of the specimen, prerequisite that greatly limits the coverage of the state space of the material. 
%This method leverages labeled experimental data to guide the inference of material model parameters. In contrast to the original EUCLID framework~\citep{flaschel2021unsupervised}, which employed an optimization-based approach using unlabeled data, the supervised scheme uses known input-output data pairs, such as stress-stretch measurements from uniaxial tension and pure shear tests, to infer the model parameters. Accordingly, this method can be regarded as the supervised counterpart to EUCLID (cf. Section~\ref{inv_problem}).

In a UT test, a displacement is applied along the axial direction of a dogbone specimen, inducing a uniform uniaxial tensile state within a sufficiently long central portion of the specimen. There, we define a region of interest (ROI) sufficiently far from the tapered connections between the clamped parts and the central region (red hatched region in Fig.~\ref{fig:Experiment_specimen}a). Aligning the $X_1$ axis with the direction of the applied load, the test records the elongation of the ROI $\delta_{1}$ and the axial reaction force $R_{1}$. Assuming ideal conditions (e.g., absence of loading misalignment or clamping imperfections), kinematic and equilibrium yield the experimental axial stretch and first Piola-Kirchhoff stress as
\begin{equation}
\hat \lambda_{1} = \frac{H + \delta_{1}}{H} \quad \text{and}\quad \hat P_{11} = \frac{R_{1}}{A}\,,
\label{eq:stretch_stress_UT}
\end{equation}
\noindent where $H$ is the height of the ROI and $A$ the undeformed cross-sectional area of the central portion of the specimen. 
%We remark that, following the adopted model, the stretch $\bar\lambda_1$ coincides with the first principal stretch. 
Thus, the experimental dataset consists of stress-stretch pairs $\left(\hat\lambda_{\mathrm{1,s}}, \hat P_{11,s}^{\text{(UT)}}\right)$ over the load steps $s = 1, \dots, n_\textsc{UT}$. 
Under UT idealized conditions for an incompressible material, the deformation gradient reads
  \be\label{eq:defGradient_UT}
   {\bm{F}}_{\text{UT}} = \begin{bmatrix}
    \lambda_{1} & 0 & 0\\
  0 & \displaystyle\frac{1}{\sqrt{ \lambda_{1}}} & 0\\
  0 & 0 & \displaystyle\frac{1}{\sqrt{ \lambda_{1}}} \\[.8em]
  \end{bmatrix}\,,
  \ee
%\noindent where $\det \left({\bs{F}}_{\text{UT}}\right) = 1$ due to incompressibility and $F_{22}=F_{33}$ due to symmetry. 
From \eqref{eq:defGradient_UT} and \eqref{eq:planeStressStress} for $i=j=1$ we obtain
\be
\label{eq:P_UT}
P_{11}^{\text{(UT)}}(\lambda_1;\, \bs{\theta}) = \left( \partderiv{\bfQ^T}{F_{11}} - \frac{\displaystyle\sqrt{\lambda_{1}}}{\lambda_{1}^2}\partderiv{\bfQ^T}{ F_{33}}  \right) \bftheta.
\ee

PS tests are conducted following the planar shear test setup suggested by \cite{treloar1944stress,jones1975properties,brown2006physical,moreira2013comparison} among others. A rectangular specimen with undeformed cross sectional area $A_{\text{PS}}$ is subjected to axial loading, again aligned with the $X_1$ axis. To produce a pure shear state for an incompressible material it is essential to consider a completely confined ROI. To this end, the rectangular specimen must have an unclamped (or active) region with a width-to-height ratio of at least 10 \citep{moreira2013comparison}, and, for proper confinement, the ROI in the center must have a width sufficiently smaller than the one of the specimen (see Fig.~\ref{fig:Experiment_specimen}b). Also in this case we record the ROI elongation $\delta_1$ and the axial reaction force $R_{1}$. Assuming perfect confinement, the relations \eqref{eq:stretch_stress_UT} still hold \citep{moreira2013comparison} and the experimental dataset again contains pairs $\left(\hat\lambda_{\mathrm{1,s}}, \hat P_{11,s}^{\text{(PS)}}\right)$ over the load steps $s = 1, \dots, n_\textsc{PS}$. 
Accounting for incompressibility and full confinement in $X_2$ direction ($F_{22}=1$), the deformation gradient reads
\be\label{eq:defGradient_PS}
\boldface{F}_{\text{PS}} = \begin{bmatrix}
\lambda_{1} & 0 & 0\\
0 & 1 & 0\\
0 & 0 & \displaystyle\frac{1}{\lambda_{1}} \\[.8em]
\end{bmatrix}.
\ee
\noindent With \eqref{eq:defGradient_PS} and \eqref{eq:planeStressStress} for $i=j=1$ we obtain
\be
\label{eq:P_PS}
P_{11}^{\text{(PS)}}(\lambda_1;\, \bs{\theta}) = \left( \partderiv{\bfQ^T}{F_{11}} - \frac{1}{\lambda_{1}^2}\partderiv{\bfQ^T}{ F_{33}}  \right) \bftheta,
\ee
\noindent Note that the accuracy of the planar tension deformation gradient approximation \eqref{eq:defGradient_PS}  deteriorates for increasing stretches; for very large stretch values the corresponding results may no longer be reliable. Therefore, we limit our tests to  moderate deformations. 

%Given the models library in Sect.~\ref{sec:problem_setting}, we aim at determining the most appropriate constitutive law that best fits the experimental data. 
When performing material model discovery based on global data, we aim to find the vector that best fits UT and PS data, i.e. so that ideally $P_{11}^{(t)}\left(\hat \lambda_{\mathrm{1},s};\, \bftheta \right)=\hat P_{11,s}^{(t)}\ \forall s= 1, \dots, n_\text{t}$ and $t=\textrm{UT},\,\textrm{PS}$. %This can be reformulated into two systems of linear equations $\mathbf{A}_{\mathrm{t}} \bftheta= \bfb_\mathrm{t}$ for $t=UT,\,PS$, where $\bfb_\mathrm{t}$ stores the measured stresses. %A similar formulation applies to the pure shear test, where stress-stretch pairs $\left(\lambda_{\mathrm{PS}}^{(l)}, P_{11}^{(l)}\right)$ yield another system, $\left(\lambda_{\mathrm{PS}}^{(l)}, P_{11}^{(l)}\right)$ lead to the linear system, , with $l = 1, \dots, n_\textsc{PS}$.
The resulting two sets of equations are combined into a single system as 
\eqn{eq:sup_system}{\mathbf{A}_{\text{sup}} \bftheta = \mathbf{b}_{\text{sup}}\,,\quad \text{with}\quad 
\mathbf{A}_{\text{sup}}=\begin{bmatrix}
\eta_{\text{UT}}\mathbf{A}_{\text{UT}} \\
\eta_{\text{PS}}\mathbf{A}_{\text{PS}}
\end{bmatrix} \quad\text{and}\quad
\mathbf{b}_{\text{sup}} = 
\begin{bmatrix}
\eta_{\text{UT}}\mathbf{b}_{\text{UT}} \\
\eta_{\text{PS}}\mathbf{b}_{\text{PS}}
\end{bmatrix}\,,}
\noindent where  $\mathbf{A}_{\text{t}}\in\Rset^{n_{\text{t}}\times n_f} $ for $t=\textrm{UT},\,\textrm{PS}$ are assembled evaluating \eqref{eq:P_UT} (for $t=\textrm{UT}$) and \eqref{eq:P_PS} (for $t=\textrm{PS}$) at the experimental stretches $\hat\lambda_{1,s}$ obtained from the corresponding tests, while $\mathbf{b}_{\text{t}}\in\Rset^{n_{\text{t}}}$ for $t=\textrm{UT},\,\textrm{PS}$ collects the associated experimental stresses $\hat P_{11}^{(t)}$. Also, $\eta_\textsc{UT}$ and $\eta_\textsc{PS}$ are weighting factors to balance the influence of the two datasets in the joint optimization.
%$\mathbf{A}_{\text{sup}}\in\Rset^{(n_{\text{UT}}+n_{\text{PS}})\times n_f} $ and $\mathbf{b}_{\text{sup}}\in\Rset^{(n_{\text{UT}}+n_{\text{PS}})}$ incorporate the concatenated contributions from both UT and PS test. , scaling factors  are applied. 
These factors are selected such that the maximum norms of the scaled measurements $\eta_{\text{UT}}\mathbf{b}_{\text{UT}}$ and $\eta_{\text{PS}}\mathbf{b}_{\text{PS}}$ are of similar magnitude, thereby preventing one dataset from dominating the fit. In this study, we adopt $\eta_\textsc{UT} = 0.35$ and $\eta_\textsc{PS} = 1$. Also in this case, the system \eqref{eq:sup_system} is overdetermined and it is relaxed to the following sparse regression problem
%As a result, the Mean Square Error (MSE), defined in  ~\eqref{eq:MSE}, is computed using $n_\textsc{LS} = n_\textsc{UT} + n_\textsc{PS}$. 
%The cost function for the supervised approach of the optimal model parameters, we solve the following $\ell_1$-regularized optimization problem:
\begin{equation}\label{eq:opt_superv}
\bftheta^*_{\lambda_{p}} = \underset{\bftheta \geq 0}{\arg\min} \left( \frac{1}{2 n_{LS}}
\sum_{s=1}^{n_{LS}}
 \left\|\bold A_{\text{sup},s} \bftheta -\bold b_{\text{sup},s} \right\|^2
+ \lambda_{p} \|\bftheta\|_1 \right)\, \quad \text{with}\quad n_{LS}=n_\textsc{UT} + n_\textsc{PS}\,.
\end{equation}
\noindent  As in Sect.~\ref{sct:unsup_id}, the objective function in \eqref{eq:opt_superv} contains a LASSO regularization; the already discussed strategies to choose $\lambda_{p}$ and to obtain the final parameter vector $\bs{\theta}_{\text{opt}}$ still apply.
%(see \cite{flaschel2023automated} for further details). The definition of

%%%%%%%%%%%%%%%%%%%%%%%%%%%%%%%%%%%%%%%%%%%%%%%%%%%%%%%%%%%%%%%%%%%%%%
%\subsection{Thresholding}
%\label{Threshold}
%
%The solution vector $\bftheta^{\text{opt}}$ obtained in \eqref{inv_problem} and \eqref{inv_problem (supervised)} may include terms that are nearly zero. These terms correspond to inactive features that have a negligible impact on the overall solution. Retaining these parameters can unnecessarily complicate the model without providing meaningful contributions. To address this, a threshold value $\theta^{\text{th}}$ is introduced, set to $10^{-6}$ in this work. Any parameter $\theta^{\text{opt}}_i$ satisfying $\left|\theta^{\text{opt}}_i \right|< \theta^{\text{th}}$, is supposed inactive and set to zero, \textit{i.e.}, $\theta^{\text{opt}}_i = 0$. This procedure simplifies the solution vector by removing insignificant terms, enhancing interpretability, and focusing the model on influential features for an efficient representation.

%%%%%%%%%%%%%%%%%%%%%%%%%%%%%%%%%%%%%%%%%%%%%%%%%%%%%%%%%%%%%%%%%%%%%%%%%%%%%%%%%%%%%%%%%
\section{Experimental campaign}
\label{sec:Exp_results}

This section describes the experimental campaign including the setup and testing protocol, the geometry of the specimens, the data acquisition and processing approach and the estimation of the noise. We also analyze the coverage of the state space for each tested specimen to highlight the role of the sample geometry.
%%%%%%%%%%%%%%%%%%%%%%%%%%%%%%%%%%%%%%%%%%%%%%%%%%%%%%%
\subsection{Experimental setup}
\label{sec:Exp_setup}

\begin{figure}[h!]
\centering
\includegraphics[width=\textwidth]{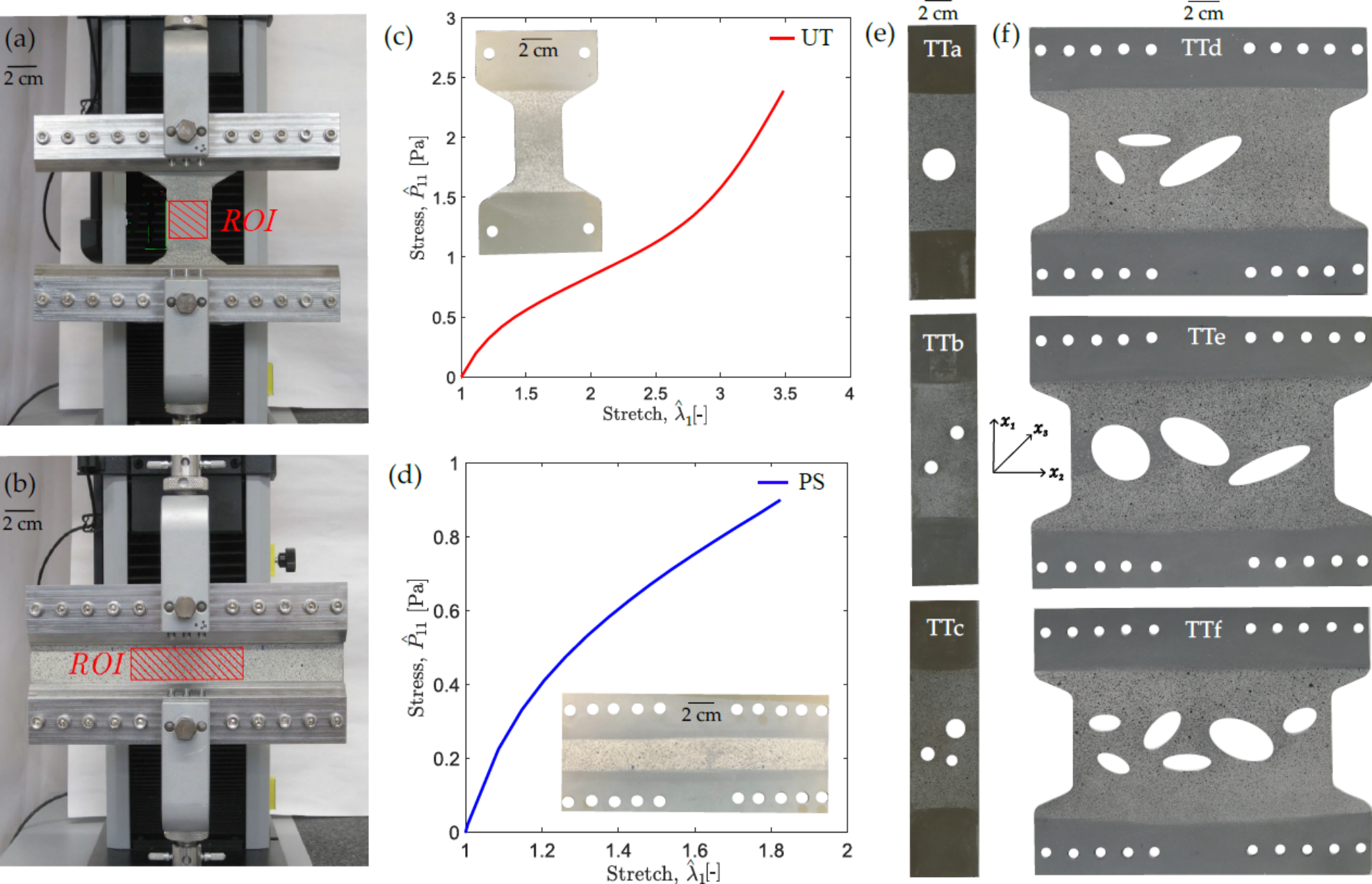}
\caption{Experimental setup and samples: (a) UT test setup with a doubly clamped dogbone specimen under tension in a mechanical testing machine. The deformation in the region of interest (ROI, highlighted with the hatched area) is analyzed. (b) PS test setup with a long rectangular specimen subjected to tension. It produces a deformation equivalent to pure shear in the ROI (hatched region). (c) Stress-stretch response from the UT test. (d) Stress-stretch response from the PS test. (e) Rectangular specimen geometries with circular holes (TTa, TTb, TTc). (f) Wide dogbone specimen geometries with elliptical holes (TTd, TTe, TTf). } 
\label{fig:Experiment_specimen} 
\end{figure}

%As aforementioned, the EUCLID approach requires two key experimental inputs, namely (i)  the overall reaction force and (ii) the full-field displacement of the region of interest of the specimen.
The data for model discovery are obtained by a series of experimental tests performed on a uniaxial universal testing machine (Figs.~\ref{fig:Experiment_specimen}a and \ref{fig:Experiment_specimen}b). The reaction forces are measured by a load cell (2519-1KN, Instron), while the displacement field is captured using a stereo DIC system (Zeiss GOM DIC SRX 12MPx/8GB, GmbH, Germany \cite{GOM_ARAMIS}) composed by two VQXT-120M.K06 8-bit cameras with a resolution of 4096 × 3000 pixels and equipped with Titanar 50 mm lenses, with illumination provided by two blue LED lamps.
To ensure high-quality measurements, the system is calibrated at the beginning of each test. %The acquisition of displacement fields is performed through a structured procedure that involved data acquisition, cleaning, and processing. A detailed description of these steps, demonstrated for a representative test, is provided in \eqref{sec:DIC_proc}.%An analog signal is used to synchronize testing machine and DIC, to ensure an accurate alignment of the two quantities during the test.

The following three sets of tests are performed 
%investigate the mechanical behavior of the selected material under different loading conditions, three different tests were performed under displacement-controlled loading. These tests were carefully designed to capture the material's response in various deformation modes and specimen geometries, providing a detailed characterization of its hyperelastic properties. The following tests were conducted:
\begin{enumerate}
\item UT on a dogbone specimen (Figs.~\ref{fig:Experiment_specimen}a,c);
\item PS on a wide rectangular specimen (Figs.~\ref{fig:Experiment_specimen}b,d);
\item tensile tests (TT) on specimens with varying geometric complexity (Figs.~\ref{fig:Experiment_specimen}e,f).
\end{enumerate}

The material used in the experiments is a natural rubber (NR-40, Brevita SIA, Latvia), and the specimens are obtained from 2.5 mm thick rubber sheets using a cutting plotter. The DIC speckle pattern needed for correlation is obtained using black and white spray paint. To avoid cracking of the paint layer at large deformations and the related deterioration of the speckle quality, we use a mix of white and black dots with an average size of 0.5 mm. Standardized sticker markers are also used to measure the displacement vector of specific points. Once prepared, the specimens are clamped between two steel tabs using a set of screws and bolts to limit slippage during the tests (Figs.~\ref{fig:Experiment_specimen}a and \ref{fig:Experiment_specimen}b). The tabs have a tapered end to avoid stress concentrations and are secured to the clamping system of the testing machine. Also, the nominal strain rate applied during the tests, i.e. the displacement rate normalized by the height of the specimen, is selected to be on the order of $10^{-2}\ \text{s}^{-1}$, a value compatible with quasi-static loading where natural rubber can be assumed as hyperelastic with negligible rate-dependent (viscoelastic) effects \citep{boyce2000constitutive, steinmann2012hyperelastic}.%The first machine, an Instron 3400 (Norwood, MA), is a single-column system equipped with a 1 kN load cell and screw-fastened steel plates, making it suitable for smaller specimens under lower loads. The second, an Instron 5900 (Norwood, MA), is equipped with a 100 kN load cell and features a hydraulic clamping system designed to hold larger specimens subjected to higher forces securely.
%

%To comprehensively investigate the mechanical behavior of the selected material under different loading conditions, three different tests were performed under displacement-controlled loading. These tests were carefully designed to capture the material's response in various deformation modes and specimen geometries, providing a detailed characterization of its hyperelastic properties. The following tests were conducted:
%\begin{enumerate}
%\item Uniaxial tension test (UT) on a dogbone specimen (\eqref{sec:UT}),
%\item Pure shear test (PS) on a wide rectangular specimen (\eqref{sec:PS}),
%\item Tensile test (TT) on specimens with varying geometric complexity (\eqref{sec:TT}).
%\end{enumerate}
%The subsequent subsections detail each type of test, including the specific loading conditions and specimen designs.
%%%%%%%%%%%%%%%%%%%%%%%%%%%%%%%%%%%%%%%%%%%%%%%%%%%%%%%%%%%%%%%%%%%%%%%%%%%%
\subsubsection{Uniaxial tension test}
\label{sec:UT}

A UT test is performed on the dogbone sample in Fig.~\ref{fig:Experiment_specimen}a. The specimen has total length $L_{\text{total}} = 140~\text{mm}$, while the central ROI has height $H = 25~\text{mm}$ and width $w = 25~\text{mm}$. The test is conducted under displacement control in quasi-static conditions at a constant rate of 40 mm/min until reaching a machine displacement of 200 mm. A DIC acquisition rate of 2 Hz is adopted, allowing to collect a total of 600 images. % These images provided a detailed spatial mapping of deformation, enabling the precise computation of strain fields that exceed the capabilities of the tensile machine's built-in sensors.
%
%A frontal view of the gripped specimen during testing is shown in \eqref{fig:Experiment_specimen}(a), with the undeformed configuration illustrated in the inset of \eqref{fig:Experiment_specimen}(c). The Region of Interest (ROI), highlighted in green in \eqref{fig:Experiment_specimen}(a), measures $25 \times 27.6~\text{mm}^2$, where the latter dimension corresponds to the initial length of the specimen, denoted as $H$. The red region shows the deformed state, with a length of $H + \delta_u$, where $\delta_u$ is the applied elongation. The stretch in the loading direction is then computed as:
%%
%\begin{equation}
%\lambda_{1} = \frac{H + \delta_{u}}{H}.
%\label{eq:stretch}
%\end{equation}
%%
%Here, $H$ is obtained by measuring the vertical distance between the average positions of the top and bottom points within the ROI. Using DIC for stretch measurement improves accuracy over the tensile machine’s built-in extensometer by eliminating potential errors due to grip slip or machine compliance. As a result, \eqref{eq:stretch} provides a reliable measure of the principal stretch in uniaxial tension. The applied stress in the elongation direction is characterized by the first Piola-Kirchhoff stress, $P_{11}$, defined as:
%%
%\begin{equation}
%P_{11} = \frac{f}{A_0},  
%  \label{eq:PK stress}
%\end{equation}
%%
%where $f$ is the force recorded by the load cell at each time step, and $A_0$ is the initial cross-sectional area of the specimen, calculated as $A_0 = \text{thickness} \times w_{\text{gauge}}$. 

The experimental $\hat P_{11}-\hat \lambda_{1}$ response obtained using \eqref{eq:stretch_stress_UT} is shown in Fig. \ref{fig:Experiment_specimen}c. The material exhibits a strongly nonlinear response, with a marked strain stiffening at stretches above $\hat \lambda_{1}\simeq2.75$. This range of behavior, arising as molecular chains in rubber approach their finite extensibility, deviates from classic statistical or Gaussian elasticity and is, therefore, termed non-Gaussian regime \citep{staverman1975gaussian}. Capturing this regime is crucial for accurately characterizing the hyperelastic properties of the material.
%%%%%%%%%%%%%%%%%%%%%%%%%%%%%%%%%%%%%%%%%%%%%%%%%%%%%%%%%%%%%%%%%%%%%%%%%

\subsubsection{Pure shear test}
\label{sec:PS}

The rectangular specimen for the PS test is designed following \citep{treloar1944stress,jones1975properties,brown2006physical,moreira2013comparison} and has overall in-plane dimensions of $210 \times 100$~mm$^2$, with an active area of $210 \times 20.3$~mm$^2$ (Fig.~\ref{fig:Experiment_specimen}b). The ROI (in red in Fig.~\ref{fig:Experiment_specimen}b) has a height $H=20.3$~mm, while the width is $w=97.4$~mm to ensure proper confinement. Since the ROI does not laterally span the full specimen width, we need to additionally assume that the axial traction is uniformly distributed across the specimen width. Accordingly, in the local-data approach the load-cell force is scaled by the ratio of the ROI width to the specimen width to obtain the reaction force acting on the ROI.

The specimen is loaded at a constant displacement rate of 20 mm/min, until reaching a total machine displacement of 50 mm. Due to the narrower ROI and smaller expected deformations, the acquisition rate is set to 4 Hz, twice that of the uniaxial tension test, resulting in a total of 600 images.  %This higher temporal resolution allowed for more precise tracking of rapid strain evolution. The stretch was computed using the same methodology described previously in \eqref{eq:stretch}.
%As discussed in \citet{moreira2013comparison}, the first Piola–Kirchhoff stress in the direction of the applied displacement, $P_{11}$, can still be calculated using \eqref{eq:PK stress}, as it provides an accurate measure of stress in pure shear deformation. 
The experimental $\hat{P}_{11}-\hat\lambda_{1}$ response, again from \eqref{eq:stretch_stress_UT}, is shown in Fig.~\ref{fig:Experiment_specimen}d.  Here the final value of the stretch is $\hat\lambda_{1}\simeq1.8$, i.e. much lower than in the UT test due to the restrictions mentioned in Sect.~\ref{inv_problem_supervised}. % Unlike in uniaxial tension, large stretches are not achievable in this setup, since the planar tension deformation gradient corresponds to pure shear only in the small-strain regime. This limitation restricts the range of achievable stress and stretch values.
%%%%%%%%%%%%%%%%%%%%%%%%%%%%%%%%%%%%%%%%%%%%%%%%%%%%%%%%%%%%%%%%%%%%%%%%%%%%

\subsubsection{Tensile tests}
\label{sec:TT}

We conduct tensile tests on specimens containing different circular and ellipsoidal holes to trigger a diverse range of multiaxial strain states and thus to enable a rich sampling of the state space of the material. Unlike UT and PS tests, these tests do not deliver stress-stretch curves, but only force-displacement curves (to be reported in later sections). 

A total of six different geometries are tested, each with progressively more intricate features. The first three specimens, shown in Fig.~\ref{fig:Experiment_specimen}e, have a length of 205 mm and a width of 50 mm, with a ROI of dimensions 60 mm (length) and 50 mm (width), and contain a single central hole (TTa), two unaligned holes of equal size (TTb), and three unaligned holes of different sizes (TTc). The second set, shown in Fig.~\ref{fig:Experiment_specimen}f, includes three wide dogbone-shaped specimens with a length of 168 mm and ROI dimensions 50 mm (length) and 170 mm (width). These specimens contain elliptical holes, namely two holes on the left and one in the middle (TTd), three holes distributed across the width (TTe) and six holes at randomly picked locations (TTf). The ROI was defined to cover the full gauge region of each specimen while excluding areas outside the physical specimen boundaries. To ensure reliable measurements near complex boundaries and hole edges, measurement subsets with high correlation residuals were automatically excluded by the DIC software at each loading step.%These designs were intentionally varied to induce different degrees of strain and stress localization, enabling the observation of complex deformation modes. However, no formal optimization procedure was applied in their selection.

Following the same procedure of the UT and PS tests, each specimen is subjected to displacement-controlled tensile testing. The displacement is applied along the $X_1$ direction at a constant rate of 20 mm/min, resulting in a maximum displacement of 100 mm over 5 minutes. With a DIC sampling rate of 2 Hz, a total of 600 images are recorded. %Due to the heterogeneity of the deformation state, simple  stress-stretch curves are not presented for these cases. Instead, the DIC displacement fields reveal complex deformation patterns, forming a rich and diverse dataset for constitutive model identification.

%As detailed in the following section, a dedicated data processing procedure was developed to transform these displacement fields into input data suitable for the EUCLID framework.
%%%%%%%%%%%%%%%%%%%%%%%%%%%%%%%%%%%%%%%%%%%%%%%%%%%%%%%%%%%%%%%%%%%%%%%%%%%%%%%%%%%%%%%%%%%

\subsection{Acquisition and processing of DIC data}
\label{sec:DIC_proc}

\begin{figure}[h!]
    \centering
    \includegraphics[width=\textwidth]{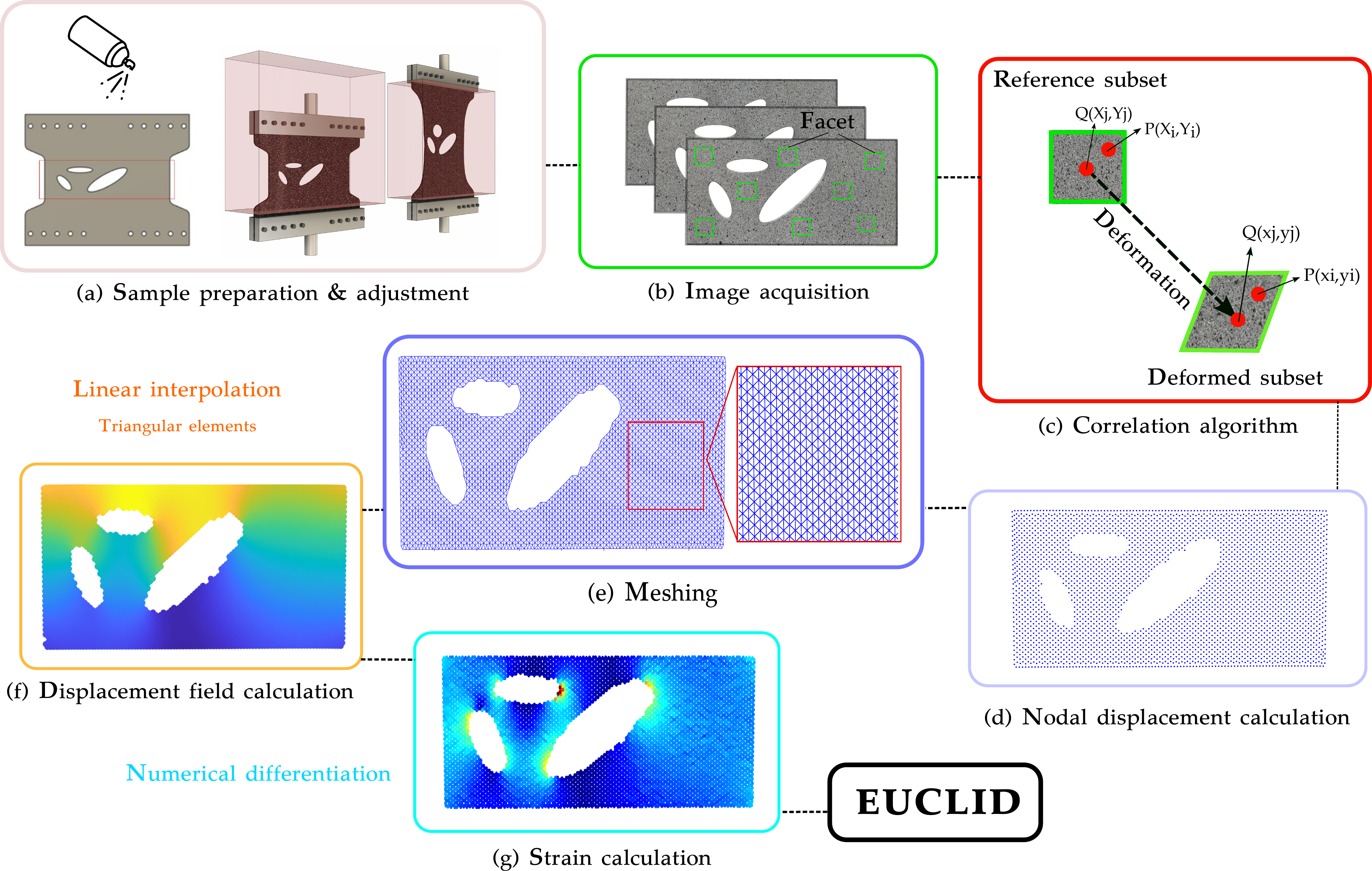}
    \caption{DIC workflow. (a) Sample preparation, including the application of a high-contrast speckle pattern and accurate positioning within the DIC system’s field of view. (b) Image acquisition using the DIC system, where a sequence of images is captured and subsets (facets) are defined across the region of interest. (c) Displacements are determined by tracking each facet between the reference and deformed images using cross-correlation techniques. (d) Following correlation, a discrete set of nodal points with corresponding displacement data is obtained. (e) A DIC supporting mesh is generated based on these nodal points, and (f) interpolation is used to reconstruct the continuous displacement field. (g) The strain field is then computed from the displacement data using differentiation of the FE interpolation. This strain field is subsequently used as input for EUCLID.} 
    \label{fig:boxSpecimens} 
  \end{figure}

This section describes the procedure used to acquire and process displacement and strain fields (Fig.~\ref{fig:boxSpecimens}).  %{The complete process is organized into several key steps, which are detailed in the following: (a) sample preparation and adjustment, (b) image acquisition, (c) application of the correlation algorithm, (d) calculation of nodal displacements, (e) meshing, (f) displacement field computation, and (g) strain calculation.}
The first step involves the creation of the speckle pattern and the installation on the testing machine as detailed in Sect.~\ref{sec:Exp_setup} (Fig.~\ref{fig:boxSpecimens}a).  During the installation, it is essential to ensure that the speckle pattern within the ROI remains entirely within the measuring volume of the DIC system throughout the test (Fig.~\ref{fig:boxSpecimens}a). For all tests apart from PS, we use a measuring volume of 260$\times$200$\times$80 mm$^3$ (length$\times$height$\times$depth), which provides an average density of 1 measuring point/mm$^2$. Due to its reduced ROI dimensions, for the PS test we use a smaller measuring volume of 130$\times$100$\times$40 mm$^3$ (length$\times$height$\times$depth) yielding a density of 4 measuring points/mm$^2$.
  
Next, a series of images is captured during the deformation process (Fig.~\ref{fig:boxSpecimens}b) and then analyzed using a cross-correlation algorithm (Fig.~\ref{fig:boxSpecimens}c) to determine the nodal displacements (Fig.~\ref{fig:boxSpecimens}d).  To this end, the reference (undeformed) image of the ROI is subdivided into small subsets (or \textit{facets}) with size $19 \times 19$ pixels with an overlap of 4 pixels. 
%Following image acquisition (see \eqref{fig:boxSpecimens}(b)), DIC analysis was performed using a subset (facet) size of $19 \times 19$ pixels with an overlap of 4 pixels. This configuration defines localized regions of the speckle-patterned surface and provides an effective compromise between spatial resolution and measurement accuracy. The selected overlap enables spatial averaging to reduce noise while maintaining sufficient data density across the sample. 
As illustrated in Fig.~\ref{fig:boxSpecimens}c, the correlation algorithm uses the pixel intensity pattern (or grayscale pattern) to identify and track the position of each facet between the reference and each deformed image. The deformed image is rescaled using a linear radiometric transformation to account for possible variations in the response of the camera sensors or in the environmental illumination. The deformation of the facets is modeled using a bilinear displacement ansatz and a bicubic subpixel intensity interpolation.
%The initial positions of the subset, such as $Q(X_{j}, Y_{j})$ and $P(X_{j}, Y_{j})$, are located through the matching of the intensity pattern. A cross-correlation algorithm is then used to determine the corresponding positions in the deformed image $Q(x_{j}, y_{j})$ and $P(x_{j}, y_{j})$, allowing accurate calculation of nodal displacements and detailed analysis of deformation (\eqref{fig:boxSpecimens}(d)).

%To minimize data loss and reduce the number of discarded nodes, careful specimen preparation is essential, especially around hole boundaries and along ROI edges. Ensuring optimal surface quality in these critical areas improves DIC tracking consistency and enhances the overall reliability of measured data. 
  
After obtaining the displacement at each node (Fig.~\ref{fig:boxSpecimens}d), a DIC supporting FE mesh composed of linear triangular elements is generated as illustrated in  Fig.~\ref{fig:boxSpecimens}e. Each node of the mesh correspond to a measuring point, namely to the center of a facet. A potential limitation of DIC is the loss of displacement data at certain nodes during the loading history. This issue is particularly relevant near the edges of the specimen (i.e., outer edges or holes). To avoid the remeshing of the ROI at each load step, only those nodes that are continuously available throughout the experiment are retained. The impact of this choice on the final results is minimal, as only a few facets are lost during a test. The definition of an interpolation mesh allows to approximate the displacement field on the whole ROI, as shown in Fig.~\ref{fig:boxSpecimens}f. The mesh characteristics for each specimen including the number of nodes and elements and the density of measuring points are summarized in Tab.~\ref{tab:meshes}. %To ensure accuracy despite the absence of edge points in the DIC, the mesh is carefully refined to avoid excessively elongated elements, particularly near the edges of the holes. This adjustment results in slightly larger hole geometries and modifies the traction-free boundary conditions. 
Once the displacement field is reconstructed, the elongation $\delta_1$ in the UT and PS tests is obtained as the difference between the average displacements in $X_1$ direction of the upper and lower rows of measuring points delimiting the ROI. From the reconstructed displacement field on the DIC supporting mesh (linear triangular elements), strains are obtained by differentiating the FE interpolant. Shape-function derivatives are evaluated at Gauss points and combined with nodal displacements to form the displacement gradient, from which the deformation gradient and strains are computed (Fig.~\ref{fig:boxSpecimens}g).

\begin{table}[h!]
\begin{center}
\begin{tabular}{c c c c} 
 \hline
 Specimen & Nodes & Elements & Measurement density \\
  & & & [$points/mm^2$] \\
 \hline
UT & 383 & 696 & 1\\
PS & 8187 & 15937 & 4\\
TTa & 2981 & 5687 & 1\\
TTb & 2852 & 5459 & 1\\
TTc & 3204 & 6116 & 1\\
TTd & 3270 & 6147 & 1\\
TTe & 2885 & 5335 & 1\\
TTf & 2839 & 5207 & 1\\
 \hline
\end{tabular}
\caption{\label{tab:meshes} Mesh nodes and elements.}
\end{center}
\end{table}
%

%To reconstruct the full-field displacement from the nodal displacements obtained through DIC, linear shape functions are employed to interpolate the displacements at each point within every triangular element. These shape functions express the displacement as a weighted sum of the displacements at the nodes. This interpolation process is carried out for each element, ensuring that the displacement field is continuous and smooth across the entire mesh, as depicted in \eqref{fig:boxSpecimens}(f).

To sum up, after the experimental tests and the outlined procedure, the necessary input data required by EUCLID are available as follows
\begin{itemize} \item as global quantities, the reaction force measured by the load cell and the elongation of the ROI $\delta_1$;
\item as local quantities, the position of the nodes, the connectivity of the associated mesh, and the displacement and strain fields. \end{itemize}
%This ensures that both macroscopic force data and local element-level kinematic information are readily accessible for further analysis and model calibration within the EUCLID framework.
%%%%%%%%%%%%%%%%%%%%%%%%%%%%%%%%%%%%%%%%%%%%%%%%%%%%%%%%%%%%%%%%%%%%%%%%%%%%%%%%%%%%%%%%%%%%
\subsection{Assessment of DIC noise}
\label{sec:noise}

\begin{figure}[h!]
    \centering
    \includegraphics[width=0.85\linewidth]{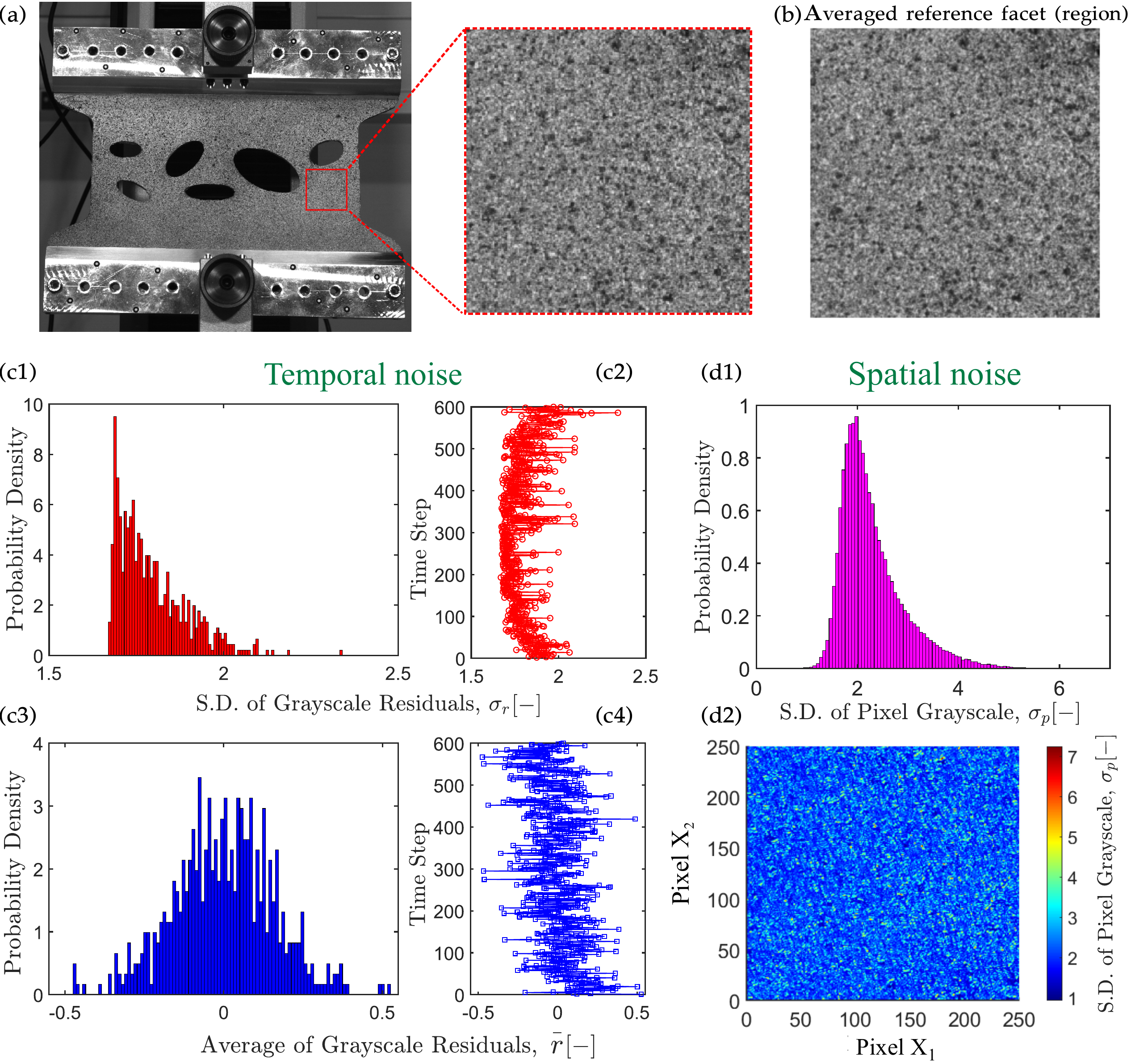}
    \caption{DIC noise assessment. (a) The TTf sample is used for noise assessment, capturing 600 still images over a 300-second duration. A confined facet of $250 \times 250$ pixels (marked by the red square) is selected for the analysis. (b) A reference average facet, reconstructed as the average of all 600 facets, is created to represent the region with the lowest noise level. (c) The temporal evolution of noise is evaluated by plotting (c1) the distribution of the standard deviation of grayscale residuals, $\sigma_{r}$, relative to the reference average facet. (c2) The temporal variation of $\sigma_{r}$ over the 600 time steps is illustrated by red markers. (c3) Distribution of the average of grayscale residuals, $\bar{r}$, over time. (c4) The temporal variation of $\bar{r}$ over 600 time steps is illustrated by blue markers. (d) The spatial distribution of noise is evaluated by plotting (d1) the distribution of the standard deviation of grayscale values for each pixel, $\sigma_{p}$. (d2) Heatmap of $\sigma_{p}$ across the selected region, with the color bar indicating the standard deviation at each pixel. }
    \label{fig:noise}
\end{figure}

In this section we detail the approach followed to quantify the noise in the DIC measurements. This noise stems from various sources such as vibrations of the system, limitations of the camera sensor, illumination variations or surrounding environment artifacts~\citep{sutton2009image}. The presence of noise introduces uncertainties in the displacement and strain field measurements, which can affect the accuracy of the analysis. We quantify the amount of noise of the adopted system by recording $n_t=600$  images of the undeformed ROI of the TTf sample, from which we extract a sample area of $250 \times 250$ pixels (Fig.~\ref{fig:noise}a). Under ideal, noise-free conditions, all pixel intensity values within this region should remain constant over time, while, in practice, noise induces random (Gaussian) variations. To obtain a nearly noise-free reference sample, we  compute pixelwise the average intensity values in the sample area over the 600 images  (Fig.~\ref{fig:noise}b). Since the standard deviation of the signal decays as $\propto1/\sqrt{\text{number of images}}$, the noise in the averaged sample area is about 4\% of the initial one. Based on this reference sample we perform two distinct analyses. The first is a temporal analysis, which investigates the evolution of noise over time, while the second is a spatial analysis, which evaluates the distribution of noise within the sample area. %This dual approach provides comprehensive insights into both the temporal stability and spatial uniformity of the image noise.

%This procedure provides a baseline for evaluating measurement uncertainty. Additionally, to avoid interpolation artifacts that arise during the correlation of grayscale patterns, noise analysis was conducted directly at the subpixel level within the grayscale data, ensuring a more reliable assessment of measurement precision. 
%
%

%To quantify the noise introduced by experimental uncertainties, we selected a specific $250 \times 250$ pixel (facet) from the initial still image, as illustrated in ~\eqref{fig:noise}(a). This region serves as a reference for all subsequent uncertainty analyses. Under ideal, noise-free conditions, all grayscale intensity values within this region should remain constant over time. However, in practice, variations arise due to the measurement noise. To assess these variations, we related each captured image to a predefined reference image. We performed two types of noise analysis: (i) a temporal analysis, which examines the propagation of noise over time by assessing intensity fluctuations across successive frames, and (ii) a spatial analysis, which investigates the distribution of noise across different pixels within the selected region. This dual approach provides comprehensive insights into both the temporal stability and spatial uniformity of the imaging system.

The temporal noise of the system is quantified by computing for each snapshot $s=1,\,...,\,n_t$ the difference between the pixel intensity values of the sample area  and those of the reference one. This difference defines the pixel value residuals, $r(\bs{X}, s)$, and reads
\begin{equation}
    r(\bs{X},s)=I(\bs{X},s)-\bar I(\bs{X})\,, \quad\text{with}\quad\bar I(\bs{X})=\frac{1}{n_t}\sum_{s=1}^{n_t} r\left(\bs{X},s\right)\,,\quad \forall s\in[1,\,n_t]\,,
\end{equation}
where $I(\bs{X},s)$ and $\bar I(\bs{X})$ are the observed pixel intensity for the $s$-th snapshot and the time-averaged reference intensity at the position $\bs{X}$, respectively. 

In Fig.~\ref{fig:noise}c1 we present the probability density histogram of the standard deviation of the snapshotwise residuals $\sigma_{r}$, defined as
\eqn{eq:snap_res}{ \sigma_r(s)=\sqrt{\frac{1}{N}\sum_{i=1}^N \left[r\left(\bs{X}_i,s\right)-\bar r(s)\right]^2}\,, \quad\text{with}\quad\bar r(s)=\frac{1}{N}\sum_{i=1}^N r\left(\bs{X}_i,s\right)\,,\quad \forall s\in[1,\,n_t]\,,}
\noindent where $\bar r(s)$ is the snapshotwise average residual, $\bs{X}_i$ is the coordinate of the $i$-th pixel and $N$ is the number of pixels in the sample area. The histogram shows the highest probability density occurring around $\sigma_{r} \approx \pm1.6$ ($\pm0.6\%$ of the image depth\footnote{The image depth refers to the bit depth of the digital image, defining the number of grayscale levels per pixel (e.g., 8 bit = 256 levels).}), with very low probability of deviation above  $\approx \pm1.9$. This is confirmed by the temporal evolution of the standard deviation in Fig.~\ref{fig:noise}c2, where we observe only a few snapshots with $\sigma_{r}$ significantly larger than 1.6. Additionally, we plot the probability density histogram of the average residuals, $\bar{r}$, in Fig.~\ref{fig:noise}c3, where we can observe a symmetric, approximately normal distribution centered around zero, with values within $\pm0.5$ ($\pm0.2\%$ of the image depth). The temporal evolution of $\bar r$ in Fig.~\ref{fig:noise}c4 further shows that the residuals are clustered around zero with only a few snapshots presenting values above $\simeq \pm 0.25$. Along with the results in Fig.~\ref{fig:noise}c1,c2, this indicates that no significant systematic bias on grayscale values is present and that the noise remains low and stable in time.

%and the evolution of $\bar{r}$ over time in ~\eqref{fig:noise}(c4). The histogram displays a The highest probability density occurs around zero,  The temporal evolution of $\bar{r}$ shows minor variations in the mean residuals over time. However, the absence of a strong trend confirms that the bias remains relatively stable throughout the measurement period.

We now shift to the analysis of the spatial noise. In Fig.~\ref{fig:noise}d1, we show the probability density histogram of the standard deviation of the pixelwise residuals $\sigma_{p}$, defined as
\eqn{eq:pixel_res}{ \sigma_p(\bs{X}_i)=\sqrt{\frac{1}{n_t}\sum_{s=1}^{n_t} \left[r\left(\bs{X}_i,s\right)-\bar r_p(\bs{X}_i)\right]^2}\,, \quad\text{with}\quad\bar r_p(\bs{X}_i)=\frac{1}{n_t}\sum_{s=1}^{n_t} r\left(\bs{X}_i,s\right)\,,\quad \forall i\in[1,\,N]\,,}
\noindent where $\bar r_p(\bs{X}_i)$ is the pixelwise average residual. The histogram shows a distribution clustered around a value of 2 (0.8\% of the image depth), with only approximately 2.5\% of pixels showing a standard deviation higher than 4. Also, Fig.~\ref{fig:noise}d2 illustrates the spatial distribution of $\sigma_{p}$ over the sample area, confirming that the majority of the pixels have a limited noise. Collectively, the results presented in Fig.~\ref{fig:noise} demonstrate that the noise for the adopted DIC system is acceptable considering the available 8-bit image depth \citep{wang2016theoretical}. 
%%%%%%%%%%%%%%%%%%%%%%%%%%%%%%%%%%%%%%%%%%%%%%%%%%%%%%%%%%%%%%%%%%%%%%%%%%%%%%%%%%%%%%%%%
\subsection{Coverage of the state space}
\label{sec:info_content}

In this section we analyze how well the geometries of the tested specimens are able to trigger different multiaxial deformation states locally as a result of the application of a global load. Here, we focus on a fixed set of predefined geometries; however, in future work, we plan to optimize the specimen geometry so as to better cover the state space of the material \citep{ghouli_topology_2025}. To evaluate the data richness of the performed experiments, we analyze the coverage of the strain invariant plane, $\left(I_{1}-3,\, I_{2}-3\right)$, along with the distribution of the principal in-plane stretches, using a similar visualization as in \citep{promma2009application, guelon2009new}. 

\begin{figure}[!]
    \centering
    \includegraphics[width=0.93\linewidth]{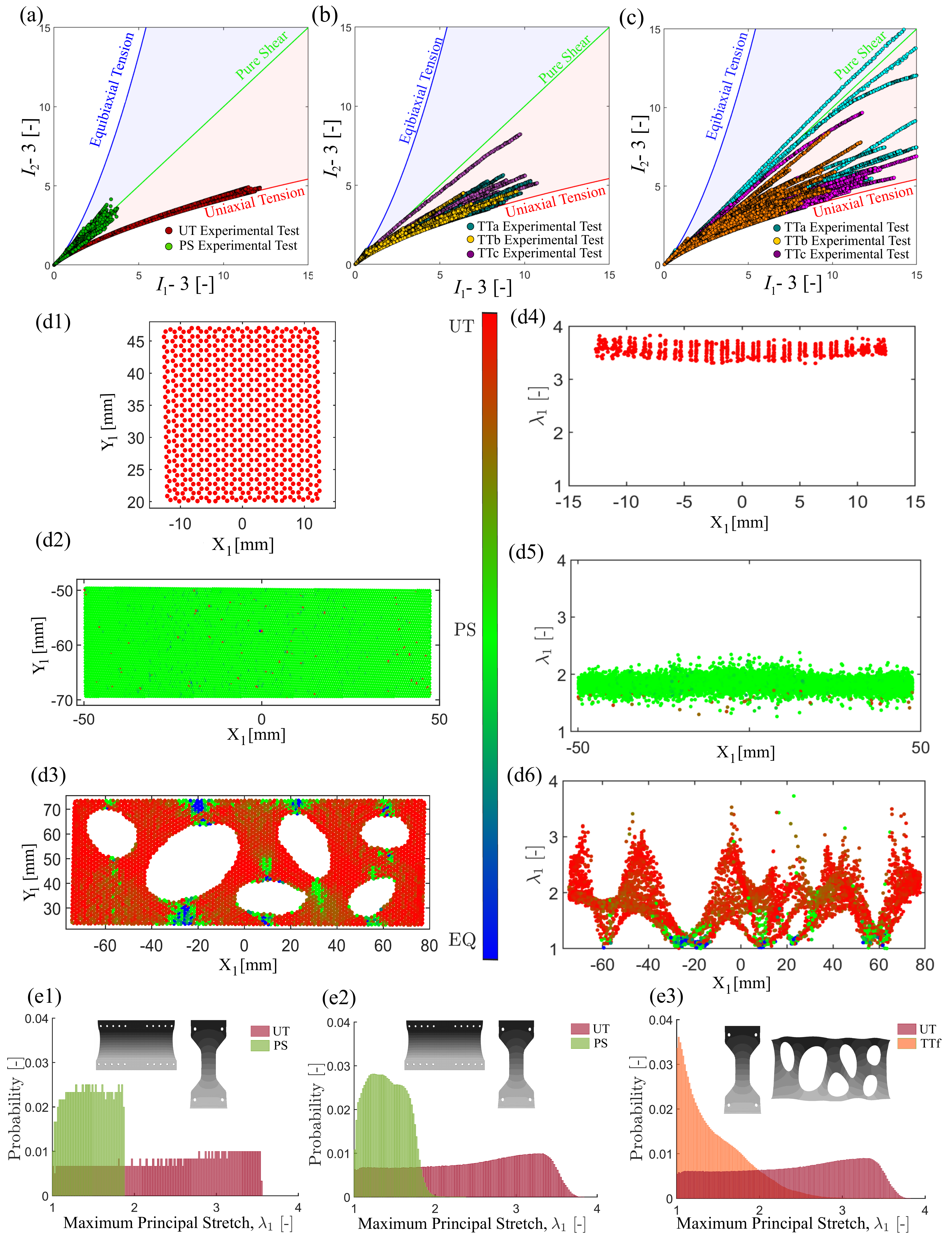}
    \caption{Test-induced strain heterogeneity across different geometries and loading conditions with a similar visualization as in \citep{promma2009application, guelon2009new}. (a) Strain invariants $(I_{1}-3)-(I_{2}-3)$ plane for UT and PS tests. Experimental data points are shown with markers alongside the theoretical boundaries for UT (red line), PS (green line), and ET (blue line). The red shaded region indicates the domain where $I_{1} > I_{2}$, while the blue shaded region represents $I_{2} > I_{1}$. (b) Strain invariants for TT tests on samples TTa, TTb, and TTc. (c) Strain invariants for TT tests on samples TTd, TTe, and TTf. (d1, d2, d3) Color-coded representation of the loading state at each Gauss point in the final loading step for the UT (d1), PS (d2), and TTf (d3) specimens. Colors indicate the proximity of each point to pure loading modes in the invariant plane. (d4, d5, d6) Corresponding maximum principal stretch $\lambda_{1}$ at each Gauss point vs. the $X_1$-axis for the UT (d4), PS (d5), and TTf (d6) specimens. Combined probability distribution of $\lambda_{1}$ from both (e1) UT and PS global data, (e2) UT and PS local data, and (e3) UT and TTf local data. Insets in (e1-e3) depict the corresponding sample geometries.}
    \label{fig:heterogenity}
\end{figure}

In Figs.~\ref{fig:heterogenity}a–c, we present the experimental data across all test setups and loading steps within the invariant plane. The experimental values of the invariants are computed from the right Cauchy–Green tensor $\mathbf{C}$ at each Gauss point of the DIC mesh (Sect.~\ref{sec:DIC_proc}) and they are presented for three test groups: UT and PS tests (Fig.~\ref{fig:heterogenity}a), tensile tests on samples TTa, TTb, and TTc (Fig.~\ref{fig:heterogenity}b) and tensile tests on samples TTd, TTe, and TTf (Fig.~\ref{fig:heterogenity}c). For reference, we report also the theoretical curves related to equibiaxial tension (ET), UT and PS  for an incompressible material. It is well known that all physically admissible deformation states lie within the region bounded by the UT and ET curves \citep{g1996genie,baaser2013reformulation}. The invariant plane is partitioned into two areas by the pure shear condition, which lies along the first bisectrix $(I_1-3) = (I_2-3)$. The region where $I_1 > I_2$ (shaded in red) corresponds to "tension-dominated" states, while the area where $I_2 > I_1$ (shaded in blue) represents "compression-dominated" states.

In Fig.~\ref{fig:heterogenity}a we can observe that data points from UT and PS tests align closely with their respective curves, leaving wide areas of the plane unexplored. On the other hand, Figs.~\ref{fig:heterogenity}b,c show that a modification of the specimen geometry through circular or elliptical holes results in a broader coverage of the invariant plane in the tension-dominated region. Also, comparing Figs.~\ref{fig:heterogenity}b and c reveals that elliptical holes introduce a larger degree of heterogeneity than circular ones, further expanding the magnitude and variety of the obtained deformation states.

To further investigate the influence of geometry on strain heterogeneity, we examine the final load step of the UT test in Fig.~\ref{fig:heterogenity}d1, of the PS test in Fig.~\ref{fig:heterogenity}d2 and of the TTf test in Fig.~\ref{fig:heterogenity}d3. These three tests are selected as representative examples, as they correspond to the simplest and most complex geometries among those examined. In all cases, we indicate the deformation state of each Gauss point using a color indicating its proximity to the ideal ET, PS, and UT curves in the ($I_1-3$, $I_2-3$) invariant space. As expected, all points of the UT sample experience a uniform uniaxial loading state, while the points of the PS sample experience a pure shear loading state. Although the majority of the points show a nearly-UT deformation state, the TTf sample displays a more heterogeneous response, with Gauss points spreading across a wider range of loading modes. Notably, this heterogeneity is concentrated within limited areas around the elliptical holes. 

To investigate the magnitude of the deformations, we plot the maximum principal stretch $\lambda_1$, i.e. the largest eigenvalue of $\mathbf{F}$, as a function of the $X_1$-coordinate in Figs.~\ref{fig:heterogenity}d4,d5, and d6 for the UT, PS, and TTf tests, respectively, at the same load steps of Figs.~\ref{fig:heterogenity}d1, d2, and d3. In the UT sample, Gauss points exhibit a nearly homogeneous stretch around $\lambda_1 \approx 4$, while in the PS sample Gauss points exhibit a nearly homogeneous stretch around $\lambda_1 \approx 2$. As expected, the TTf sample displays substantial spatial variation in stretch magnitude, consistently with its non-uniform deformation field. However, its maximum $\lambda_1$ values are significantly lower than those of the UT test. %, while those under pure shear, equibiaxial tension, or mixed modes exhibit lower stretches. This further underscores how geometric features, such as holes, introduce significant local deviations in both loading mode and deformation intensity.
Moreover, the points experiencing deformation states different than UT correspond to even lower $\lambda_1$ values, mostly below 1.5.

Next, we examine the distribution of $\lambda_1$ across various loading scenarios. Fig.~\ref{fig:heterogenity}e1 presents the probability density histogram of the global $\lambda_1$ values over all loading steps for the UT and PS tests, while Figure~\ref{fig:heterogenity}(e2) presents the corresponding local distributions. Both distributions indicate predominantly homogeneous deformation in each sample, with the strain remaining nearly uniform across the region of interest (ROI) and scaling consistently with the applied displacement according to \cref{eq:stretch_stress_UT}. The main difference lies at the upper end of the stretch domain—reaching values close to 2 for the PS test and up to 4 for the UT test. Overall, the global and local datasets exhibit comparable behavior.    

Figure~\ref{fig:heterogenity}(e3) presents the probability density histogram of the global $\lambda_1$ values across all loading steps for the UT and TTf tests. The contrast between the two specimens representing simple and complex geometries is clearly visible in the figure. The TTf sample highlights a non-uniform probability distribution with high probability to obtain low $ \lambda_1$ values and only a few highly stretched points. The UT specimen primarily contributes to the higher-stretch region, whereas the TTf specimen introduces more significant variability at low or intermediate stretch levels. This comparison suggests that a complicated specimen may not necessarily outperform a simple specimen in terms of state space coverage for the purpose of material model discovery (or also simply of parameter identification of an a priori given model); rather, exploiting the synergy of the two (which offer respectively diversity of deformation states and a wide stretch range) may be the best strategy. We will return on this aspect in later sections. 

%Notably, the tail diminishes in the high-stretch region, due to two factors: (i) the limited number of high-stress concentration zones near the elliptical holes and (ii) experimental challenges that may result in the loss of extreme stretch data. 

%differences between the more uniform but larger deformation of the UT sample and the diverse but limited one in the TTf sample, 
%demonstrating thus how geometric complexity affects strain magnitude and heterogeneity. %Additionally, the impact of increased heterogeneity in complex samples is significantly influenced by their quantity, which is crucial in material characterization and will be further elaborated in the next section.

%%%%%%%%%%%%%%%%%%%%%%%%%%%%%%%%%%%%%%%%%%%%%%%%%%%%%%%%%%%%%%%%%%%%%%%%%%%%%%%%%%%%%%%
\section{Results of material model identification and discovery on experimental data}
\label{sec:Num_results}
In this section we present the results obtained using EUCLID based on both local and global data to discover the constitutive law of the tested natural rubber. Also, we compare the obtained results with those of the classical parameter identification approach where the functional form of the constitutive law is assumed a priori. %Both supervised and unsupervised calibration strategies are employed to ensure a comprehensive assessment. 

\subsection{Preliminaries}
In the following study, we have a few objectives:
\begin{itemize}
    \item \textit{Compare model discovery and parameter identification}: we aim to compare the results of model discovery (i.e. model selection + parameter identification) via EUCLID with those of model identification (i.e. parameter identification on an a priori chosen model, as in the traditional paradigm).

    \item \textit{Compare discovery (or identification) based on global and on local data}: based on the observations in Sect. \ref{sec:info_content}, it is not clear whether local data obtained from one complicated specimen can be a viable alternative to global data from multiple tests performed on simple specimens.
    
    \item \textit{Assess generalization and predictive ability}: we aim at assessing how well the discovered (or identified) constitutive laws generalize to unseen loading conditions. This is examined at both global and local level comparing the experimental results with those predicted using FE computations adopting the discovered (or identified) constitutive laws. Together, these comparisons assess the ability of the discovered (or identified) models to capture both global and local responses of the material beyond the calibration regime.
    
    \item \textit{Evaluate performance:} performance metrics are defined that allow to quantitatively assess the accuracy of the obtained constitutive laws in reproducing the experimental evidence.
\end{itemize}

To evaluate the ability of a constitutive law to reproduce an experimental quantity we use the  relative $\mathcal{L}^2$ error, which reads
\begin{equation}\mathcal{L}^2(\bullet) = \frac{\norm{ (\hat\bullet) - (\bullet)}_2}{\norm{(\hat\bullet)}_2}, 
\label{eq: l^2}\end{equation} 
\noindent where $(\hat\bullet)$ and $(\bullet)$ are the vectors storing the experimental and numerically obtained quantities, respectively. Whenever relevant, we also use the local relative error between experimental results and numerical predictions, namely
\begin{equation}
\label{local_error}
 \epsilon_{\mathrm{rel}}(\xi(\bs{X})) 
      = \frac{\left|\hat\xi(\bs{X}) - \xi(\bs{X})\right|}
          {\hat\xi(\bs{X})} \,,
\end{equation}
\noindent where $\hat\xi(\bs{X})$ and $\xi(\bs{X})$ are respectively the experimental and predicted observed scalar quantities at position $\bs{X}$.

\subsection{Material model identification and discovery using global data from UT and PS tests}
\label{sec: UT-PS-global}
First, we deploy stress–stretch data pairs obtained from UT and PS tests (Sect.~\ref{inv_problem_supervised})  to identify or discover the constitutive law of natural rubber. For classical parameter identification, we assume as functional form alternatively the first-, second-, and third-order GMR models, the Gent–Thomas model and the 1-term and 2-term Ogden models. 
%Higher-order models such as a fourth-order GMR formulation or Ogden models with more than two terms are intentionally excluded to mitigate the risk of overfitting. 
The generic functional forms of these models are stated in \eqref{eq:GMR_GT}, \eqref{eq:GT} and \eqref{eq:ogden}. For model discovery with EUCLID, no a-priori selection of the functional form is needed as the entire library \eqref{eq:W} is used.
%— to evaluate and compare the performance of the EUCLID framework against the conventional approach of calibrating individual material models. By leveraging these combined datasets, we aim to assess each method’s ability to capture the material's mechanical behavior across different loading conditions. Our primary objective is to identify a material model that not only provides an accurate representation of the global stress–stretch response but also demonstrates predictive capabilities when applied to various loading scenarios and sample geometries composed of the same material. 

Classical parameter identification of an a priori selected model is performed through a  nonlinear least-squares solver (implemented in the MATLAB function \texttt{lsqcurvefit}) which adopts the trust-region reflective algorithm. Unlike the GMR and GT models, the Ogden model introduces nonlinear dependencies on exponents, making the minimization problem non-convex. For Ogden models, we treat both the coefficient ($\mu$) and the exponent ($\beta$) as unknowns, and perform multiple minimization rounds with random initial parameter guesses to find the best of possibly multiple local minima. In all cases, non-negativity of the parameters is enforced by specifying zero as the lower bound for all variables. Solver settings include a maximum of 8000 iterations, up to 60,000 function evaluations, a function tolerance of $10^{-12}$, and a step size tolerance of $10^{-8}$. For the EUCLID version using global data we follow the procedure outlined in Sect.~\ref{inv_problem_supervised}.

\begin{figure}[h!]
    \centering
    \includegraphics[width=1\linewidth]{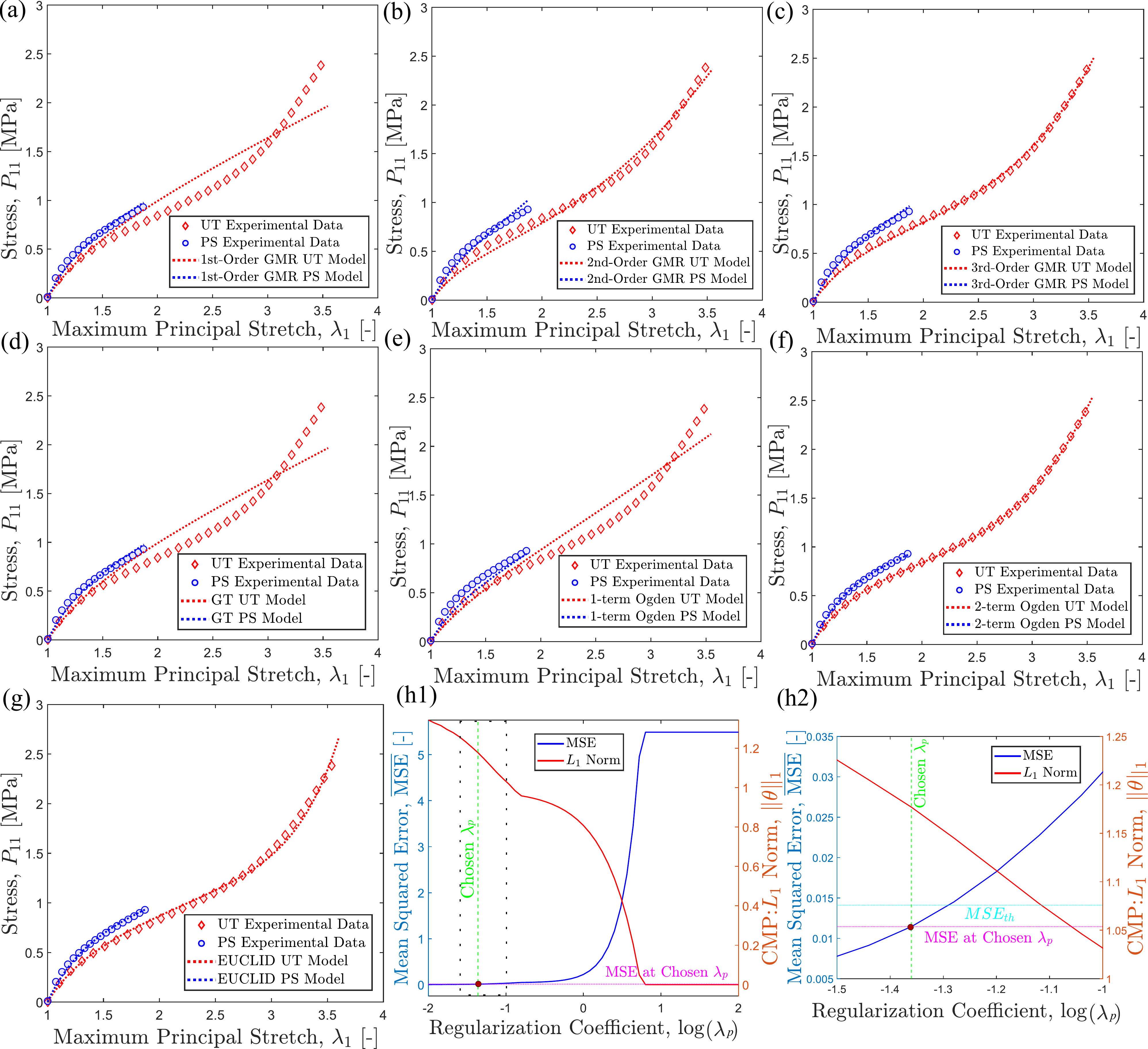}
    \caption{Comparison of the a priori chosen models upon parameter identification and the model discovered by EUCLID, whereby identification/discovery uses combined UT and PS global data. Panels (a)–(g) show the stress-stretch response in UT and PS tests for different material models, compared with experimental data. Specifically, panels (a)–(f) show results for a priori chosen models: (a) 1st-order GMR, (b) 2nd-order GMR, (c) 3rd-order GMR, (d) GT, (e) 1-term Ogden, and (f) 2-term Ogden. Panel (g) shows the response obtained through automated model discovery using EUCLID. (h1) Pareto analysis for the automated selection of the hyperparameter $\lambda_{p}$, showing the MSE and the $L_1$ norm of $\boldsymbol{\theta}$ as functions of $\lambda_{p}$. (h2) A close-up around the automatically selected hyperparameter, with the chosen threshold for MSE and the selected solution indicated by cyan and green dashed lines, respectively.}
    \label{fig:UT_PS_GLOBAL}
\end{figure}

In Figs.~\ref{fig:UT_PS_GLOBAL}a-f, we show the  curves obtained with classical parameter identification (dotted lines) along with experimental data points (represented by markers) for UT and PS tests, while the obtained models and the accuracy metric $\mathcal{L}^2(P_{11})$ are summarized in Tab.~\ref{Table2}. The comparison shows that for the PS test, characterized by relatively small deformations ($\lambda \leq 1.9$), all the hyperelastic models provide a reasonable fit to the experimental data, with the 2-term Ogden model exhibiting the lowest error, i.e. $\mathcal{L}^2(P_{11})=0.06\%$. In contrast, the UT test involves large deformations, during which the hyperelastic stress–strain curve shifts from a Gaussian to a non-Gaussian regime \citep{treloar1975physics}. Among the calibrated models, both the second- and third-order GMR models and the 2-term Ogden model perform satisfactorily, with the 2-term Ogden model again achieving an optimal fit with $\mathcal{L}^2(P_{11}) = 0.00\%$. Thus, the 2-term Ogden model is the one (among the a priori chosen ones) that most accurately captures the global behavior of the material across the UT and PS loading scenarios.

%The calibrated strain energy functions for each model, along with their corresponding relative $\mathcal{L}^2$ error metrics for the UT and PS stress–stretch curves, are summarized in \eqref{Table2}. The relative $\mathcal{L}^2$ error is defined as 
%% 
%\begin{equation} \text{Relative } \mathcal{L}^2 \text{ Error} = \frac{\norm{ P^{\text{exp}}_{11} - P^{\text{pre}}_{11}}_2}{\norm{P^{\text{exp}}_{11}}_2}, 
%\label{eq: l^2}\end{equation} 
%% 
%where $P^{\text{exp}}_{11}$ and $P^{\text{pre}}_{11}$ denote the experimental measurements and the predicted values, respectively. To calibrate the strain energy density for each individual model, we employed MATLAB’s nonlinear least-squares solver (\texttt{lsqcurvefit}) to minimize the discrepancy between the experimental data and the model-predicted stress–stretch relationship. 
 %
%

%Analysis indicates that for the PS test—characterized by relatively small deformations ($\lambda \leq 1.9$)—all the hyperelastic models provide a reasonable fit to the experimental data, with the 2-term Ogden model exhibiting the lowest error ($\mathcal{L}^2=0.0006$). In contrast, the uniaxial tensile test involves large deformations, during which the hyperelastic stress–strain curve shifts from a Gaussian regime (small deformation) to a non-Gaussian regime (large deformation) \citep{treloar1975physics}. This transition introduces significant calibration challenges due to the increased nonlinearity in the material response. Among the calibrated models, both the second- and third-order generalized Mooney–Rivlin models and the 2-term Ogden model performed best, with the 2-term Ogden model again achieving an optimal fit ($\mathcal{L}^2 = 0.0000$). This consistency underscores the 2-term Ogden model’s capability to accurately capture the global behavior across different loading scenarios.
%
\begin{table}[ht]
\small  % Reduces the font size
\centering
\caption{Strain energy density of the identified or discovered material models using UT and PS global data.}  % Add your table title here
\label{Table2}  % Add your label here
\begin{tabular}{lp{7.5cm}cc}
  \toprule
  \multirow{ 2}{*}{\textbf{Identification/Discovery}} &  {\textbf{Strain energy density }} & \textbf{UT} & \textbf{PS}\\
  &$W$ [\si{Pa}] & \textbf{$\mathcal{L}^2(P_{11})$} [\%] & \textbf{$\mathcal{L}^2(P_{11})$} [\%]\\
  \midrule
  1st-order GMR & $W = 0.2836(I_1 - 3) + 1.9076 \cdot 10^{-9}(I_2 - 3)$ & 1.93\% & 0.28\% \\
  \midrule
  2nd-order GMR & 
  $\begin{aligned}
    W = & 0.1130(I_1 - 3) + 0.1518(I_2 - 3) \\
    & + 0.0092(I_1 - 3)^2 + 2.4029 \cdot 10^{-14}(I_2 - 3)^2  \\
    & + 3.9418 \cdot 10^{-14}(I_1 - 3)(I_2 - 3)
  \end{aligned}$ & 0.23\% & 0.65\% \\
  \midrule
  3rd-order GMR & 
  $\begin{aligned}
    W = & 0.1739(I_1 - 3) + 0.1071(I_2 - 3) \\
    & + 2.2221 \cdot 10^{-14}(I_1 - 3)^2 + 2.2205 \cdot 10^{-14}(I_1 - 3)(I_2 - 3) \\
    & + 2.2207 \cdot 10^{-14}(I_2 - 3)^2 + 0.0005(I_1 - 3)^3 \\
    & + 2.2242 \cdot 10^{-14}(I_1 - 3)^2(I_2 - 3) \\
    & + 2.2205 \cdot 10^{-14}(I_1 - 3)(I_2 - 3)^2 + 2.2235 \cdot 10^{-14}(I_2 - 3)^3
  \end{aligned}$ & 0.04\% & 0.32\% \\
  \midrule
  GT & $W = 0.2836(I_1 - 3) + 1.4219 \cdot 10^{-9} \ln(I_2 / 3)$ & 1.93\% & 0.28\% \\
  \midrule
  1-term Ogden & $W = {0.1873}(\lambda_1^{2.2787} + \lambda_2^{2.2787} + \lambda_3^{2.2787} - 3)$ & 1.19\% & 1.36\%· \\
  \midrule
  2-term Ogden & 
  $\begin{aligned}
    W = & 4.8045 \cdot 10^{-4}(\lambda_1^{5.9509} + \lambda_2^{5.9509} + \lambda_3^{5.9509} - 3) \\
    & + 2.2700 (\lambda_1^{0.7355} + \lambda_2^{0.7355} + \lambda_3^{0.7355} - 3)
  \end{aligned}$ & 0.00\% & 0.06\% \\
  \midrule
  EUCLID & $W = 0.2180(I_1 - 3) + 2.4532\cdot 10^{-6}(I_1 - 3)^5 + 0.2538 \ln(I_2 / 3)$ & 0.10\% & 0.09\% \\
  \bottomrule
\end{tabular}
\end{table}

Next, we apply EUCLID to the same dataset; we report the resulting curves in Fig.~\ref{fig:UT_PS_GLOBAL}g and the discovered model, along with its accuracy metric, in  Tab.~\ref{Table2}. As detailed in Sect.~\ref{sct:unsup_id}, tuning of the sparsity promoting parameter $\lambda_{p}$ is performed through a Pareto analysis, which is illustrated in Figs.~\ref{fig:UT_PS_GLOBAL}h1, h2.  In this study, we examine 41 values of $\lambda_{p}$ evenly distributed on a logarithmic scale from $10^{-2}$ to $10^2$. As $\lambda_{p}$ increases, the MSE \eqref{eq:MSE} rises while the MCP \eqref{eq:MCP} decreases, indicating a shift from accurate but complex models to simpler but less accurate ones. For $\lambda_{p} \gtrsim 10$, the parameter vector collapses to zero, causing the MCP value to drop to zero while the MSE stabilizes. Following Sect.~\ref{sct:unsup_id}, an optimal value of $\lambda_{p}$ is selected to balance model accuracy and simplicity following \eqref{eq:MSE_th} while setting $\gamma = 0.002$. This leads to a value of $\lambda_{p}=10^{-1.44}$.
As targeted, EUCLID is able to discover in one-shot (i.e. with no need for iterative exploration of multiple individual models) a  model that integrates terms from both the GMR and GT formulations. This "composite" model yields for both UT and PS data an accuracy very close to that of the best individually calibrated model (i.e. the 2-term Ogden model).

\begin{figure}[h!]
    \centering
    \includegraphics[width=1\linewidth]{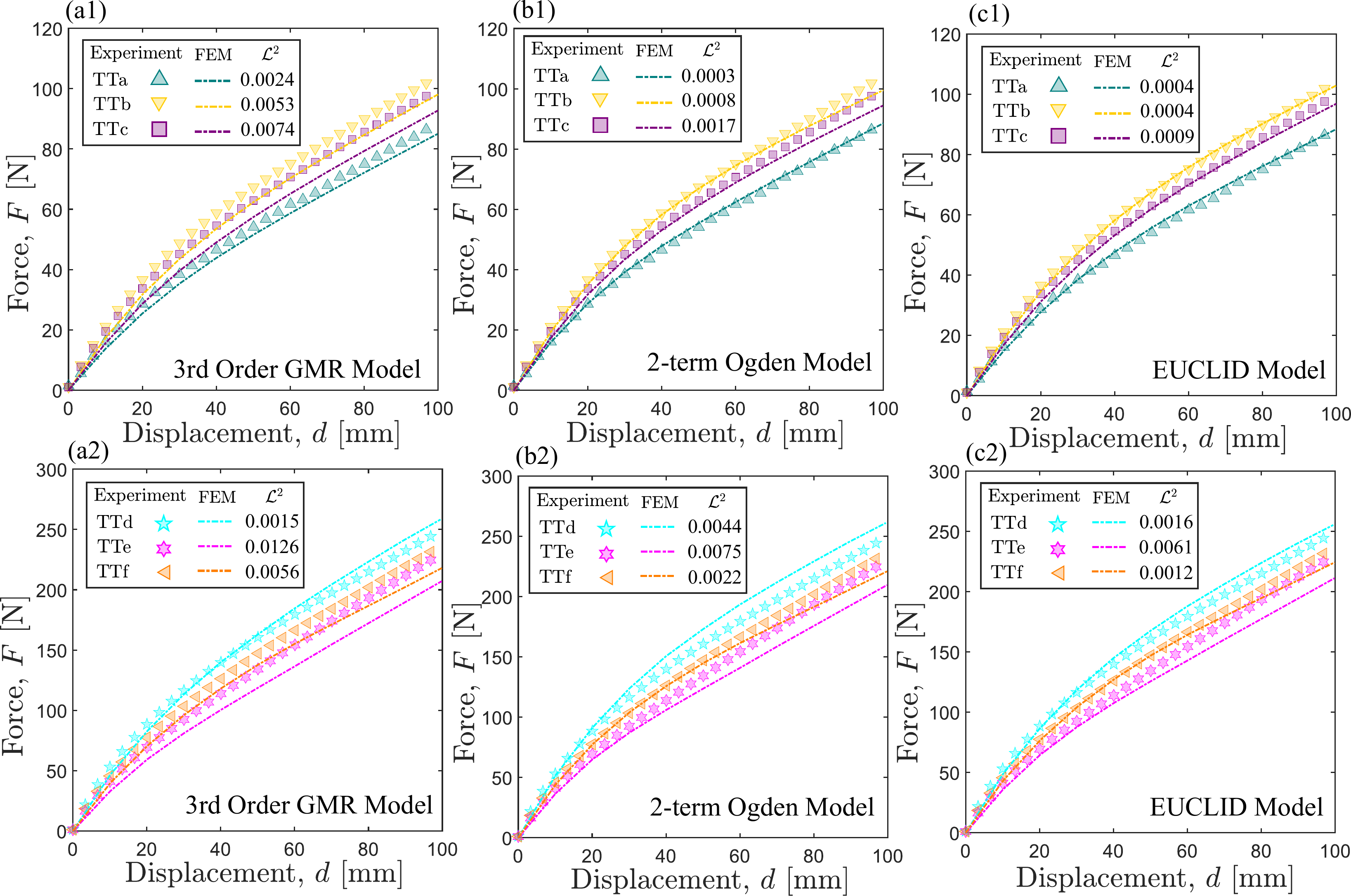}
    \caption{Prediction of the global response of TT sample geometries by the models identified/discovered using global data from UT and PS tests. Experimental force-displacement data are compared with FE predictions from different constitutive models across various sample geometries. Subfigures (a1)–(c1) present results for simpler geometries (TTa, TTb, TTc), while subfigures (a2)–(c2) show results for more complex geometries (TTd, TTe, TTf). The models include the 3rd-order GMR model (a1, a2), the 2-term Ogden model (b1, b2), and the EUCLID-discovered model (c1, c2). Experimental data points are shown as markers and FE predictions as dashed lines.}
    \label{fig:UT_PS_GLOBAL_Prediction}
\end{figure}

We now aim at assessing the ability of the best obtained models to predict the global response of previously unseen sample geometries. To this end, we compare the experimental force–displacement ($F$–$\delta_1$) curves of the TT samples with their numerical predictions obtained using FE computations with the third-order GMR, the 2-term Ogden and the EUCLID-discovered models. Note that the numerical computations are driven by applying Dirichlet boundary conditions derived from the DIC measurements obtained during the tests. This choice was made to enable a direct comparison between measured and simulated response while minimizing additional modeling uncertainties. These boundary conditions are synchronized with the machine displacement recorded by the universal testing machine, which is used to represent the applied displacement history. All FE analyses presented in the remainder of this paper follow this procedure; further details about the numerical computations are reported in \ref{sec:FE}. 
Results are given in Fig.~\ref{fig:UT_PS_GLOBAL_Prediction} where, for better readability, we separate the specimens with circular from those with elliptical holes. 
Although all three retained models accurately capture the UT and PS global responses, their abilities to predict the unseen sample responses show some (mild) differences. In nearly all cases, the lowest errors are obtained by EUCLID, followed by the 2-term Ogden and then by the third-order GMR model. 
%For the simpler geometries (TTa, TTb, TTc), the 2-term Ogden model (Fig.~\ref{fig:UT_PS_GLOBAL_Prediction}c1) shows lower $\mathcal{L}^2(F)$ errors ($0.03\%$, $0.08\%$, and $0.17\%$ for TTa, TTb, and TTc, respectively) compared to the third-order GMR models (Fig.~\ref{fig:UT_PS_GLOBAL_Prediction}a1,b1). However, the model identified using the EUCLID approach outperforms all of them (Fig.~\ref{fig:UT_PS_GLOBAL_Prediction}d1), achieving errors of $0.03\%$, $0.04\%$, and $0.09\%$ for TTa, TTb, and TTc, respectively. 
For the more complex geometries (TTd, TTe, TTf), all models experience increased difficulty in matching the experimental data than for the simplest geometries, likely due to larger local strain variations and experimental noise.

\begin{figure}[]
    \centering
    \includegraphics[width=1\linewidth]{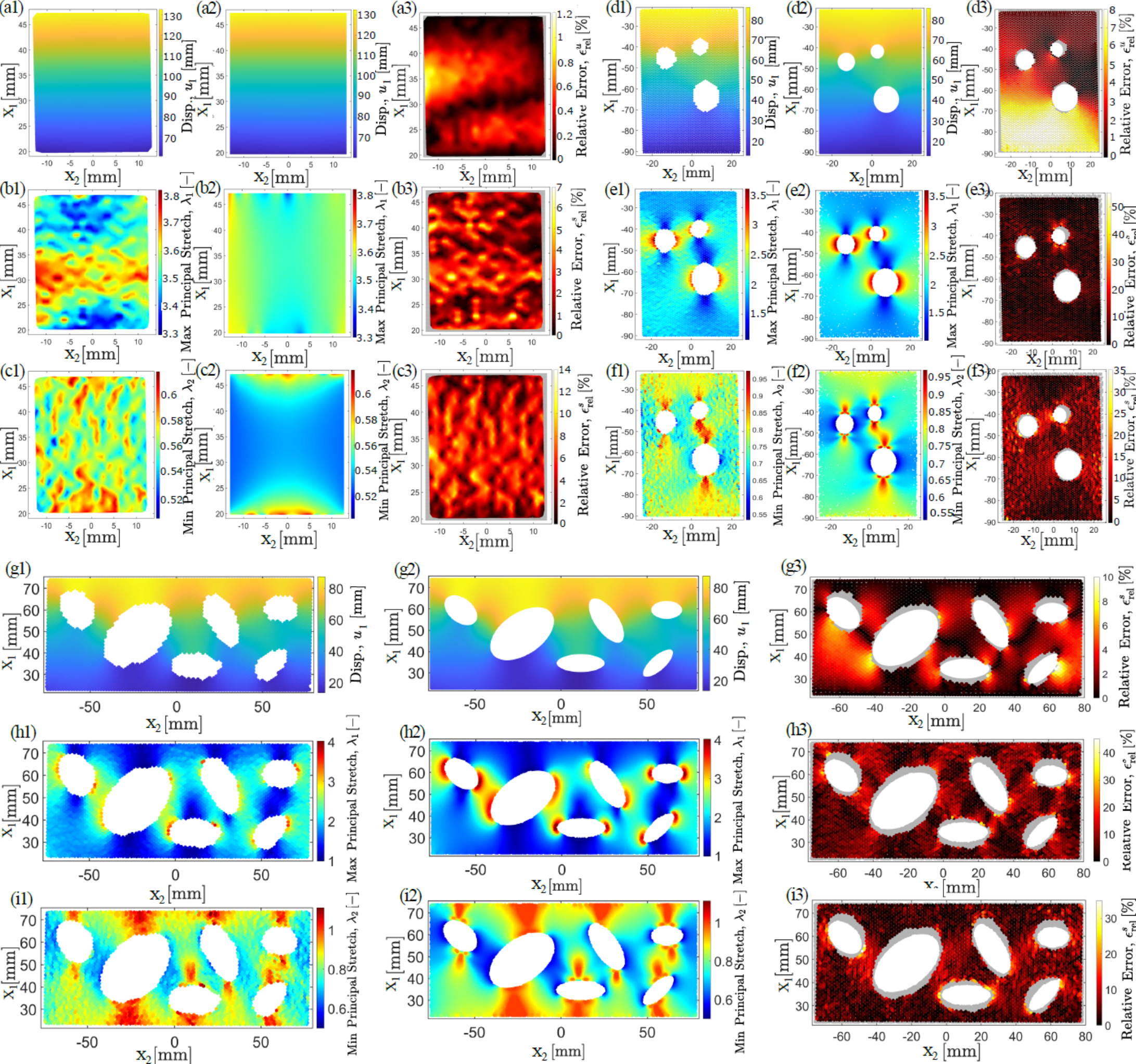}
    \caption{Comparison of experimental results and FE predictions for displacement and strain data obtained with the model discovered by EUCLID using UT and PS global data. Each panel corresponds to a different specimen type: UT (a–c), TTc (d–f), and TTf (g–i). The first row in each set (a, d, g) represents the displacement field in the $X_1$-directions, $u_1$. The second row (b, e, h) shows the maximum principal stretch field, $\lambda_1$, while the third row (c, f, i) presents the minimum principal stretch field, $\lambda_2$. For each dataset, the first column (a1, b1, c1, ...) displays the experimental data, the second column (a2, b2, c2, ...) the FE prediction, and the third column (a3, b3, c3, ...)  the corresponding relative error field, $\epsilon_{rel}$. Gray areas in the error maps indicate regions of the FE mesh where experimental measurements are unavailable, highlighting areas without measurable deviations.}
    \label{fig:UT_PS_GLOBAL_Prediction_local}
\end{figure}

We finally assess the accuracy of the model discovered with EUCLID in predicting the local behavior of the specimens. Fig.~\ref{fig:UT_PS_GLOBAL_Prediction_local} compares the experimental and FE predictions of the first displacement component  $u_1$  and of the maximum and minimum in-plane principal stretches, $\lambda_1$ and $\lambda_2$, for three representative specimens, namely UT, TTc, and TTf. Within each specimen group and for each of the investigated quantities, we illustrate the experimental field, its numerical counterpart and the heat map of the local relative error between the two.
%$\eps_{rel}(u_1(\bs{X}))$ (Figs.~\ref{fig:UT_PS_GLOBAL_Prediction_local}a3,d3,g3). The second row shows the experimental maximum principal stretch $\lambda_1$ (Figs.~\ref{fig:UT_PS_GLOBAL_Prediction_local}b1,e1,h1), the predicted one (Figs.~\ref{fig:UT_PS_GLOBAL_Prediction_local}b2,e2,h2) and the heat map of the local error $\eps_{rel}(\lambda_1(\bs{X}))$. Similarly, the third row shows the experimental values of $\lambda_2$ (Figs.~\ref{fig:UT_PS_GLOBAL_Prediction_local}c1,f1,i1), the predicted one  (Fig.~\ref{fig:UT_PS_GLOBAL_Prediction_local}c2,f2,i2) and the heat map of the error $\eps_{rel}(\lambda_2(\bs{X}))$  (Figs.~\ref{fig:UT_PS_GLOBAL_Prediction_local}c3,f3,i3). 
Overall, the reasonable agreement in displacement and principal stretches indicates that the discovered model accurately captures also the local behavior across a variety of geometries, from the simple UT specimen to the more complex and unseen TTf geometry. As expected, the errors are lower for the UT sample, which is used for calibration, compared to the unseen and more intricate TTc and TTf geometries. The highest errors are primarily localized near the hole boundaries. This is likely due to two distinct factors. First, the DIC precision close to the boundaries decreases and, second, the stress state is markedly multiaxial, a condition not involved in the data used for discovery. Away from the holes, the error is much lower, with isolated peaks compatible with oscillations due to the measurement uncertainty.
%
%The average relative errors for these specimens are low, with the minimum values found for the UT sample $\overline{\epsilon_{\mathrm{rel}}} = 0.28 \pm 0.22\%$ for $u_1$, $2.13 \pm 1.48\%$ for $\lambda_1$, and $4.57 \pm 2.64\%$ for $\lambda_1$. While errors increase slightly for the more intricate, these discrepancies  or other geometric discontinuities. These results underscore the robustness of the EUCLID framework in predicting both displacement and local stretch fields. Although full-field measurements are generally more sensitive to kinematic effects than to the details of the constitutive model, the strong correlation between experimental data and FE predictions further highlights EUCLID’s ability to capture complex mechanical behavior with high fidelity.
%

We conducted an additional study (not shown here) where we compare the results obtained using the full dataset with those obtained through a uniform subsampling involving one half and $1/12$ of the full dataset. Although the identified parameters vary slightly, in all cases EUCLID detects a material model with similar fitting accuracy. This comparison suggests that the temporal resolution does not have a strong effect on the results obtained by EUCLID, at least in the analyzed data range.

%
%%%%%%%%%%%%%%%%%%%%%%%%%%%%%%%%%%%%%%%%%%%%%%%%%%%%%%%%%%%%%%%%%%%%%%%%%%%%%%%%%%
\subsection{Material model identification and discovery using local data from UT and PS tests}
\label{sec: UT-PS-local}
In the following, we assess the performance of classical parameter identification and EUCLID starting from local data (full-field displacement and reaction force measurements), using the approach in Sect.~\ref{sct:unsup_id}. %Unlike the supervised approach, which relies on explicit stress-strain data pairs, the original EUCLID method uses . 
In this section, we start by using the local data from the UT and PS tests. The models considered for parameter identification are the same of Sect.~\ref{sec: UT-PS-global}. To identify the unknown parameters of an a priori chosen model (except Ogden), we start by excluding from the library \cref{eq:W} all the terms apart from those of the chosen model. This reduced library is then used in \cref{eq:eqb_test} to obtain the objective function, which is minimized using a sequential quadratic programming algorithm (implemented in the MATLAB function \texttt{fmincon}). The non-negativity constraint on the coefficients is enforced by specifying zero as the lower bound for all variables, ensuring that each coefficient remains greater than or equal to zero throughout the optimization. The solver settings include a maximum of 3000 iterations, an optimality tolerance of $10^{-6}$, and a step size tolerance of $10^{-6}$.  

For the Ogden model, the optimization is performed via MATLAB's \texttt{lsqnonlin} trust-region reflective algorithm with bound constraints ($-50 \le \mu, \beta \le 50$), using a step-size tolerance of $10^{-3}$, and limits of 1000 iterations and 5000 function evaluations. This approach allows for the simultaneous identification of coefficients and exponents from local full-field and reaction force data, despite the non-convexity of the problem. To enhance robustness and find the best out of possibly multiple local minima, the optimization is performed 100 times with randomly initialized coefficients. The best solution is then selected based on the lowest value of the objective function.

\begin{figure}[h!]
    \centering
    \includegraphics[width=1\linewidth]{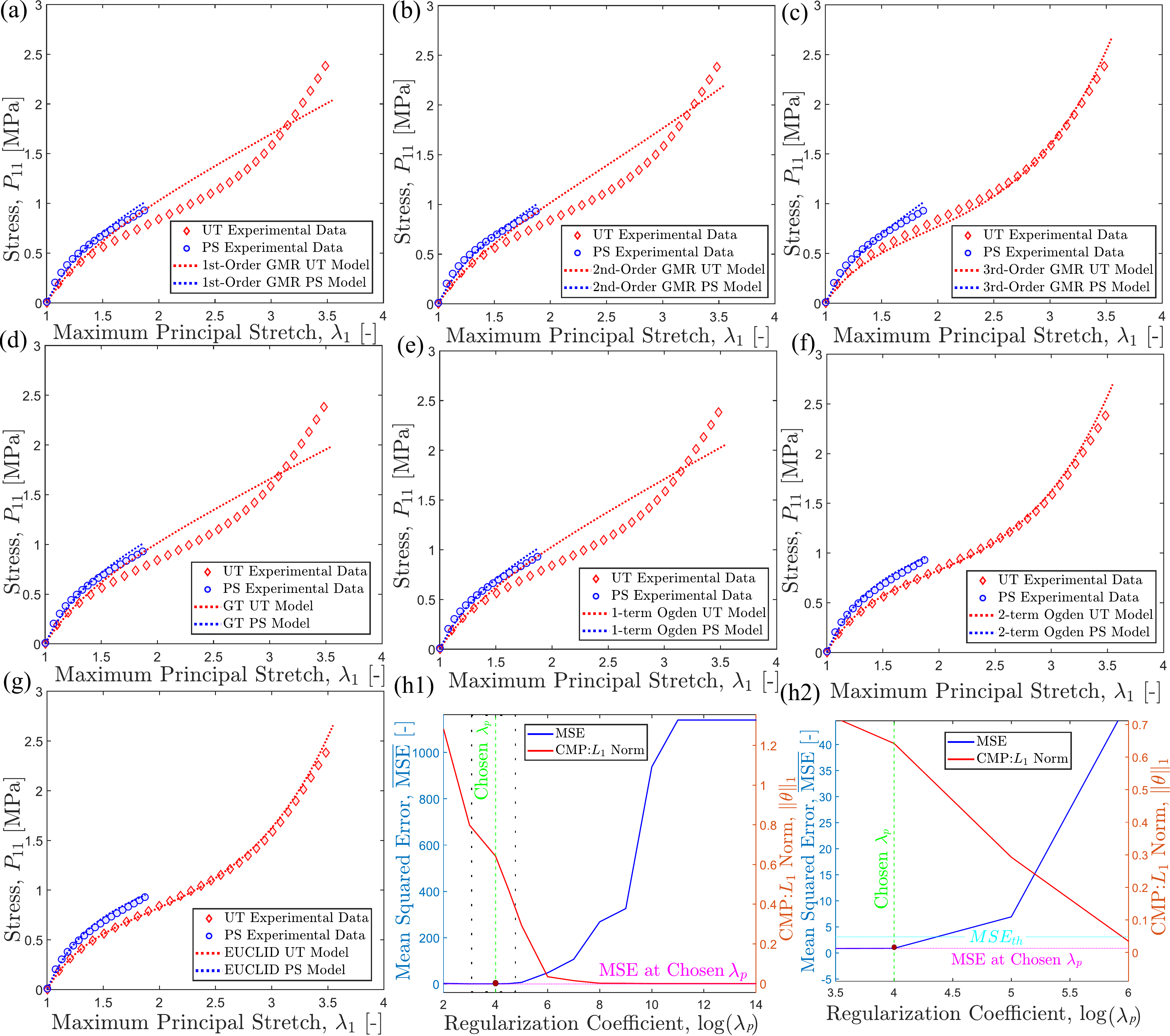}
    \caption{Comparison of the a priori chosen models upon parameter identification and the model discovered by EUCLID, whereby identification/discovery uses combined UT and PS local data. Panels (a)–(g) show the stress-stretch response in UT and PS tests for different material models, compared with experimental data. Specifically, panels (a)–(f) show results for a priori chosen models: (a) 1st-order GMR, (b) 2nd-order GMR, (c) 3rd-order GMR, (d) GT, (e) 1-term Ogden, and (f) 2-term Ogden. Panel (g) shows the response obtained through automated model discovery using EUCLID. (h1) Pareto analysis for the automated selection of the hyperparameter $\lambda_{p}$, showing the MSE and the $L_1$ norm of $\boldsymbol{\theta}$ as functions of $\lambda_{p}$. (h2) A close-up around the automatically selected hyperparameter, with the chosen threshold for MSE and the selected solution indicated by cyan and green dashed lines, respectively.}
    \label{fig:UT_PS_Local}
\end{figure}

The obtained models are summarized in Tab.~\ref{Table3} along with the corresponding $\mathcal{L}^2(P_{11})$ errors for UT and PS tests. In principle, the first comparison to be made would be with the local data, since these are now the data used for identification or discovery. However, to simplify the structure of the presentation, we keep the same sequence of results as in the previous section and start from the comparison between the experimental and predicted $P_{11}-\lambda_{1}$ curves, see Fig.~\ref{fig:UT_PS_Local}a-g. Of the a priori chosen models, now only the third-order GMR and 2-term Ogden models exhibit a good agreement with the experimental data  (Fig.~\ref{fig:UT_PS_Local}c,f). These results are consistent with those obtained using global data, except that in that case also the second-order GMR had an acceptable performance. %Among the models tested, the 2-term Ogden model achieves the best overall agreement with both UT and PS global data, with $\mathcal{L}^2=0.0017$ for UT and $\mathcal{L}^2=0.0025$ for PS.

\begin{table}[ht]
\small  % Reduces the font size
\centering
\caption{Strain energy density of the identified or discovered material models using UT and PS local data.}  % Add your table title here
\label{Table3}  % Add your label here
\begin{tabular}{lp{6.5cm}cc}
  \toprule
  \multirow{ 2}{*}{\textbf{Identification/Discovery}} &  {\textbf{Strain energy density }} & \textbf{UT} & \textbf{PS}\\
  &$W$ [\si{Pa}] & \textbf{$\mathcal{L}^2(P_{11})$} [\%] & \textbf{$\mathcal{L}^2(P_{11})$} [\%]\\
  \midrule
  1st-order GMR & $W = 0.2938(I_1 - 3)$ &2.04\% & 0.48\% \\
  \midrule
  2nd-order GMR & 
  $\begin{aligned}
    W = & 0.2842(I_1 - 3) + 0.0016(I_1 - 3)^2
  \end{aligned}$ & 1.79\% & 0.38\% \\
  \midrule
  3rd-order GMR & 
  $\begin{aligned}
    W = & 0.1354(I_1 - 3) + 0.1528(I_2 - 3) \\
    & + 3.7369 \cdot 10^{-6}(I_1 - 3)^2(I_2 - 3)  \\
    & + 8.6685 \cdot 10^{-5}(I_1 - 3)^2 + 6.5870 \cdot 10^{-4}(I_1 - 3)^3
  \end{aligned}$ & 0.18\% & 0.46\% \\
  \midrule
  GT & $W = 0.2842(I_1 - 3) + 0.0445 \ln(I_2 / 3)$ & 2.03\% & 0.49\% \\
  \midrule
  1-term Ogden & $W = {0.2881}(\lambda_1^{2.0160} + \lambda_2^{2.0160} + \lambda_3^{2.0160} - 3)$ & 2.03\% & 0.48\% \\
  \midrule
  2-term Ogden &  $W = {0.0006}(\lambda_1^{5.9364} + \lambda_2^{5.9364} + \lambda_3^{5.9364} - 3)   +{10.9138}(\lambda_1^{0.3435} + \lambda_2^{0.3435} + \lambda_3^{0.3435} - 3) $& 0.17\% & 0.25\% \\
  \midrule
  EUCLID &
  $\begin{aligned} 
  W = & \ 0.0814(I_1 - 3) + 0.0006(I_1 - 3)^3 \\
  & + 0.0917(\lambda_{1}^{2}+ \lambda_{2}^{2}+ \lambda_{3}^{2}- 3)  + 0.4682 \ln(I_2 / 3)
  \end{aligned}$ & 0.12\% & 0.23\% \\
  \bottomrule
\end{tabular}
\end{table}

Concerning model discovery with EUCLID, for the Pareto analysis we use $\gamma = 0.002$ and vary $\lambda_{p}$ over the range $10^{2}$ to $10^{14}$ in steps of $10^{1}$. The results are illustrated in Figs.~\ref{fig:UT_PS_Local}h1,h2 and the obtained value is $\lambda_{p}=10^4$. Interestingly, the model discovered by EUCLID combines the 2-term Ogden model and the logarithmic term from the GT model, and yields the lowest $\mathcal{L}^2(P_{11})$ error for both UT and PS tests. %The second advantage of EUCLID becomes apparent when comparing the unsupervised EUCLID model with a model extracted using supervised data. Although it might not be entirely fair to compare the unsupervised identified model with a supervised counterpart, as it is already discovered based on the same dataset with which it is compared, the unsupervised model developed by EUCLID achieves a comparable level of accuracy. Despite this, the unsupervised version of the EUCLID framework also successfully predicts the global response for both types of tests, demonstrating its robustness and efficiency in identifying constitutive models. This consistency underscores the effectiveness of the EUCLID approach in discovering accurate constitutive models solely from unsupervised data, reinforcing its potential applicability in experimental scenarios where stress-strain measurements are unavailable.

\begin{figure}[h!]
    \centering
    \includegraphics[width=\linewidth]{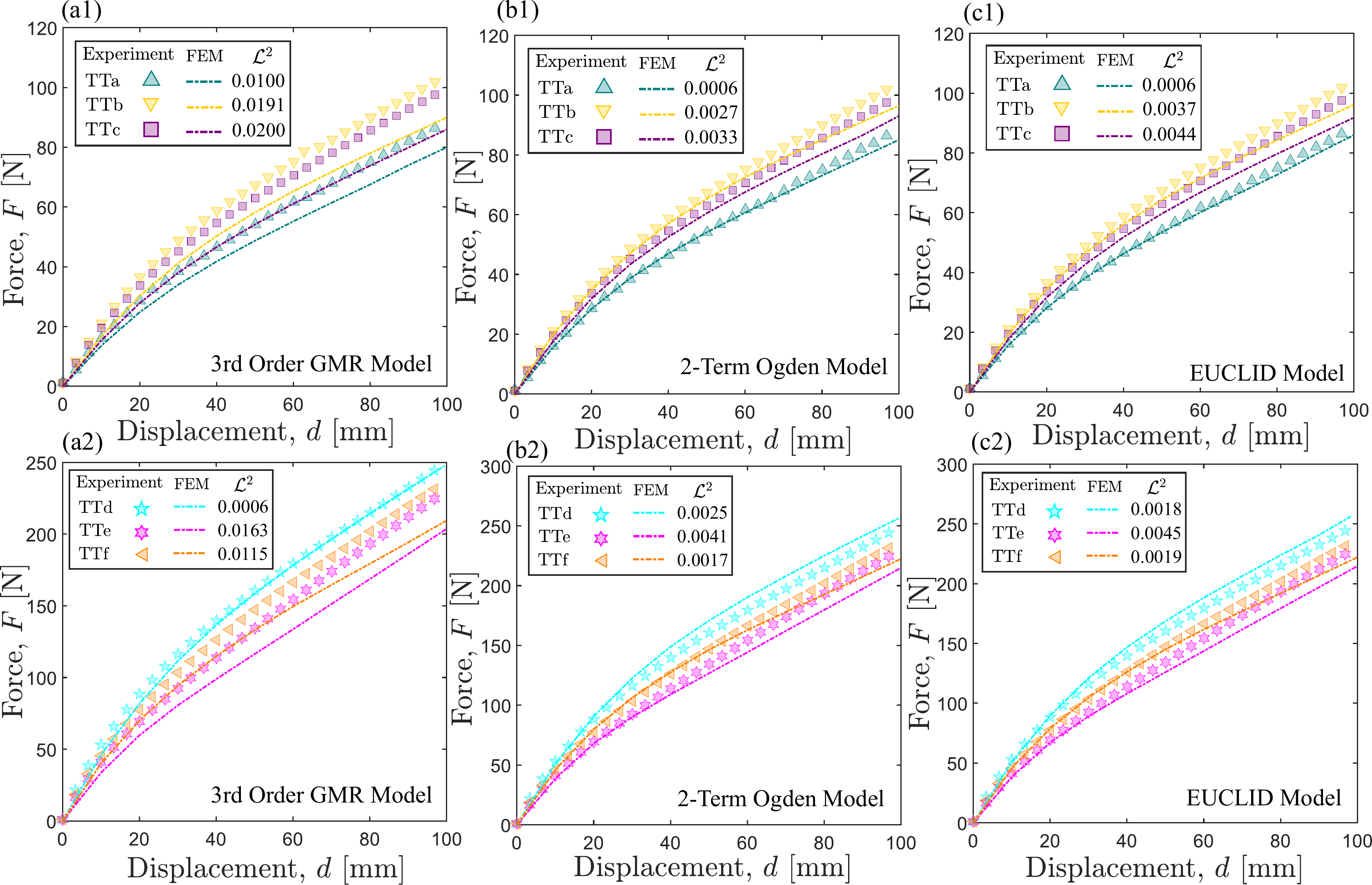}
    \caption{Prediction of the global response of TT sample geometries by the models identified/discovered using local data from UT and PS tests. Experimental force-displacement data are compared with FE predictions from different constitutive models across various sample geometries. Subfigures (a1)–(c1) present results for simpler geometries (TTa, TTb, TTc), while subfigures (a2)–(c2) show results for more complex geometries (TTd, TTe, TTf). The models include the 3rd-order GMR model (a1, a2), the 2-term Ogden model (b1, b2), and the EUCLID-discovered model (c1, c2). Experimental data points are shown as markers and FE predictions as dashed lines.}
    \label{fig:UT_PS_Local_Prediction}
\end{figure}

We now assess the capability of the best identified models and of the discovered model to predict the global force-displacement ($F-\delta_1$) curves of previously unseen tests, see Figs.~\ref{fig:UT_PS_Local_Prediction}a1,b1,c1 and Figs.~\ref{fig:UT_PS_Local_Prediction}a2,b2,c2  for the specimens containing circular and  elliptical holes, respectively.
Here the 2-term Ogden and the EUCLID-discovered models deliver quite reasonable predictions for both sets of geometries, whereas the 3rd-order GMR model is less accurate. Similar considerations as in Sect.~\ref{sec: UT-PS-global} apply in this case.

% The best performance is observed for the TTa sample with a prediction error of $\mathcal{L}^2(F) = 0.0006$, while the largest error occurs for TTe with $\mathcal{L}^2(F) = 0.0045$. EUCLID performs particularly well for TTa, TTd, and TTf, but its prediction for TTe shows a noticeable underestimation of stiffness at higher displacements, indicating limitations in capturing the full nonlinear response for certain geometries. The third-order generalized Mooney-Rivlin model (a1-a2) consistently underestimates the force response, making discrepancies more pronounced in more complex geometries (a2). The 2-term Ogden model (b1–b2) offers improved performance over GMR, especially in capturing nonlinear stiffening. However, it still exhibits visible deviations, particularly at larger displacements. Overall, while EUCLID shows improved generalization and robustness compared to fixed-form models, its accuracy is not uniform across all cases. These results highlight the importance of model flexibility and the need for continued refinement when applying data-driven frameworks to more complex loading conditions and geometries.

\begin{figure}[h!]
    \centering
    \includegraphics[width=\linewidth]{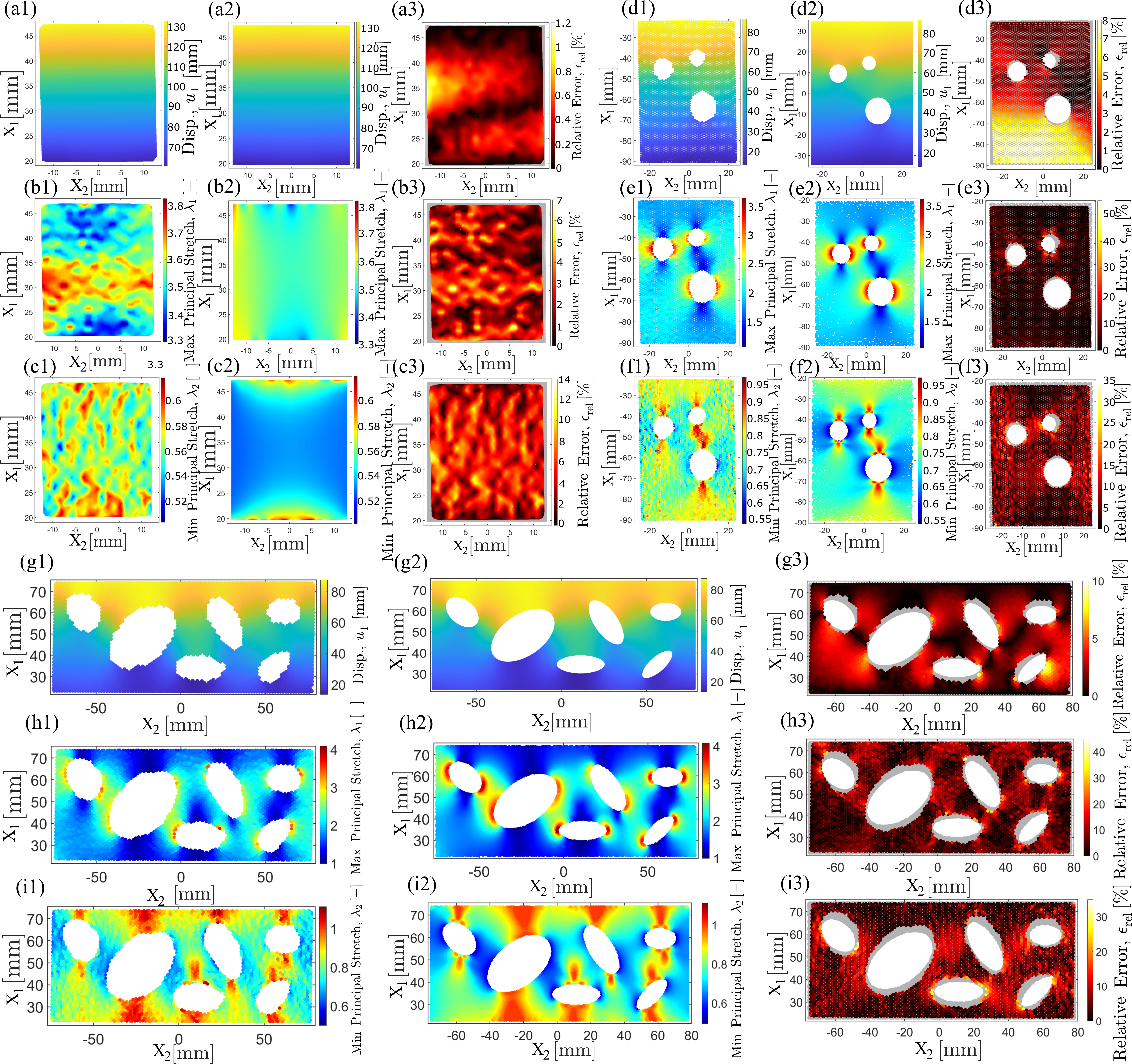}
    \caption{Comparison of experimental results and FE predictions for displacement and strain data obtained with the model discovered by EUCLID using UT and PS local data. Each panel corresponds to a different specimen type: UT (a–c), TTc (d–f), and TTf (g–i). The first row in each set (a, d, g) represents the displacement field in the $1$-direction, $u_1$. The second row (b, e, h) shows the maximum principal stretch field, $\lambda_1$, while the third row (c, f, i) presents the minimum principal stretch field, $\lambda_2$. For each dataset, the first column (a1, b1, c1, ...) displays the experimental data, the second column (a2, b2, c2, ...)  the FE prediction, and the third column (a3, b3, c3, ...)  the corresponding relative error field, $\epsilon_{rel}$. Gray areas in the error maps indicate regions of the FE mesh where experimental measurements are unavailable, highlighting areas without measurable deviations.}
    \label{fig:UT_PS_Local_Prediction_local}
\end{figure}

The ability of the model discovered by EUCLID to reproduce the local response of the material is assessed in Fig.~\ref{fig:UT_PS_Local_Prediction_local} for three representative specimen geometries (UT, TTc, and TTf), using the same presentation scheme as in the previous section. %There, the experimental data are compared with the results of FE analyses obtained using the identified model. For each specimen, the first row shows the experimental and numerical $u_1$ displacement fields respectively in Figs.~\ref{fig:UT_PS_Local_Prediction_local}a1,d1,g1 and Figs.~\ref{fig:UT_PS_Local_Prediction_local}a2,d2,g2, while the heat maps of the local relative $\eps_{rel}(u_1(\bs{X}))$ is shown in Figs.~\ref{fig:UT_PS_Local_Prediction_local}a3,d3,g3. The second row shows the experimental maximum principal stretch $\lambda_1$ (Figs.~\ref{fig:UT_PS_Local_Prediction_local}b1,e1,h1), the predicted one (Figs.~\ref{fig:UT_PS_Local_Prediction_local}b2,e2,h2) and the heat map of the local error $\eps_{rel}(\lambda_1(\bs{X}))$ (Figs.~\eqref{fig:UT_PS_Local_Prediction_local}b3,e3,h3). Similarly, the third row shows the experimental values of $\lambda_2$ (Figs.~\ref{fig:UT_PS_Local_Prediction_local}c1,f1,i1), the predicted one  (Figs.~\ref{fig:UT_PS_Local_Prediction_local}c2,f2,i2) and the heat map of the error $\eps_{rel}(\lambda_2(\bs{X}))$  (Figs.~\ref{fig:UT_PS_Local_Prediction_local}c3,f3,i3). 
Overall, the good agreement between experimental data and model predictions demonstrates that the EUCLID-discovered model effectively captures the local mechanical response across all tested geometries. Compared to the results obtained in Sect.~\ref{sec: UT-PS-global}, here the local behavior (especially that of the more complex specimens) is more accurate, which is not too surprising since the objective function of the optimization is based on local data. %The obtained results demonstrate thus that the adoption of local data can enhance the accuracy of the predictions.

%%%%%%%%%%%%%%%%%%%%%%%%%%%%%%%%%%%%%%%%%%%%%%%%%%%%%%%%%
\subsection{Material model identification and discovery using local data from UT and TTf tests}
\label{sec: UT-TTf-local}
Finally, we investigate the performance of model identification and discovery based on local data from the UT and TTf results. The UT geometry is selected as a simple test giving a rather homogeneous stress states distribution but allowing for large stretches (Fig.~\ref{fig:heterogenity}d1,d2). Conversely, the TTf specimen offers a diverse pool of stress states but with a limited magnitude (Fig.~\ref{fig:heterogenity}d3,d4). 

\begin{figure}[h!]
    \centering
    \includegraphics[width=\linewidth]{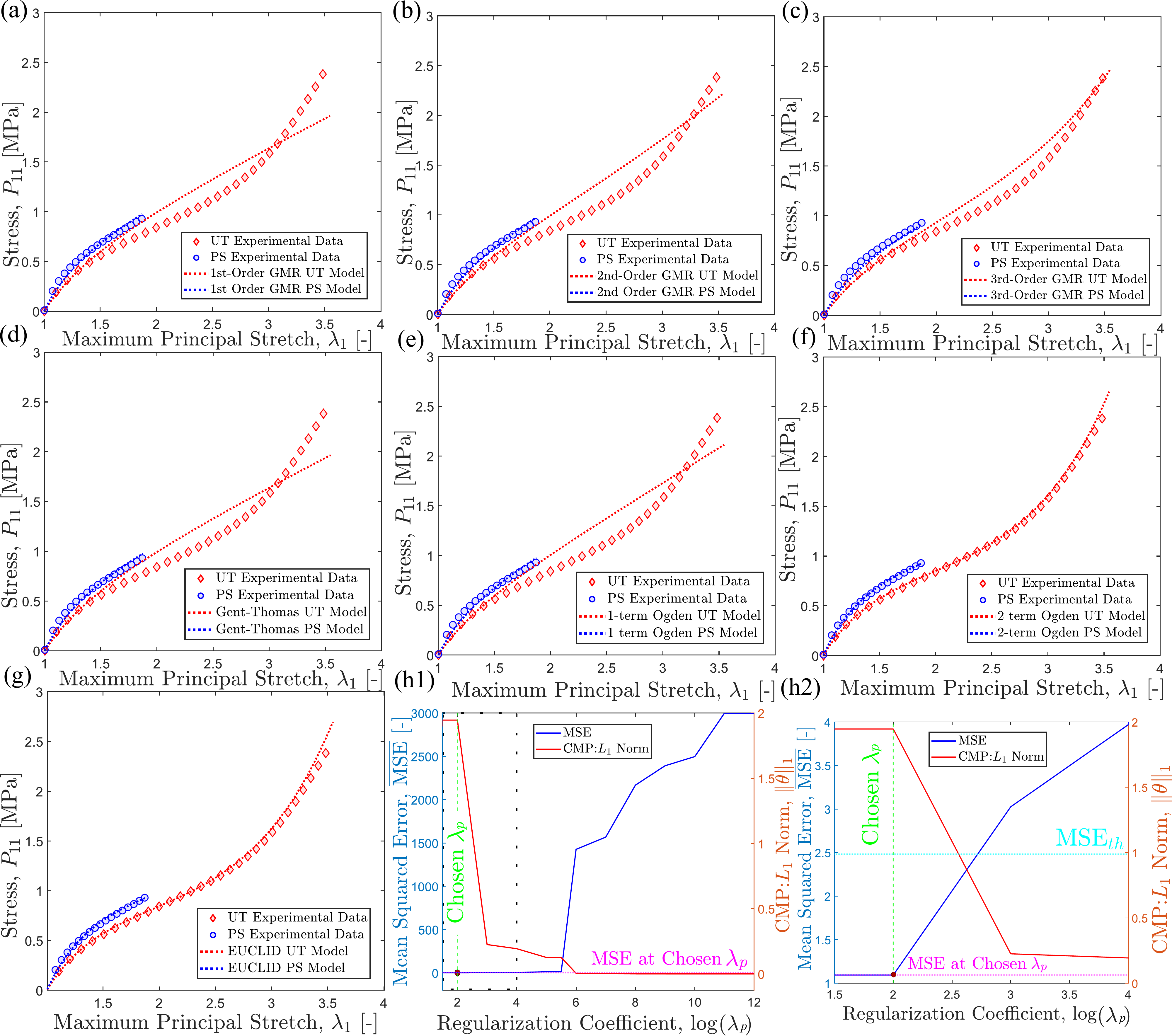}
    \caption{Comparison of the a priori chosen models upon parameter identification and the model discovered by EUCLID, whereby identification/discovery uses combined UT and TTf local data.. Panels (a)–(g) show the stress-stretch response in UT and PS tests for different material models, compared with experimental data. Specifically, panels (a)–(f) show results for a priori chosen models: (a) 1st-order GMR, (b) 2nd-order GMR, (c) 3rd-order GMR, (d) GT, (e) 1-term Ogden, and (f) 2-term Ogden. Panel (g) shows the response obtained through automated model discovery using EUCLID. (h1) Pareto analysis for the automated selection of the hyperparameter $\lambda_{p}$, showing the MSE and the $L_1$ norm of $\boldsymbol{\theta}$ as functions of $\lambda_{p}$. (h2) A close-up around the automatically selected hyperparameter, with the chosen threshold for MSE and the selected solution indicated by cyan and green dashed lines, respectively.}
    \label{fig:UT_TTf_Local}
\end{figure}

\begin{figure}[h!]
    \centering
    \includegraphics[width=0.7\linewidth]{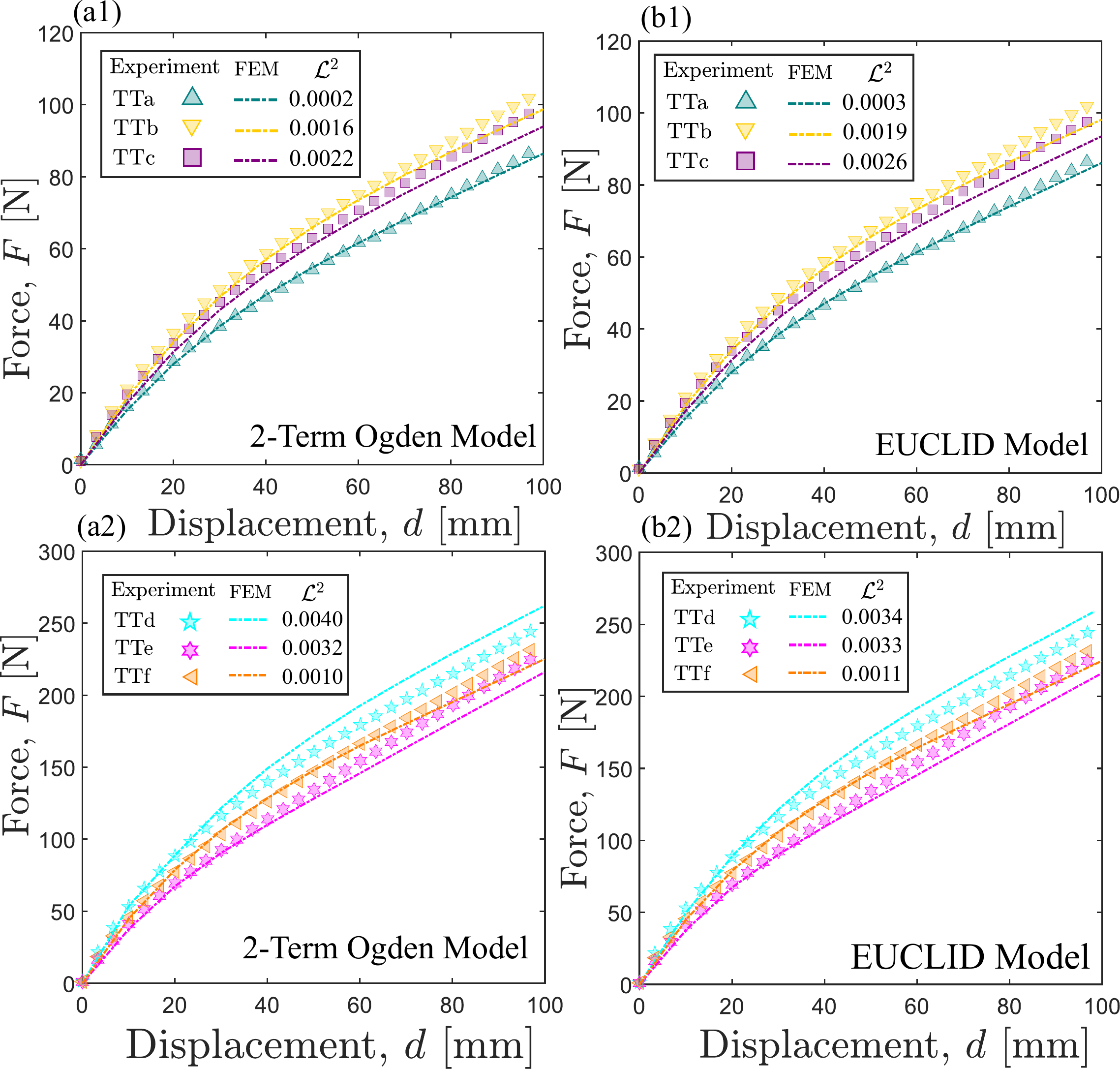}
    \caption{Prediction of the global response of TT sample geometries by the models identified/discovered using local data from UT and TTf tests. Experimental force-displacement data are compared with FE predictions from different constitutive models across various sample geometries. Subfigures (a1) and (b1) present results for simpler geometries (TTa, TTb, TTc), while subfigures (a2) and (b2) show results for more complex geometries (TTd, TTe, TTf). The models include the 2-term Ogden model (a1, a2) and the EUCLID-discovered model (b1, b2). Experimental data points are shown as markers and FE predictions as dashed lines.}
    \label{fig:UT_TTf_Local_Prediction}
\end{figure}

\begin{figure}[h!]
    \centering
    \includegraphics[width=\linewidth]{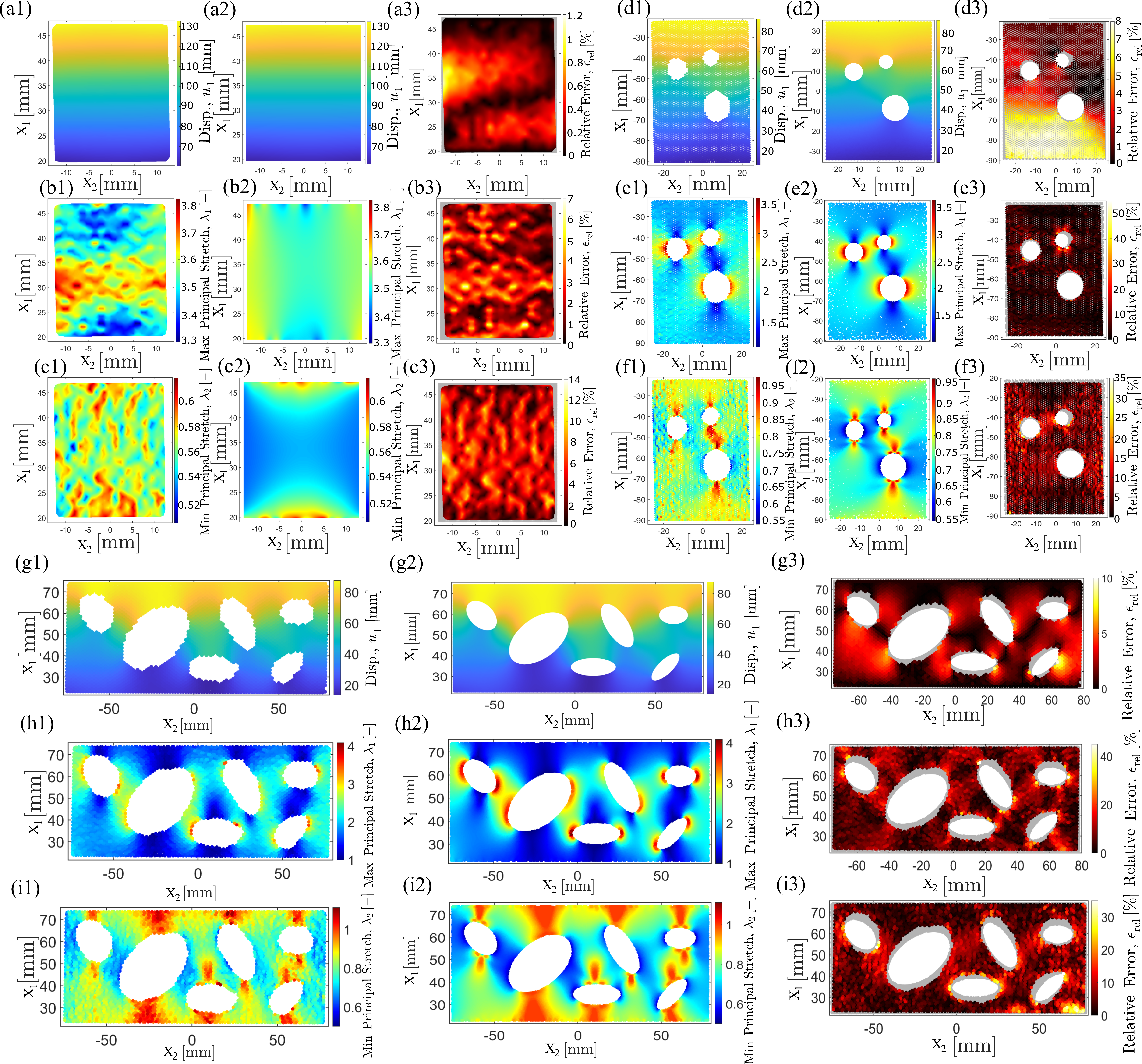}
    \caption{Comparison of experimental results and FE predictions for displacement and strain data obtained with the model  discovered by EUCLID using UT and TTf local data. Each panel corresponds to a different specimen type: UT (a–c), TTc (d–f), and TTf (g–i). The first row in each set (a, d, g) represents the displacement field in the $1$-directions, $u_1$. The second row (b, e, h) shows the maximum principal stretch field, $\lambda_1$, while the third row (c, f, i) presents the minimum principal stretch field, $\lambda_2$. For each dataset, the first column (a1, b1, c1, ...) displays the experimental data, the second column (a2, b2, c2, ...) the FE prediction, and the third column (a3, b3, c3, ...) the corresponding relative error field, $\epsilon_{rel}$. Gray areas in the error maps indicate regions of the FE mesh where experimental measurements are unavailable, highlighting areas without measurable deviations.}
    \label{fig:UT_TTf_Local_Prediction_local}
\end{figure}

The results for this case are shown in Tab.~\ref{Table4} and Figs.~\ref{fig:UT_TTf_Local}-~\ref{fig:UT_TTf_Local_Prediction_local}. Once again, we start from the comparison between the experimental and predicted $P_{11}-\lambda_{1}$ curves. Of the a priori chosen models, now only the 2-term Ogden model exhibits a good agreement with the experimental data. Thus, these results are even more restrictive than those obtained with local data from the UT and PS tests, since in that case also the third-order GMR had an acceptable performance. For EUCLID, we vary $\lambda_{p}$ from $10^{1}$ to $10^{12}$ in increments of $10^{1}$ while setting $\gamma = 0.0005$, and obtain an optimum for $\lambda_{p} = 10^{2}$. Interestingly, EUCLID discovers a 2-term Ogden model with parameters similar, but not identical, to those obtained with conventional parameter identification for the same type of model. This slight discrepancy is due to the discretization of the Ogden exponents in the EUCLID library, whereas conventional identification treats the exponents as continuous parameters. To further assess the robustness of the discovery process, we repeat the EUCLID analysis using a reduced feature library that purposefully excludes the Ogden terms. With no access to Ogden terms, EUCLID is able to discover an alternative model with slightly worse, but still very high predictive accuracy (see the last row in Tab.~\ref{Table4} and the plot in \ref{sec:EUCLID wo Ogden UTTTf}). Since the global response of the UT and PS specimens is accurately predicted only by the 2-term Ogden model and by the EUCLID-discovered model (which is also, for the full library case, of the 2-term Ogden type), only these two models are used to predict the global response of the TT specimens, both obtaining reasonably accurate results. Finally, only the model discovered by EUCLID is assessed on the prediction of the local response of all specimens, giving again reasonably accurate results. 

%Compared to previous benchmarks (Sects.~\ref{sec: UT-PS-global} and \ref{sec: UT-PS-local}), the discrepancy between model predictions and experimental measurements is more pronounced for the assumed models (Figs.~\ref{fig:UT_TTf_Local}a-f and Tab.~\ref{Table4}). Apart for the 2-term Ogden model (Fig.~\ref{fig:UT_TTf_Local}f) that achieves $\mathcal{L}^2(P_{11})$ errors equal to $0.08\%$ for UT and $0.07\%$ for PS, the other constitutive forms fail in reproducing the strain-stiffening behavior of the UT test (Figs.~\ref{fig:UT_TTf_Local}a-e). 
%
\begin{table}[ht]
\small  % Reduces the font size
\centering
\caption{Strain energy density of the identified or discovered material models using UT and TTf local data.}  % Add your table title here
\label{Table4}  % Add your label here
\begin{tabular}{lp{6.5cm}cc}
  \toprule
  \multirow{ 2}{*}{\textbf{Identification/Discovery}} &  {\textbf{Strain energy density }} & \textbf{UT} & \textbf{PS}\\
  &$W$ [\si{Pa}] & \textbf{$\mathcal{L}^2(P_{11})$} [\%] & \textbf{$\mathcal{L}^2(P_{11})$} [\%]\\
  \midrule
  1st-order GMR & $W = 0.2831(I_1 - 3)$ &  1.93\% & 0.28\% \\
  \midrule
  2nd-order GMR & 
  $\begin{aligned}
    W = & 0.2738(I_1 - 3) + 0.0023(I_1 - 3)^2
  \end{aligned}$ & 1.50\% & 0.35\% \\
  \midrule
  3rd-order GMR & 
  $\begin{aligned}
    W = & 0.2630(I_1 - 3) + 0.0003(I_2 - 3)^3
  \end{aligned}$ & 0.81\% &  0.65\% \\
  \midrule
  GT & 
  $\begin{aligned}
  W = 0.2835(I_1 - 3) 
  \end{aligned}$& 1.93\% & 0.28\% \\
  \midrule
  1-term Ogden  & $\begin{aligned}
    & W = {0.2442}(\lambda_1^{2.1241} + \lambda_2^{2.1241} + \lambda_3^{2.1241} - 3) \end{aligned}$ & 1.74\% & 0.35\% \\
   \midrule
  2-term Ogden & $W = {0.0002}(\lambda_1^{6.5881} + \lambda_2^{6.5881}+ \lambda_3^{6.5881} - 3)  
  +{1.4939}(\lambda_1^{0.9109} + \lambda_2^{0.9109} + \lambda_3^{0.9109} - 3) $ & 0.08\% & 0.07\% \\
   \midrule
  EUCLID &
  $\begin{aligned} 
  W = & 1.9458(\lambda_{1}^{0.8}+ \lambda_{2}^{0.8}+ \lambda_{3}^{0.8}- 3) \\
  & + 0.0003(\lambda_{1}^{6.4}+ \lambda_{2}^{6.4}+ \lambda_{3}^{6.4}- 3)
  \end{aligned}$ & 0.16\% & 0.07\% \\
     \midrule
EUCLID w/o Ogden&
 $\begin{aligned} 
  W = & 0.2157(I_1 - 3)+ 0.00004(I_1 - 3)^4+\\
  & + 0.2771 \text{ln}(I_{2}/3)
  \end{aligned}$ & 0.17\% & 0.11\% \\
  \bottomrule
\end{tabular}
\end{table}
%

%The capabilities to generalize to cases not accounted for during calibration is illustrated in Fig.~\ref{fig:UT_TTf_Local_Prediction}. Also in this case, only those models able to reproduce the TU and PS tests are accounted for here, namely the 2-term Ogden model  the one obtained through the EUCLID approach. As expected, the very similar model form and parameters lead to similar results, which, in general, are in good agreement with the experimental evidence for the specimens with both circular holes (Fig.~\ref{fig:UT_TTf_Local_Prediction}a1,b1) and elliptical defects (Fig.~\ref{fig:UT_TTf_Local_Prediction}a2,b2). Comparing the accuracy in Fig.~\ref{fig:UT_TTf_Local_Prediction} with those in Fig.~\ref{fig:UT_PS_GLOBAL_Prediction}, we observe a general improvement except for the TTd specimen for which an overly rigid response in the late stage of the test is observed. However, the obtained predictions are satisfactory, with the model identified by the EUCLID procedure performing slightly better. As aforementioned, the improvement in the accuracy is due to the wider variety of stress states covered by the TTf specimen thanks to the higher number of elliptical holes introduced. Finally, we remark again that the EUCLID approach is able to obtain such level of accuracy without the need to test different functional forms.

%

\subsection{Summary of results}

To complement the results discussed so far, in Table~\ref{tab:synthetic_errors} we report the overall global and local prediction errors for the best performing models obtained with the three dataset options used in the previous subsections. Global performance is quantified by the mean $\mathcal{L}^2$ error of $P_{11}$ averaged across all specimens; local performance is summarized by relative errors in $u_1$, $\lambda_1$, and $\lambda_2$ normalized over all quadrature points and specimens.

When using global UT and PS data, EUCLID attains the lowest global error ($\mathcal{L}^2=0.16\%$), followed by the 2-term Ogden ($0.22\%$) and the 3rd-order GMR ($0.48\%$). For the UT and PS local dataset the 2-term Ogden and EUCLID perform comparably ($\mathcal{L}^2=0.24\%$ and $0.26\%$, respectively), while the 3rd-order GMR shows substantially larger global error ($1.05\%$). With the UT and TTf local data, the 2-term Ogden and EUCLID again achieve nearly identical global errors (respectively $0.17\%$ and $0.19\%$). Across all scenarios the local-field errors in $u_1$, $\lambda_1$ and $\lambda_2$ remain modest and of similar magnitude for the best-performing methods.

A few observations emerge here. First, somewhat surprisingly, a model identified or discovered based on global data has a comparable level of accuracy in local predictions than the same model identified or discovered based on local data. Overall, the lowest prediction errors in terms of both global and local metrics for a given model tend to be obtained when the model is identified or discovered based on the UT+TTf local data, possibly because this dataset offers a combination of a  large range of principal stretches and a large diversity of stress states (Figure \ref{fig:heterogenity}e3). Such diversity could be enhanced by adopting complex loading setups such as biaxial tension or combined tension and torsion, which, however, require dedicated machines not widely available in standard experimental labs. Conversely, the common uniaxial loading test on complex geometries adopted here can be performed with basic equipment and the presented results seem to indicate that the achieved coverage of the state space is enough to satisfactorily discover the constitutive law of natural rubber. In terms of model selection, comparing the results obtained from the three considered datasets (UT+PS global data, UT+PS local data and UT+TTf local data), the UT+TTf local dataset also appears as the most demanding one for model selection, followed by the UT+PS local and then by the UT+PS global datasets. In this hierarchy of datasets, models that perform well with a more demanding dataset also do with the less demanding one(s), but not vice versa. Finally, comparing the performance of the pre-selected models, the 2-term Ogden model emerges as the most robust one, since it performs well for identification from all datasets. However, EUCLID offers two important advantages: on one hand, it matches or outperforms the predictive accuracy of the 2-term Ogden model without requiring an a priori choice of the model form; on the other hand, it is computationally more efficient as it amounts to a convex minimization, whereas fitting a multi-term Ogden law requires nonlinear, non-convex optimization (ideally to be performed multiple times with different initial guesses).

\begin{table}[ht]
\centering
\small
\caption{Synthetic error measures for the best performing models obtained from the datasets used so far. Shown are mean $\mathcal{L}^2$ errors of $P_{11}$ averaged across all specimens, and relative errors for $u_1$, $\lambda_1$, and $\lambda_2$ normalized over all quadrature points and specimens.}
\label{tab:synthetic_errors}
\begin{tabular}{llcccc}
\toprule
\textbf{Data scenario} & \textbf{Model} & $\mathcal{L}^2(P_{11})[\%]$ & $\epsilon_{rel}(u_1$)[\%] & $\epsilon_{rel}(\lambda_1$)[\%] & $\epsilon_{rel}(\lambda_2$)[\%] \\
\midrule
UT and PS global data & 3rd-order GMR  & 0.48 & 3.91  & 3.60  & 5.26  \\
                     & 2-term Ogden   & 0.22  & 3.90  & 3.78  & 4.71  \\
                     & EUCLID         & 0.16  & 4.11  & 3.94  & 5.13  \\
\midrule
UT and PS local data  & 3rd-order GMR  & 1.05  & 3.82  & 3.05  & 5.34  \\
                       & 2-term Ogden   & 0.24  & 4.12  & 3.73  & 4.88  \\
                       & EUCLID         & 0.26  & 3.93  & 3.59  & 5.13  \\
\midrule
UT and TTf local data & 2-term Ogden  & 0.17  & 3.78  & 3.66 & 4.76 \\
                      & EUCLID        & 0.19  & 3.89  & 3.57  & 4.87  \\
\bottomrule
\end{tabular}
\end{table}
%%%%%%%%%%%%%%%%%%%%%%%%%%%%%%%%%%%%%%%%%%%%%%%%%%%%%%%%%%%%%
\section{Conclusions}
\label{sec:con}

We evaluate the performance of EUCLID, a recently introduced framework for the automated discovery of constitutive laws, using experimental data. Mechanical tests are conducted on natural rubber specimens with geometries ranging from simple to complex, providing both global (force–elongation) and local (full-field displacement) measurements. These data are then used to derive constitutive laws through two alternative approaches: (i) the traditional identification of unknown parameters within preselected material models, and (ii) the EUCLID framework, which integrates model selection and parameter identification into a single automated model-discovery process. The obtained results lead to the following main conclusions:
\begin{itemize}
\item{For a given model, the lowest prediction errors—in both global and local metrics—are typically achieved when identification or discovery is based on the UT+TTf local dataset, likely due to its broad range of principal stretches and diverse stress states;}

\item{In terms of model selection difficulty, the datasets can be ranked as follows: UT+TTf local (most demanding), followed by UT+PS local, followed by UT+PS global (least demanding). Models that perform well on a more demanding dataset also perform well on the less demanding ones, but not vice versa;}

\item{Among the pre-selected models, the 2-term Ogden model proves to be the most robust, performing consistently well across all datasets;}

\item{EUCLID also performs consistently well across all datasets. Due to its greater flexibility, it matches or surpasses the predictive accuracy of the 2-term Ogden model without requiring a priori model selection. Moreover, it is computationally more efficient, since it relies on convex minimization, whereas fitting a multi-term Ogden model involves nonlinear, non-convex optimization that must typically be repeated with different initial guesses.}

Overall, the present assessment on complex experimental data confirms the performance of EUCLID as expected by the previous investigations largely based on artificially generated data. 

%\item[-] the EUCLID approach is able to automatically identify accurate, predictive and parsimonious strain energy density functions by selecting relevant features from a large set of possible candidates without requiring multiple trial-and-error calibrations.  

%\item[-] The comparison between experimental and numerical results shows that the models identified using the EUCLID framework are capable of capturing the local and global mechanical response of the material in previously unseen loading conditions. 

%\item[-] Differently than traditional calibration approaches, the model identified using the EUCLID approach is non-trivial in the sense that can combine any term from different classic generalized theories included in the adopted model library. This eliminates the need of the iterative and time consuming process involving the manual selection and calibration of different models and their evaluation to define the most suited to represent the material behavior.

%\item[-] The obtained results highlight how the accuracy of the identified models strongly depends on how well the experimental training dataset is able to cover the state space of the material. In particular, both diversity and magnitude of the stretches reached during the tests are of paramount importance to reliably predict the non-Gaussian hyperelastic regime.
\end{itemize}

 %These findings underscore EUCLID's potential as a versatile, unsupervised, and data-efficient framework for constitutive model discovery. Building on these results, future work will include extending the approach to other material classes, such as elastoplastic and viscoelastic materials, where numerical validation already exists. In addition, the integration of 3D displacement and strain fields obtained from Digital Volume Correlation (DVC) will enable the application of EUCLID to more complex loading scenarios and materials exhibiting three-dimensional behavior.

%%%%%%%%%%%%%%%%%%%%%%%%%%%%%%%%%%%%%%%%%%%%%%%%%%%%%%%%%%%%%%%%%%%%%%%%%%%

\section*{Acknowledgements}
Veit Schönherr (ETH Zurich - Computational Mechanics group) is gratefully acknowledged for his invaluable support in designing and conducting the experimental tests.

\appendix
\section{Finite element simulations}
\label{sec:FE}

FE simulations were carried out using the commercial software \textsc{ABAQUS/STANDARD} to predict the global force-displacement response and the local displacement and stretch fields of samples with complex geometries. 
%While closed-form solutions exist for classical loading modes such as uniaxial tension and pure shear, the complex nature of the present samples necessitated numerical modeling. 
The 2D geometry of each sample was obtained by importing the corresponding CAD model into \textsc{ABAQUS}, and simulations were conducted under the assumption of plane-stress conditions with a constant thickness of 2.5 mm. The material was modeled as hyperelastic, with the constitutive behavior defined via a user-defined material subroutine (\texttt{UHYPER}). This allowed for a flexible definition of strain energy density functions as combinations of classical terms, according to the models discovered by EUCLID.

The computational domain was truncated to match the height of the region of interest used in the DIC measurements. Meshing was performed using four-node quadrilateral dominated elements (a combination of CPS3 and CPS4I elements). A mesh convergence study was conducted to ensure that the results were insensitive to further mesh refinement. It was observed that directly adopting the DIC mesh led to non-convergent or inaccurate results due to insufficient mesh quality and resolution, hence a refined and regularized mesh was employed in the FE analysis. Displacement-controlled loading was applied by prescribing the vertical and horizontal displacements on the top and bottom edges of the ROI. These displacement boundary conditions were extracted from the DIC measurements and mapped onto the FE mesh using interpolation and extrapolation techniques to ensure geometric and kinematic consistency. Dirichlet boundary conditions were applied to enforce the DIC-derived displacements, allowing for a one-to-one comparison between experimental and simulated deformation fields. 
This modeling choice allows for a one-to-one, local comparison between the experimental and simulated deformation> We thus isolate the capability of the approach to discover the constitutive model of the material from additional, and often hard-to-quantify, modeling errors arising from idealized and unrealistic boundary conditions which neglect, e.g., grip geometry, clamp contact/friction, imperfect alignment, out-of-plane motion and small geometric deviations. 
We used the \texttt{General Static} solver.

This simulation setup enabled the evaluation of the identified/discovered models in replicating both the global mechanical response and the local deformation patterns, as captured experimentally through full-field DIC measurements. The reaction force, corresponding to the force measured experimentally by the testing machine, was computed in the simulations as the sum of the vertical reaction forces at all nodes along the top edge of the ROI. This approach is consistent with the equilibrium condition of the system and enables a direct comparison between simulated and measured global responses.

%%%%%%%%%%%%%%%%%%%%%%%%%%%%%%%%%%%%%%%%%
\section{Local predictions using the 2-term Ogden model}
\label{sec:2TermOgden_local}

\begin{figure}[h!]
    \centering
    \includegraphics[width=\linewidth]{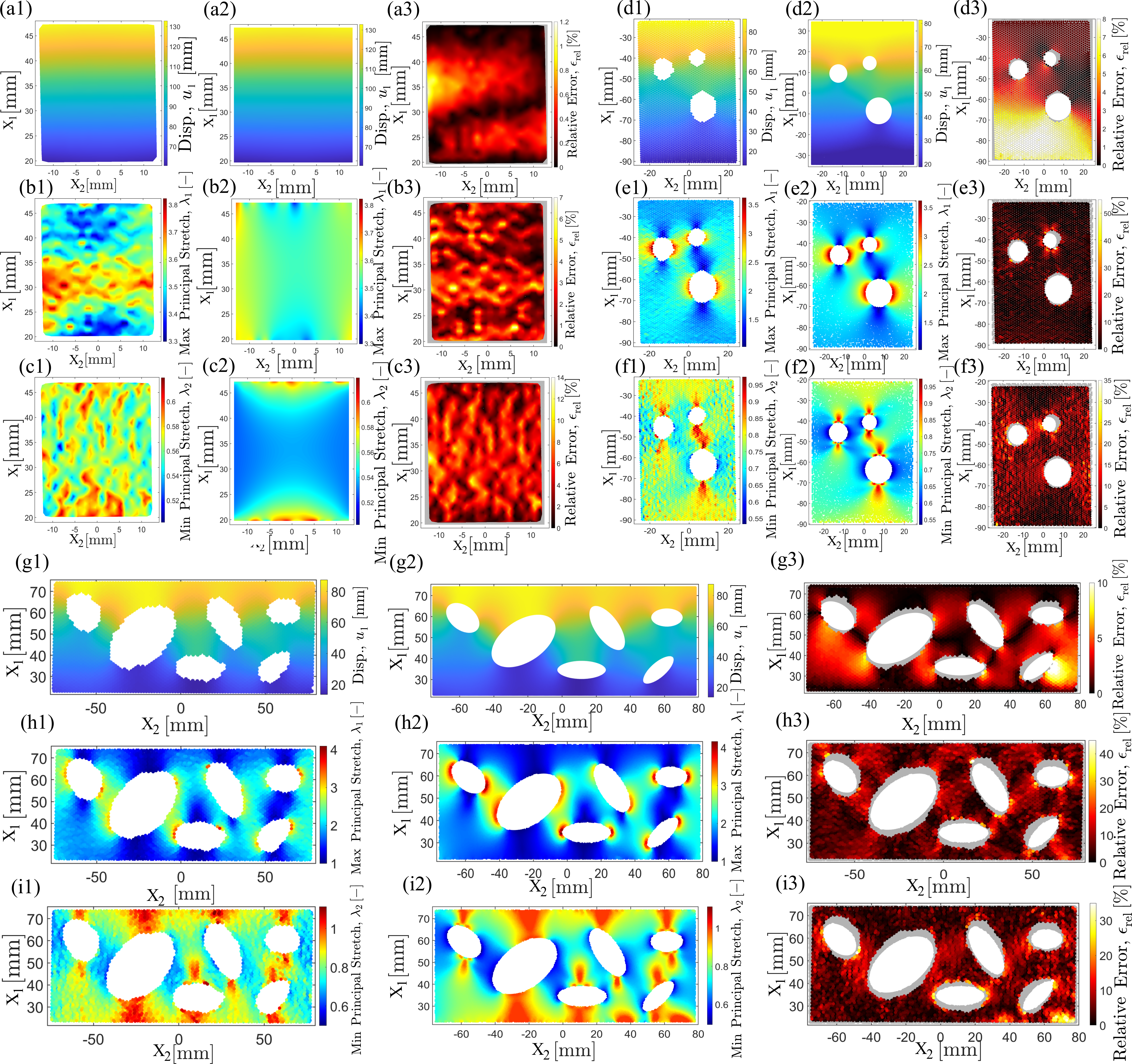}
    \caption{Comparison of experimental results and FE predictions for displacement and strain data obtained with the 2-term Ogden model identified using UT and PS global data. Each panel corresponds to a different specimen type: UT (a–c), TTc (d–f), and TTf (g–i). The first row in each set (a, d, g) represents the displacement field in the $1$-directions, $u_1$. The second row (b, e, h) shows the maximum principal stretch field, $\lambda_1$, while the third row (c, f, i) presents the minimum principal stretch field, $\lambda_2$. For each dataset, the first column (a1, b1, c1, ...) displays the experimental data, the second column (a2, b2, c2, ...) the FE prediction, and the third column (a3, b3, c3, ...) the corresponding relative error field, $\epsilon_{rel}$. Gray areas in the error maps indicate regions of the FE mesh where experimental measurements are unavailable, highlighting areas without measurable deviations.}
    \label{fig:UT_PS_Global_Prediction_local_og2}
\end{figure}

\begin{figure}[h!]
    \centering
    \includegraphics[width=\linewidth]{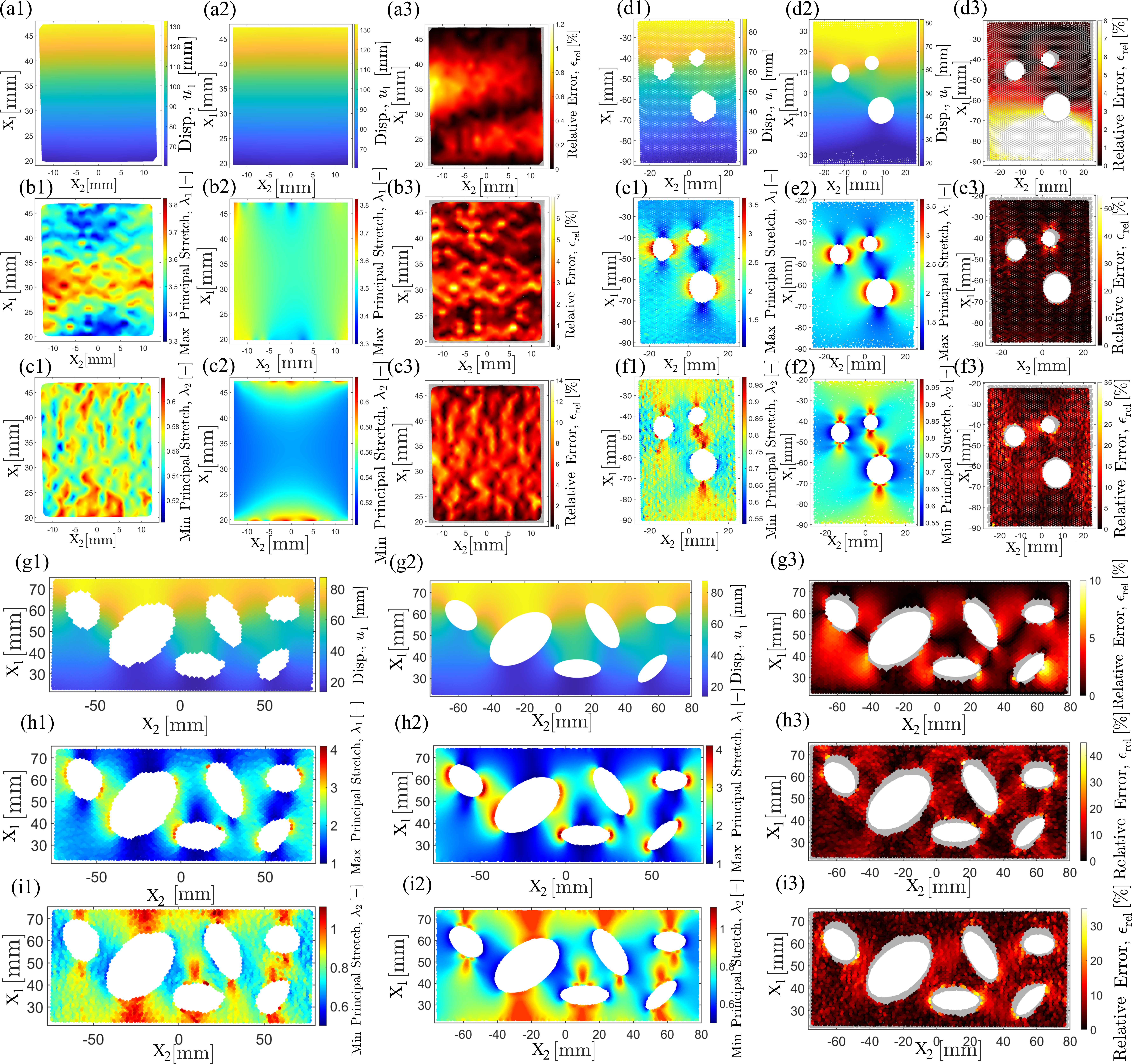}
    \caption{Comparison of experimental results and FE predictions for displacement and strain data obtained with the model discovered by 2-term Ogden using UT and PS local data. Each panel corresponds to a different specimen type: UT (a–c), TTc (d–f), and TTf (g–i). The first row in each set (a, d, g) represents the displacement field in the $1$-directions, $u_1$. The second row (b, e, h) shows the maximum principal stretch field, $\lambda_1$, while the third row (c, f, i) presents the minimum principal stretch field, $\lambda_2$. For each dataset, the first column (a1, b1, c1, ...) displays the experimental data, the second column (a2, b2, c2, ...) the FE prediction, and the third column (a3, b3, c3, ...) the corresponding relative error field, $\epsilon_{rel}$. Gray areas in the error maps indicate regions of the FE mesh where experimental measurements are unavailable, highlighting areas without measurable deviations.}
    \label{fig:UT_PS_Local_Prediction_local_og2}
\end{figure}

\begin{figure}[h!]
    \centering
    \includegraphics[width=\linewidth]{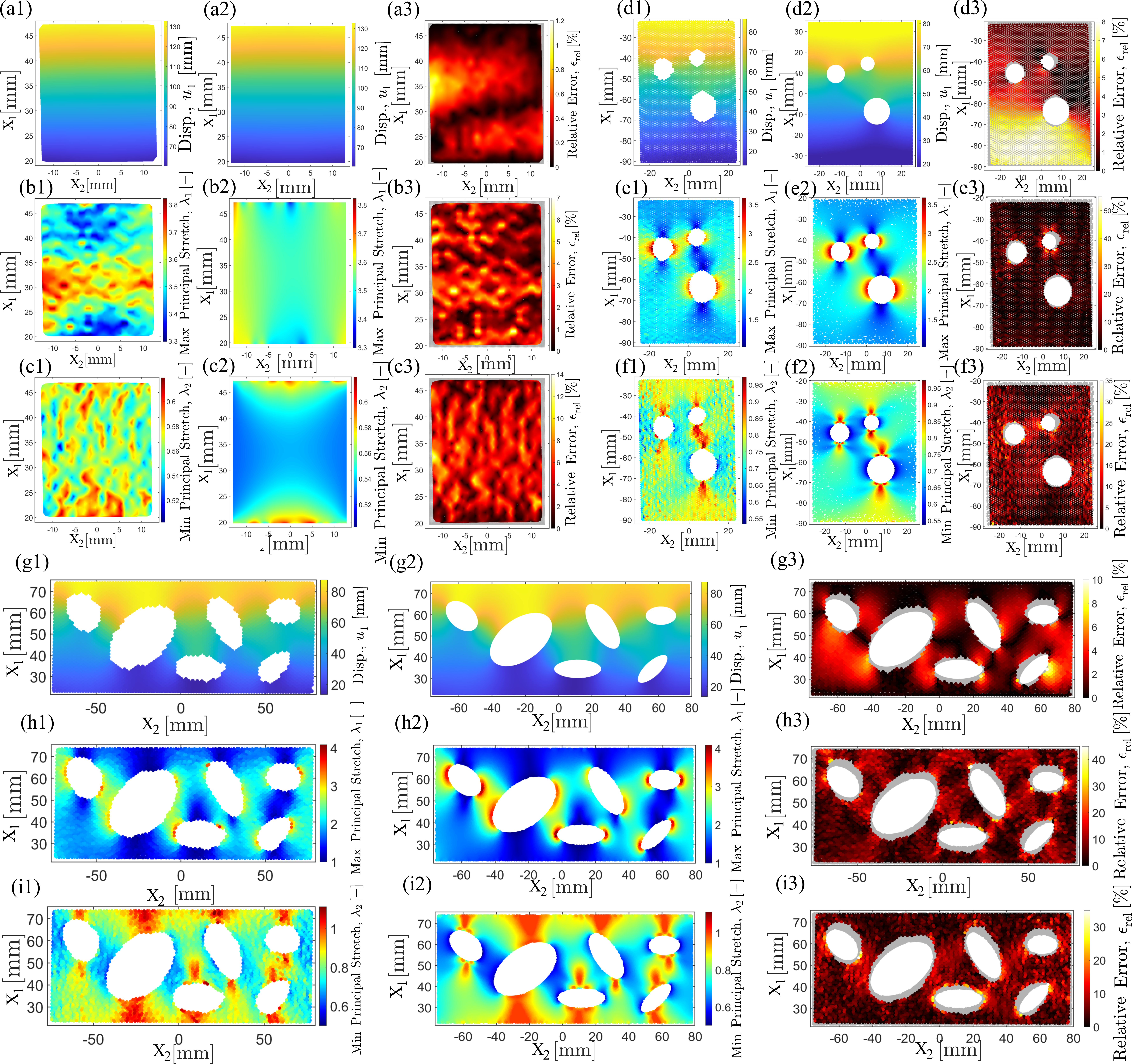}
    \caption{Comparison of experimental results and FE predictions for displacement and strain data obtained with the model discovered by 2-term Ogden using UT and TTf local data. Each panel corresponds to a different specimen type: UT (a–c), TTc (d–f), and TTf (g–i). The first row in each set (a, d, g) represents the displacement field in the $1$-directions, $u_1$. The second row (b, e, h) shows the maximum principal stretch field, $\lambda_1$, while the third row (c, f, i) presents the minimum principal stretch field, $\lambda_2$. For each dataset, the first column (a1, b1, c1, ...) displays the experimental data, the second column (a2, b2, c2, ...) the FE prediction, and the third column (a3, b3, c3, ...) the corresponding relative error field, $\epsilon_{rel}$. Gray areas in the error maps indicate regions of the FE mesh where experimental measurements are unavailable, highlighting areas without measurable deviations.}
    \label{fig:UT_TTf_Local_Prediction_local_og2}
\end{figure}

Since the 2-term Ogden model was found to be the best performing among the chosen fixed models, for completeness, we also evaluate its predictions in terms of local quantities. As follows, the 2-term Ogden models identified using three datasets—(i) UT and PS global data, (ii) UT and PS local data, and (iii) UT and TTf local data—are assessed for their ability to reproduce the local response of some specimens, as shown in \cref{fig:UT_PS_Global_Prediction_local_og2}, \cref{fig:UT_PS_Local_Prediction_local_og2}, and \cref{fig:UT_TTf_Local_Prediction_local_og2}. This assessment follows the same procedure adopted for the EUCLID models in the main text to ensure a consistent comparison, see Figures \ref{fig:UT_PS_GLOBAL_Prediction_local}, \ref{fig:UT_PS_Local_Prediction_local} and \ref{fig:UT_TTf_Local_Prediction_local}. The predicted fields exhibit good agreement with the experimental data, and a level of accuracy comparable to that of EUCLID.

%%%%%%%%%%%%%%%%%%%%%%%%%%%%%%%%%%%%%%%%%%%%%%%%%%%%%%%%%%%%%%%%%%%%%
\section{EUCLID discovery using UT and TTf local data with Ogden terms removed from the library}
\label{sec:EUCLID wo Ogden UTTTf}

\begin{figure}[h!]
    \centering
    \includegraphics[width=0.5\linewidth]{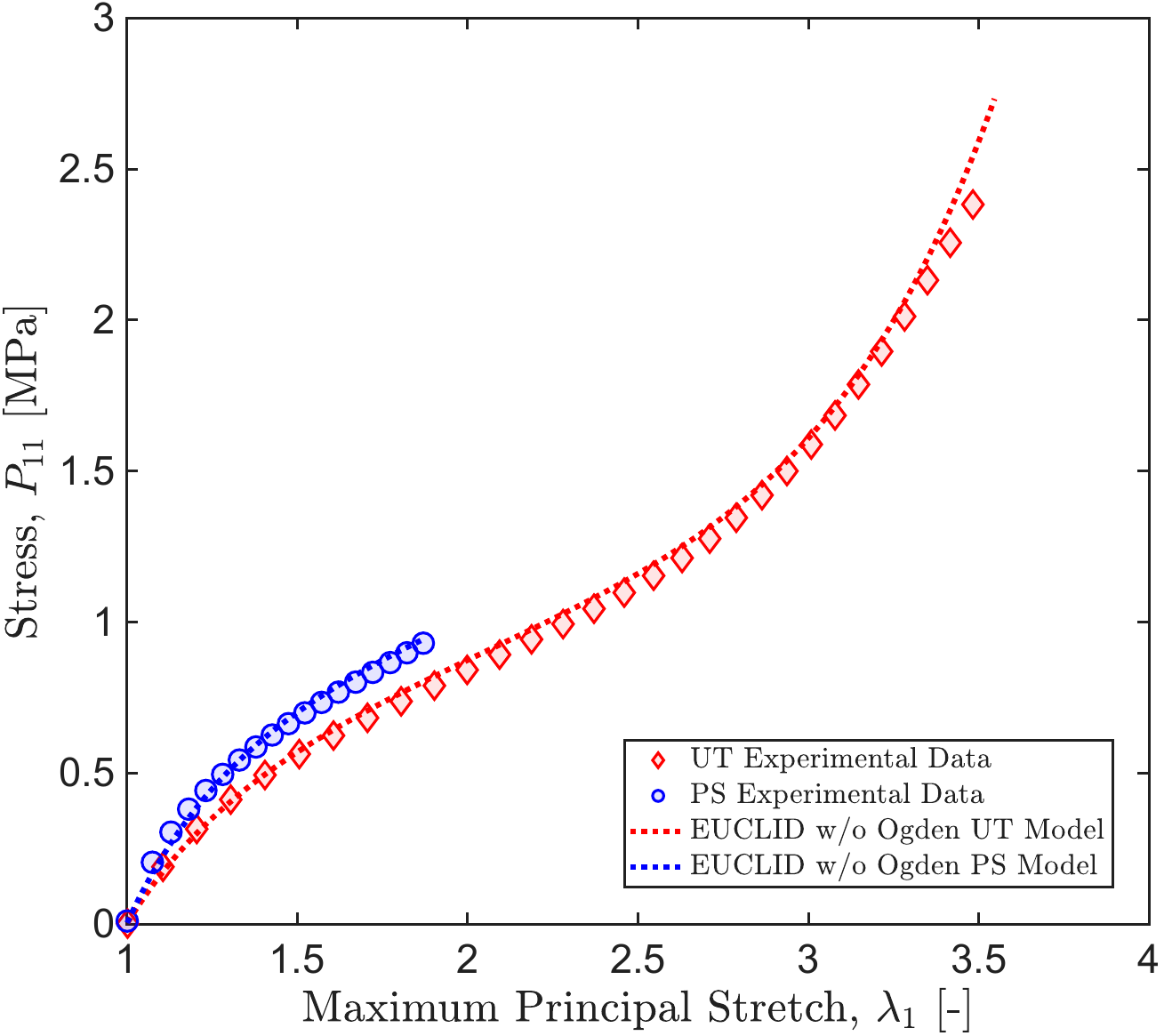}
    \caption{Comparison of stress-stretch responses in UT and PS tests between experimental data and EUCLID predictions obtained from UT and TTf local data, with the Ogden model excluded from the feature library.}
    \label{fig:UTTTf_wo_Ogden}
\end{figure}

The EUCLID discovery is repeated using a reduced feature library in which the Ogden terms are purposefully excluded, relying on UT and TTf local data. In the absence of Ogden terms, EUCLID discovers an alternative constitutive model, which however is capable of reproducing the experimental stress–stretch responses of UT and PS tests with similar accuracy as with the originally discovered 2-term Ogden model, see Fig. \ref{fig:UTTTf_wo_Ogden}. This result highlights the robustness of the discovery process with respect to the choice of library components.

\section*{Ethical Approval} Not applicable.

\section*{Consent to Participate and Consent to Publish} Not applicable.

\section*{Competing Interests} The authors declare that they have no known competing financial interests or personal relationships that could have appeared to influence the work reported in this paper.

\section*{Author Contributions} Arefeh Abbasi: Conceptualization, Data curation,
Formal analysis, Investigation, Methodology,
Software, Visualization, Writing - Original
Draft, Writing - Review and Editing. Maurizio Ricci: Conceptualization, Data curation,
Formal analysis, Investigation, Methodology,
Software, Visualization, Writing - Original
Draft. Pietro Carrara: Conceptualization, Methodology,
Investigation, Writing - Review and Editing. Moritz Flaschel: Conceptualization, Writing - Review and Editing. Siddhant Kumar: Conceptualization, Writing - Review and Editing. Sonia Marfia: Conceptualization, Writing - Review and Editing. 
Laura De Lorenzis: Conceptualization, Methodology, Supervision, Writing - Review and Editing, Funding acquisition.

\section*{Funding} This study was funded by the Swiss National Science Foundation through grant 200021 204316 “Unsupervised data-driven discovery of material laws”.

\section*{Availability of Data and Materials}
The data generated during the current study are available at \url{https://doi.org/10.3929/ethz-c-000788267}, while the relevant codes can be accessed at \url{https://euclid-code.github.io/}.

\bibliography{Bib}

\begin{thebibliography}{10}
\expandafter\ifx\csname url\endcsname\relax
  \def\url#1{\texttt{#1}}\fi
\expandafter\ifx\csname urlprefix\endcsname\relax\def\urlprefix{URL }\fi
\expandafter\ifx\csname href\endcsname\relax
  \def\href#1#2{#2} \def\path#1{#1}\fi

\bibitem{Ogden2004}
R.~W. Ogden, G.~Saccomandi, I.~Sgura, {Fitting hyperelastic models to experimental data}, Computational Mechanics 34~(6) (2004) 484--502.
\newblock \href {https://doi.org/10.1007/s00466-004-0593-y} {\path{doi:10.1007/s00466-004-0593-y}}.

\bibitem{marckmann_comparison_2006}
G.~Marckmann, E.~Verron, \href{https://meridian.allenpress.com/rct/article/79/5/835/93139/Comparison-of-Hyperelastic-Models-for-RubberLike}{Comparison of {Hyperelastic} {Models} for {Rubber}-{Like} {Materials}}, Rubber Chemistry and Technology 79~(5) (2006) 835--858.
\newblock \href {https://doi.org/10.5254/1.3547969} {\path{doi:10.5254/1.3547969}}.
\newline\urlprefix\url{https://meridian.allenpress.com/rct/article/79/5/835/93139/Comparison-of-Hyperelastic-Models-for-RubberLike}

\bibitem{Ricker2023}
A.~Ricker, P.~Wriggers, \href{https://doi.org/10.1007/s11831-022-09865-x}{{Systematic Fitting and Comparison of Hyperelastic Continuum Models for Elastomers}}, Vol.~30, Springer Netherlands, 2023.
\newblock \href {https://doi.org/10.1007/s11831-022-09865-x} {\path{doi:10.1007/s11831-022-09865-x}}.
\newline\urlprefix\url{https://doi.org/10.1007/s11831-022-09865-x}

\bibitem{Mooney1940}
M.~Mooney, {A theory of large elastic deformation}, Journal of Applied Physics 11~(9) (1940) 582--592.
\newblock \href {https://doi.org/10.1063/1.1712836} {\path{doi:10.1063/1.1712836}}.

\bibitem{Rivlin1948}
R.~S. Rivlin, {Large elastic deformations of isotropic materials IV. further developments of the general theory}, Philosophical Transactions of the Royal Society of London. Series A, Mathematical and Physical Sciences 241~(835) (1948) 379--397.
\newblock \href {https://doi.org/10.1098/rsta.1948.0024} {\path{doi:10.1098/rsta.1948.0024}}.

\bibitem{gent_forms_1958}
A.~N. Gent, A.~G. Thomas, \href{http://doi.wiley.com/10.1002/pol.1958.1202811814}{Forms for the stored (strain) energy function for vulcanized rubber}, Journal of Polymer Science 28~(118) (1958) 625--628.
\newblock \href {https://doi.org/10.1002/pol.1958.1202811814} {\path{doi:10.1002/pol.1958.1202811814}}.
\newline\urlprefix\url{http://doi.wiley.com/10.1002/pol.1958.1202811814}

\bibitem{Ogden1972}
R.~W. Ogden, \href{https://royalsocietypublishing.org/doi/10.1098/rspa.1972.0026}{{Large deformation isotropic elasticity – on the correlation of theory and experiment for incompressible rubberlike solids}}, Proceedings of the Royal Society of London. A. Mathematical and Physical Sciences 326~(1567) (1972) 565--584.
\newblock \href {https://doi.org/10.1098/rspa.1972.0026} {\path{doi:10.1098/rspa.1972.0026}}.
\newline\urlprefix\url{https://royalsocietypublishing.org/doi/10.1098/rspa.1972.0026}

\bibitem{avril2008overview}
S.~Avril, M.~Bonnet, A.-S. Bretelle, M.~Gr{\'e}diac, F.~Hild, P.~Ienny, F.~Latourte, D.~Lemosse, S.~Pagano, E.~Pagnacco, F.~Pierron, Overview of identification methods of mechanical parameters based on full-field measurements, Experimental Mechanics 48~(4) (2008) 381--402.

\bibitem{Roux2020}
S.~Roux, F.~Hild, {Optimal procedure for the identification of constitutive parameters from experimentally measured displacement fields}, International Journal of Solids and Structures 184 (2020) 14--23.
\newblock \href {https://doi.org/10.1016/j.ijsolstr.2018.11.008} {\path{doi:10.1016/j.ijsolstr.2018.11.008}}.

\bibitem{fuhg2024review}
J.~N. Fuhg, G.~Anantha~Padmanabha, N.~Bouklas, B.~Bahmani, W.~Sun, N.~N. Vlassis, M.~Flaschel, P.~Carrara, L.~De~Lorenzis, A review on data-driven constitutive laws for solids, Archives of Computational Methods in Engineering (2024) 1--43.

\bibitem{Roemer}
U.~Römer, S.~Hartmann, J.-A. Tröger, D.~Anton, H.~Wessels, M.~Flaschel, L.~De~Lorenzis, Reduced and all-at-once approaches for model calibration and discovery in computational solid mechanics, Applied Mechanics Reviews 77~(4) (2025) 040801.
\newblock \href {https://doi.org/10.1115/1.4066118} {\path{doi:10.1115/1.4066118}}.

\bibitem{kirchdoerfer2016data}
T.~Kirchdoerfer, M.~Ortiz, Data-driven computational mechanics, Computer Methods in Applied Mechanics and Engineering 304 (2016) 81--101.

\bibitem{conti2018data}
S.~Conti, S.~M{\"u}ller, M.~Ortiz, Data-driven problems in elasticity, Archive for Rational Mechanics and Analysis 229 (2018) 79--123.

\bibitem{nguyen2018data}
L.~T.~K. Nguyen, M.-A. Keip, A data-driven approach to nonlinear elasticity, Computers \& Structures 194 (2018) 97--115.

\bibitem{carrara2020data}
P.~Carrara, L.~De~Lorenzis, L.~Stainier, M.~Ortiz, Data-driven fracture mechanics, Computer Methods in Applied Mechanics and Engineering 372 (2020) 113390.

\bibitem{huang2020}
D.~Z. Huang, K.~Xu, C.~Farhat, E.~Darve, \href{https://linkinghub.elsevier.com/retrieve/pii/S0021999120302655}{{Learning constitutive relations from indirect observations using deep neural networks}}, Journal of Computational Physics 416 (2020) 109491.
\newblock \href {https://doi.org/10.1016/j.jcp.2020.109491} {\path{doi:10.1016/j.jcp.2020.109491}}.
\newline\urlprefix\url{https://linkinghub.elsevier.com/retrieve/pii/S0021999120302655}

\bibitem{zhong2022explainable}
X.~Zhong, B.~Gallagher, S.~Liu, B.~Kailkhura, A.~Hiszpanski, T.~Y.-J. Han, Explainable machine learning in materials science, npj computational materials 8~(1) (2022) 204.

\bibitem{thakolkaran2022nn}
P.~Thakolkaran, A.~Joshi, Y.~Zheng, M.~Flaschel, L.~De~Lorenzis, S.~Kumar, Nn-euclid: Deep-learning hyperelasticity without stress data, Journal of the Mechanics and Physics of Solids 169 (2022) 105076.

\bibitem{xu2023small}
P.~Xu, X.~Ji, M.~Li, W.~Lu, Small data machine learning in materials science, npj Computational Materials 9~(1) (2023) 42.

\bibitem{flaschel2021unsupervised}
M.~Flaschel, S.~Kumar, L.~De~Lorenzis, Unsupervised discovery of interpretable hyperelastic constitutive laws, Computer Methods in Applied Mechanics and Engineering 381 (2021) 113852.

\bibitem{flaschel2022discovering}
M.~Flaschel, S.~Kumar, L.~De~Lorenzis, Discovering plasticity models without stress data, npj Computational Materials 8~(1) (2022) 91.

\bibitem{marino2023automated}
E.~Marino, M.~Flaschel, S.~Kumar, L.~De~Lorenzis, Automated identification of linear viscoelastic constitutive laws with euclid, Mechanics of Materials 181 (2023) 104643.

\bibitem{flaschel2023automated}
M.~Flaschel, S.~Kumar, L.~De~Lorenzis, Automated discovery of generalized standard material models with euclid, Computer Methods in Applied Mechanics and Engineering 405 (2023) 115867.

\bibitem{flaschel2023}
M.~Flaschel, Automated discovery of material models in continuum solid mechanics, Ph.D. thesis, ETH Zurich (2023).

\bibitem{joshi2022bayesian}
A.~Joshi, P.~Thakolkaran, Y.~Zheng, M.~Escande, M.~Flaschel, L.~De~Lorenzis, S.~Kumar, Bayesian-euclid: Discovering hyperelastic material laws with uncertainties, Computer Methods in Applied Mechanics and Engineering 398 (2022) 115225.

\bibitem{kissas_language_2024}
G.~Kissas, S.~Mishra, E.~Chatzi, L.~De~Lorenzis, \href{https://linkinghub.elsevier.com/retrieve/pii/S0045782524003098}{The language of hyperelastic materials}, Computer Methods in Applied Mechanics and Engineering 428 (2024) 117053.
\newblock \href {https://doi.org/10.1016/j.cma.2024.117053} {\path{doi:10.1016/j.cma.2024.117053}}.
\newline\urlprefix\url{https://linkinghub.elsevier.com/retrieve/pii/S0045782524003098}

\bibitem{linka_best--class_2024}
K.~Linka, E.~Kuhl, \href{https://linkinghub.elsevier.com/retrieve/pii/S2352431624000610}{Best-in-class modeling: {A} novel strategy to discover constitutive models for soft matter systems}, Extreme Mechanics Letters 70 (2024) 102181.
\newblock \href {https://doi.org/10.1016/j.eml.2024.102181} {\path{doi:10.1016/j.eml.2024.102181}}.
\newline\urlprefix\url{https://linkinghub.elsevier.com/retrieve/pii/S2352431624000610}

\bibitem{flaschel_non-smooth_2025}
M.~Flaschel, T.~Hastie, E.~Kuhl, Non-smooth optimization meets automated material model discovery, arXiv:2507.10196 [cs] (Jul. 2025).
\newblock \href {https://doi.org/10.48550/arXiv.2507.10196} {\path{doi:10.48550/arXiv.2507.10196}}.

\bibitem{urrea}
J.-H. Urrea-Quintero, D.~Anton, L.~De~Lorenzis, H.~Wessels, Automated constitutive model discovery by pairing sparse regression algorithms with model selection criteria (2025).

\bibitem{flaschel2023tissue}
M.~Flaschel, H.~Yu, N.~Reiter, J.~Hinrichsen, S.~Budday, P.~Steinmann, S.~Kumar, L.~De~Lorenzis, Automated discovery of interpretable hyperelastic material models for human brain tissue with euclid, arXiv preprint arXiv:2305.16362 (2023).

\bibitem{holzapfel2002nonlinear}
G.~A. Holzapfel, Nonlinear solid mechanics: a continuum approach for engineering science (2002).

\bibitem{hartmann2001parameter}
S.~Hartmann, Parameter estimation of hyperelasticity relations of generalized polynomial-type with constraint conditions, International Journal of Solids and Structures 38~(44-45) (2001) 7999--8018.

\bibitem{tibshirani_regression_1996}
R.~Tibshirani, \href{http://doi.wiley.com/10.1111/j.2517-6161.1996.tb02080.x}{Regression {Shrinkage} and {Selection} via the {Lasso}}, Journal of the Royal Statistical Society: Series B (Methodological) 58~(1) (1996) 267--288.
\newblock \href {https://doi.org/10.1111/j.2517-6161.1996.tb02080.x} {\path{doi:10.1111/j.2517-6161.1996.tb02080.x}}.
\newline\urlprefix\url{http://doi.wiley.com/10.1111/j.2517-6161.1996.tb02080.x}

\bibitem{treloar1944stress}
L.~R. Treloar, Stress-strain data for vulcanized rubber under various types of deformation, Rubber Chemistry and Technology 17~(4) (1944) 813--825.

\bibitem{jones1975properties}
D.~Jones, L.~Treloar, The properties of rubber in pure homogeneous strain, Journal of Physics D: Applied Physics 8~(11) (1975) 1285.

\bibitem{brown2006physical}
R.~Brown, Physical testing of rubber, Springer Science \& Business Media, 2006.

\bibitem{moreira2013comparison}
D.~Moreira, L.~Nunes, Comparison of simple and pure shear for an incompressible isotropic hyperelastic material under large deformation, Polymer Testing 32~(2) (2013) 240--248.

\bibitem{GOM_ARAMIS}
{GOM GmbH}, Aramis digital image correlation software, \url{https://www.gom.com} (2022).

\bibitem{boyce2000constitutive}
M.~C. Boyce, E.~M. Arruda, Constitutive models of rubber elasticity: a review, Rubber chemistry and technology 73~(3) (2000) 504--523.

\bibitem{steinmann2012hyperelastic}
P.~Steinmann, M.~Hossain, G.~Possart, Hyperelastic models for rubber-like materials: consistent tangent operators and suitability for treloar’s data, Archive of Applied Mechanics 82~(9) (2012) 1183--1217.

\bibitem{staverman1975gaussian}
A.~Staverman, The gaussian chain in the theory of rubber elasticity, in: Journal of Polymer Science: Polymer Symposia, Vol.~51, Wiley Online Library, 1975, pp. 45--56.

\bibitem{sutton2009image}
M.~A. Sutton, J.-J. Orteu, H.~Schreier, Image correlation for shape, motion and deformation measurements: basic concepts, theory and applications (2009).

\bibitem{wang2016theoretical}
Y.~Wang, P.~Lava, P.~Reu, D.~Debruyne, Theoretical analysis on the measurement errors of local 2d dic: part i temporal and spatial uncertainty quantification of displacement measurements, Strain 52~(2) (2016) 110--128.

\bibitem{ghouli_topology_2025}
S.~Ghouli, M.~Flaschel, S.~Kumar, L.~De~Lorenzis, \href{https://linkinghub.elsevier.com/retrieve/pii/S0022509625001863}{A topology optimisation framework to design test specimens for one-shot identification or discovery of material models}, Journal of the Mechanics and Physics of Solids 203 (2025) 106210, publisher: Elsevier BV.
\newblock \href {https://doi.org/10.1016/j.jmps.2025.106210} {\path{doi:10.1016/j.jmps.2025.106210}}.
\newline\urlprefix\url{https://linkinghub.elsevier.com/retrieve/pii/S0022509625001863}

\bibitem{promma2009application}
N.~Promma, B.~Raka, M.~Grediac, E.~Toussaint, J.-B. Le~Cam, X.~Balandraud, F.~Hild, Application of the virtual fields method to mechanical characterization of elastomeric materials, International journal of solids and Structures 46~(3-4) (2009) 698--715.

\bibitem{guelon2009new}
T.~Gu{\'e}lon, E.~Toussaint, J.-B. Le~Cam, N.~Promma, M.~Grediac, A new characterisation method for rubber, Polymer testing 28~(7) (2009) 715--723.

\bibitem{g1996genie}
C.~G'Sell, A.~Coupard, G{\'e}nie M{\'e}canique des caoutchoucs et des {\'e}lastom{\`e}res thermoplastiques, Apollor, 1996.

\bibitem{baaser2013reformulation}
H.~Baaser, C.~Hopmann, A.~Schobel, Reformulation of strain invariants at incompressibility, Archive of Applied Mechanics 83 (2013) 273--280.

\bibitem{treloar1975physics}
L.~G. Treloar, The physics of rubber elasticity (1975).

\end{thebibliography}

\end{document}